%% file: main.tex
\documentclass[sigconf,nonacm]{acmart}

\usepackage{xurl}

\AtBeginDocument{%
  \providecommand\BibTeX{{%
    \normalfont B\kern-0.5em{\scshape i\kern-0.25em b}\kern-0.8em\TeX}}}

\setcopyright{acmcopyright}
\copyrightyear{2022}
\acmYear{2022}
\acmConference[AFT '22]{4th ACM Conference on Advances in Financial Technologies}{September 19--21, 2022}{San Francisco, CA, USA}
\acmBooktitle{4th ACM Conference on Advances in Financial Technologies (AFT '22), September 19--21, 2022, San Francisco, CA, USA}
\acmPrice{15.00}
\acmDOI{10.1145/3558535.3559782}
\acmISBN{978-1-4503-9861-9/22/09}

\input{macros}

\usepackage[group-separator={,}]{siunitx}
\usepackage{tikz}
\usepackage{pgfplots}
\usepackage{booktabs}
\usepackage{hyperref}
\usepackage{algorithm2e}
\usepackage{seqsplit}

\usepackage{caption}
\usepackage{subcaption}

\usepackage{graphicx}

\begin{document}

\author{Rainer St\"u{}tz}
%\email{stuetz@csh.ac.at}
\orcid{0000-0001-9244-1441}
\affiliation{%
  \institution{Complexity Science Hub Vienna}
  \city{Vienna}
  \country{Austria}
}
\author{Johann Stockinger}
\affiliation{%
  \institution{TU Wien}
  \city{Vienna}
  \country{Austria}
}
\author{Pedro Moreno-Sanchez}
\orcid{0000-0003-2315-7839}
\affiliation{%
  \institution{IMDEA Software Institute}
  \city{Madrid}
  \country{Spain}
}
\author{Bernhard Haslhofer}
%\email{haslhofer@csh.ac.at}
\orcid{0000-0002-0415-4491}
\affiliation{%
  \institution{Complexity Science Hub Vienna}
  \city{Vienna}
  \country{Austria}
}
\author{Matteo Maffei}
\affiliation{%
  \institution{TU Wien, Christian Doppler Lab Blockchain Technologies for the Internet of Things}
  \city{Vienna}
  \country{Austria}
}

% \affiliation{%

\renewcommand{\shortauthors}{St\"u{}tz et al.}

\title{Adoption and Actual Privacy of Decentralized CoinJoin Implementations in Bitcoin}
\input{sections/0_abstract}
\keywords{Cryptoassets, Mixing, CoinJoin}
\maketitle              % typeset the header of the contribution
%
\input{sections/1_introduction}
\input{sections/2_background}
\input{sections/3_coinjoin_detection}
\input{sections/4_ecosystem}
\input{sections/5_security_privacy}

\input{sections/6_discussion}

\input{sections/7_conclusions}

\section{Acknowledgments}

This work was partly supported by the Austrian security research programme
KIRAS of the Federal Ministry of Agriculture, Regions and Tourism (BMLRT)
under the project KRYPTOMONITOR (879686);
by grant IJC2019-041599-I/MCIN/AEI/10.13039/501100011033 and European Union
NextGenerationEU/PRTR; by Madrid regional government as part of the program
S2018/TCS-4339 (BLOQUES-CM) co-funded by EIE Funds of the European Union; by
the project High-Assurance Crytography (HACRYPT); by SCUM Project
(RTI2018-102043-B-I00) MCIN/AEI/10.13039/501100011033/ERDF A way of making
Europe; by the European Research Council (ERC) under the Horizon 2020 research
(grant 771527-BROWSEC); by the Austrian Research Promotion Agency (FFG) through
the COMET K1 SBA and COMET K1 ABC; by the Vienna Business Agency through the
project Vienna Cybersecurity and Privacy Research Center (VISP), and by the
Christian Doppler Lab Blockchain Technologies for the Internet of Things
(CDL-BOT).

\bibliographystyle{ACM-Reference-Format}
\bibliography{references}

\appendix

\input{sections/8_appendix}

\end{document}

%% file: sections/0_abstract.tex
% !TeX root = ../main.tex

\begin{abstract}
% Goal
We present a first measurement study on the adoption and actual privacy of two popular decentralized CoinJoin implementations, Wasabi and Samourai, in the broader Bitcoin ecosystem.
% Contributions
	% 1. Coin Join Detection
By applying highly accurate (> 99\%) algorithms we can effectively detect \nWasabiTotal Wasabi and \nSamouraiTotal Samourai transactions within the block range \numrange{530500}{725348} (2018-07-05 to 2022-02-28).
	% 2. Ecosystem analysis
We also found a steady adoption of these services with a total value of mixed coins of \nTotalMixedUSD USD and average monthly mixing amounts of \nWasabiMixedRecentUSD USD) for Wasabi and \nSamouraiMixedRecentUSD USD for Samourai. Furthermore, we could trace \nExchangesReceivedUSDLevelOne USD directly received by cryptoasset exchanges and \nExchangesReceivedUSDLevelTwo USD indirectly received via two hops.
	% 3. Anonymity impact
Our analysis further shows that the traceability of addresses during the pre-mixing and post-mixing narrows down the anonymity set provided by these coin mixing services. It also shows that the selection of addresses for the CoinJoin transaction can harm anonymity. 
% What is the implication of our work
Overall, this is the first paper to provide a comprehensive picture of the adoption and privacy of distributed CoinJoin transactions. Understanding this picture is particularly interesting in the light of ongoing regulatory efforts that will, on the one hand, affect compliance measures implemented in cryptocurrency exchanges and, on the other hand, the privacy of end-users.
\end{abstract}

%% file: sections/1_introduction.tex
% !TeX root = ../main.tex

\section{Introduction}
\label{sec:introduction}

% Motivation | Privacy
Privacy in financial transactions and traceability of funds are two inherently contradictory goals in cryptoasset ecosystems like Bitcoin. On the one hand, we observe the increasing adoption of privacy-focused wallets like Wasabi or Samourai to make transactions from several senders unlinkable by combining them into a single transaction using a well-known coin swapping technique called \emph{CoinJoin}. These wallets fulfill the increasing need for privacy, which is otherwise not given by default in mainstream cryptocurrencies like Bitcoin because it is possible to effectively de-anonymize end-users and trace funds~\cite{Meiklejohn2013,Androulaki2013,Reid2013}. 

% Motivation | Need for traceability
On the other hand, current regulatory efforts expand the traceability of funds requirement to cryptoassets and impose that obligation on virtual asset service providers, like exchanges. For instance, the Financial Action Task Force (FATF)~\cite{FATF:2021} explicitly identifies mixing and tumbling services as risk factors for money laundering and terrorism financing. It also points out that Virtual Asset Service Providers (VASPs) such as cryptoasset exchanges need to apply preventive measures such as customer due diligence and the obligation to obtain, hold, and transmit originator and beneficiary information (``travel rule'') for transactions above a threshold of USD/EUR \num{1000}. Also, the Council of the European Union recently proposed that this regulation applies for transfers to or from ``unhosted wallets'' as long as at least one VASP provider is involved~\cite{EU:2021a}. The target of this proposal, which could become part of the broader European Regulation on Markets in Crypto Assets (MiCA)~\cite{EU:2020a},  includes transfers between privacy-preserving wallets, like Wasabi or Samourai, and virtual asset providers, such as cryptoasset exchanges.

% Goals
However, little empirical evidence is available on the adoption and actual privacy guarantees of CoinJoin wallet implementations when considering them part of the broader Bitcoin ecosystem. Therefore, we would like to examine this area of tension more closely and contribute empirical evidence to the discussion. In particular, this work focuses on the transactions generated by two user-friendly wallet implementations with built-in \emph{decentralized} CoinJoin functionality, Wasabi and Samourai~\cite{Ghesmati:2022b}. 
While the role of \emph{centralized} mixing services like JoinMarket, where a trusted third party matches CoinJoin participants, has been studied in the past~\cite{Moeser:2017a}, decentralized wallet implementations have not yet been the focus of a comprehensive measurement study.

% Specific objectivs & challenges
Given the current regulatory efforts, we are particularly interested in the extent to which transactions generated by such wallets flow into cryptoasset exchanges and how this is evolving in light of ongoing regulatory efforts. We also want to better understand the actual privacy of CoinJoin transactions given the information leaked on-chain, thereby helping users assess the anonymity gain of these services, often smaller than the a priori perceived one. However, before answering these questions, we must detect the transactions created by Wasabi and Samourai wallets as accurately as possible. For example, we could use available threshold heuristics proposed by \Ficsor, the creator of Wasabi wallet~\cite{Ficsor2019,Ficsor2020}. However, these heuristics were defined ad-hoc and never evaluated systematically. Therefore, little is known about their accuracy compared to other approaches. In this state of affairs, developing detection methods that can be assessed systematically, preferably using ground-truth data as input, would be interesting. 

% Contributions, methods, and findings
\textbf{Our contributions}.
% Section 3 | Detection Heuristics
First, as discussed in Section~\ref{sec:detection}, we propose two highly effective algorithmic methods to detect on-chain CoinJoin transactions. For Wasabi, we show that \Ficsor's threshold heuristics and a new machine learning approach trained on historical ground-truth data are both highly accurate (\SI{99}{\percent}) and that the choice between them depends on the application area. For Samourai, we introduce a simple deterministic alternative to known heuristics, which traverses the chain starting from genesis mixes and yields Samourai transactions with \SI{100}{\percent} accuracy. We applied these methods on the Bitcoin blockchain and identified \num{32251} Wasabi and \num{223597} Samourai transactions within the block range \num{530500} (2018-07-05) and \num{725345} (2022-02-28), which we make publicly available for other researchers.

% Section 4 Ecosystem Analysis
Second, we analyzed how the number of transactions and the amount of mixed BTC evolved and found a steady adoption of Wasabi since Nov. 2018 and Samourai since Jan 2020. We also found that the total value of mixed coins is \nTotalMixedBTC BTC (\nTotalMixedUSD USD). Wasabi was used to mix \nWasabiMixedBTC BTC (\nWasabiMixedUSD USD) and Samourai for \nSamouraiMixedBTC BTC (\nSamouraiMixedUSD USD). Recently, the monthly mixing amounts were on average \nWasabiMixedRecentBTC BTC (\nWasabiMixedRecentUSD USD) for Wasabi and \nSamouraiMixedRecentBTC BTC (\nSamouraiMixedRecentUSD USD) for Samourai. Furthermore, we could identify \nWasabiTXsToExchangeLevelOne Wasabi and
\nSamouraiTXsToExchangeLevelOne Samourai transactions that were directly
accepted by exchange entities and \nWasabiTXsToExchangeLevelTwo and
\nSamouraiTXsToExchangeLevelTwo transactions that were accepted indirectly via
two hops. In terms of mixed coins, this sums up to
\nExchangesReceivedBTCLevelOne directly received BTC
(\nExchangesReceivedUSDLevelOne USD) and \nExchangesReceivedBTCLevelTwo
indirectly received BTC (\nExchangesReceivedUSDLevelTwo USD). These results show that acceptance of CoinJoins by cryptoasset exchanges is a living practice and not a phenomenon of the unregulated past.

% Section 5 Anonymity Study
Third, we proposed methods to quantify the gap between the anonymity perceived by users of Wasabi and Samourai wallets and the often smaller anonymity guarantees provided by such services. Unlike in an ideal world
where pre- and post-mixed addresses are fresh and unlinkable to any other, the traceability of addresses during the pre-mixing and post-mixing narrows down the anonymity set provided by these coin mixing services. For instance, for the Wasabi wallet, we observe that the ideally possible anonymity set of almost $75K$ addresses in Jan'21 is largely reduced to less than $25K$. Finally, we also observed biases in the selection of addresses for the CoinJoin transactions themselves, also harming anonymity.

% Implications (summary of discussion section)
Our work has several implications. First, it shows privacy-seeking end-users that third parties can detect CoinJoin transactions generated by Wasabi and Samourai, two popular implementations of decentralized CoinJoin, through relatively simple algorithmic methods. For end-users, this implies that having CoinJoin in the transaction history chain could raise issues when cashing out at cryptocurrency exchanges implementing stringent compliance measures requiring proof of origin of funds. Furthermore, our privacy analysis demonstrates that the anonymity guarantees (i.e., the anonymity set) offered by Wasabi and Samourai is much smaller than expected as effective deanonymization techniques are possible by inspecting pre-mixing and post-mixing transactions. 
%Finally, our work offers a picture of the current use of CoinJoin implementations and an expandable framework for future, more comprehensive analyses.
%Our anonymity analysis also sheds light on the anonymity losses suffered by participants of Wasabi and Samourai 
%wallets due to the traceability of their pre-mixed and post-mixed addresses. \matteo{I would drop these two last paragraphs, as they do not say much and are repetitive}

% Reproducibility
Our CoinJoin detection algorithms and analytics procedures were implemented in Python 3.7, Apache Spark 3.1.2/Scala 2.12, and \textsf{R} 4.2.0, respectively. We make our scripts along with the collected data available for further research in the following GitHub repository: \url{\giturl}

%% file: sections/2_background.tex
% !TeX root = ../main.tex

\section{Background and Related Work}
\label{sec:background}

\subsection{Coin Swapping, Mixing Services, and CoinJoin}

Swapping coins of different users is a well-known method to disrupt existing de-anonymization methods, such as the multiple-input or co-spent heuristics~\cite{Meiklejohn:2013aa}, and complicates the traceability of funds. In the early days of Bitcoin, centralized mixers such as ``Bitcoin Fog'' or ``Bitmixer'' offered coin swapping as a service.

\begin{figure*}
    \centering
    \begin{subfigure}{.5\textwidth}
      \centering
      \resizebox{.8\columnwidth}{!}{
        \includegraphics{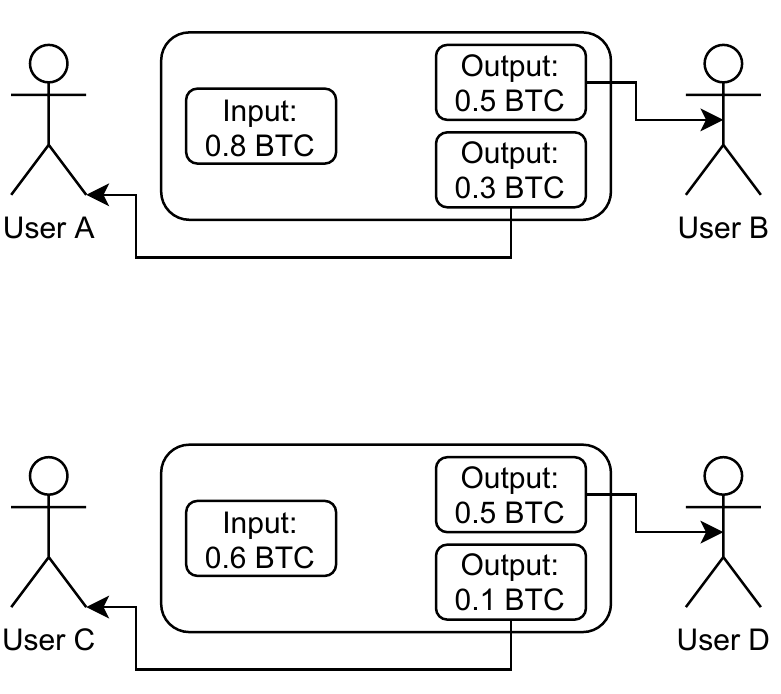}
      }
      \caption{Two regular Bitcoin transactions.}
      \label{fig:regular_tx}
    \end{subfigure}%
    \begin{subfigure}{.5\textwidth}
      \centering
      \resizebox{.8\columnwidth}{!}{
        \includegraphics{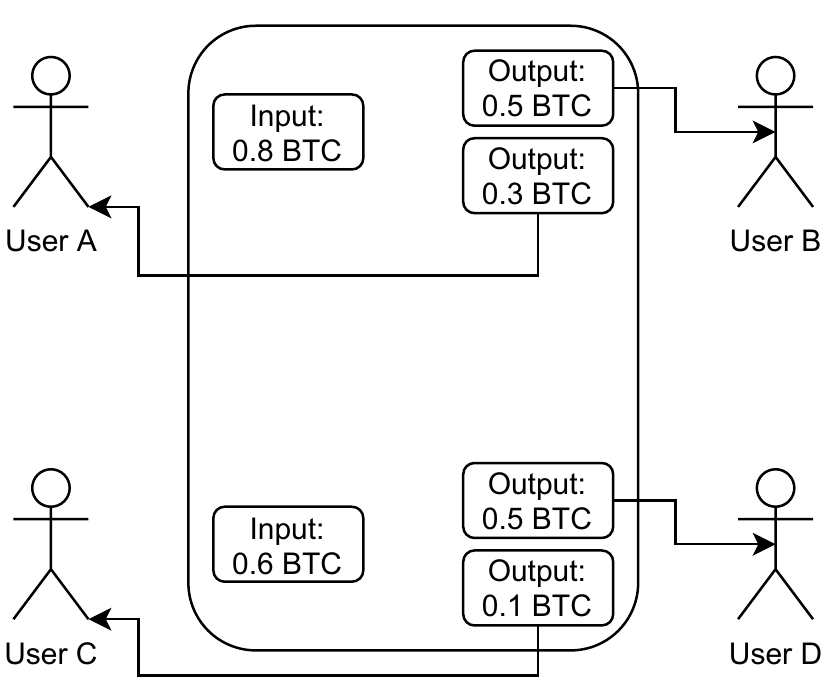}
      }
      \caption{A CoinJoin transaction.}
      \label{fig:coinjoin_tx}
    \end{subfigure}
    \caption{The basic idea of a CoinJoin transaction based on \cite{Maxwell2013}. A regular Bitcoin transaction (left) represents a relationship between two users (A and B; C and D). A CoinJoin transaction (right) combines inputs signed by several users (A and C). It assigns values to outputs controlled by multiple distinct users (B and D). For a third party, it becomes increasingly difficult to link the various outputs to individual users as the number of participants in a CoinJoin transaction increases.}
    \label{fig:coinjoin}
    \Description{TODO}
\end{figure*}

CoinJoin is a specific mixing method for combining transactions from multiple senders into one transaction~\cite{Maxwell2013}, as illustrated in Figure~\ref{fig:coinjoin}. Centralized mixing services or decentralized mixing protocols are possible implementation strategies for CoinJoins. An example implementation for centralized CoinJoin is JoinMarket\footnote{\url{https://github.com/JoinMarket-Org/joinmarket-clientserver}}. Early examples of decentralized mixing protocols making use of the CoinJoin technique include CoinShuffle~\cite{Ruffing:2014a}, CoinParty~\cite{Ziegeldorf:2015a}, CoinShuffle++~\cite{Ruffing:2017b}, and ValueShuffle~\cite{Ruffing:2017a}. CoinJoin has also become an integral feature of the cryptocurrency Dash~\cite{DeuberS21}.

In a decentralized setting, the challenge lies in coordinating and matching Bitcoin users that want to participate in a CoinJoin transaction and the input and output addresses they want to include. A central coordination server often handles the coordination task and naturally becomes a prime entry point to breach privacy. The users' privacy can be further improved by splitting the outputs into a fixed set of standard denominations, which obscure the relationship between individual inputs and outputs. CoinJoin is now an integral privacy feature of Bitcoin wallets like Wasabi or Samourai, which are the focus of our investigation.

\subsection{Empirical Analysis of Mixing Services}\label{subsec:empirical_studies}

Back in 2013, M\"{o}ser et al.~\cite{Moeser:2013s} analyzed transactions of three first-generation, centralized mixing services. They already concluded that enforcement of the KYC principle appears unlikely if mixing services are involved. Later, M\"{o}ser and B\"{o}hme~\cite{Moeser:2017a} conducted a longitudinal measurement study of JoinMarket, an online market designed to match Bitcoin users wishing to participate in CoinJoin transactions. They~\cite{Moeser:2017b} also studied second-generation mixing techniques not requiring users to trust in a single entity that might steal their coins. More recently, Pakki et al.~\cite{Pakki:2021a} explored the Bitcoin mixer ecosystem and analyzed qualitatively how existing mixing services adopt academia's proposed security features. Finally, Wu et al.~\cite{Wu:2021a} proposed a generic model for mixing services and a method for identifying mixing transactions and estimating the profit.

Our work complements this line of research by investigating CoinJoins generated by two second-generation, decentralized CoinJoin implementations. It focuses on two wallets that managed to bring that second-generation mixing technique closer to the end-user: Wasabi\footnote{\url{https://github.com/zkSNACKs/WalletWasabi}} and Samourai\footnote{\url{https://code.samourai.io/whirlpool/Whirlpool}}. Both are available as open-source software and backed by an active developer community. Contrary to previous studies, which either relied on a small set of manually created CoinJoin transactions or ad-hoc heuristics with unknown accuracy, we base our analysis on highly accurate CoinJoin transaction detection algorithms. In addition, we systematically evaluated them against a comprehensive ground-truth dataset.

\subsection{Wasabi and Samourai CoinJoins}

CoinJoin forms the basis for the ZeroLink framework~\cite{Ficsor2017a}, which in turn serves as the foundation for both Wasabi and Samourai wallets. ZeroLink defines three conceptual wallets: a pre-mix wallet for unmixed coins, a post-mix wallet for mixed coins, and a mixing technique that moves coins from the pre-mix into the post-mix wallet. While ZeroLink is compatible with various on-chain mixing protocols, it also defines Chaumian CoinJoin~\cite{Maxwell2013} used by both Wasabi and Samourai. It allows participants of these two wallets to construct transactions collaborating with each other and aided by a central coordinator without revealing the links between inputs and outputs.

\begin{figure*}
    \centering
    \begin{subfigure}{.5\textwidth}
      \centering
      \resizebox{.8\columnwidth}{!}{
        \includegraphics{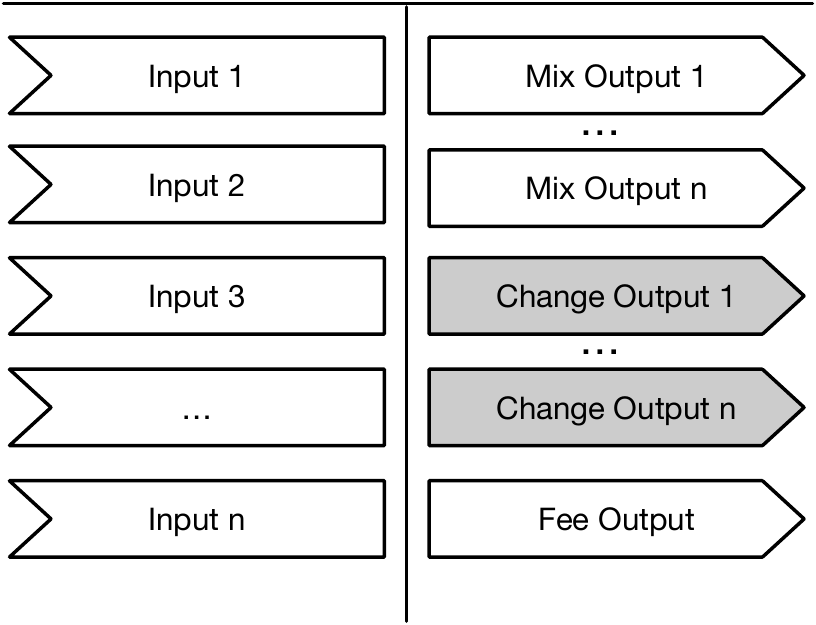}
      }
      \caption{Wasabi CoinJoin transaction.}
      \label{fig:wasabi_coinjoin}
    \end{subfigure}%
    \begin{subfigure}{.5\textwidth}
      \centering
      \resizebox{.8\columnwidth}{!}{
        \includegraphics{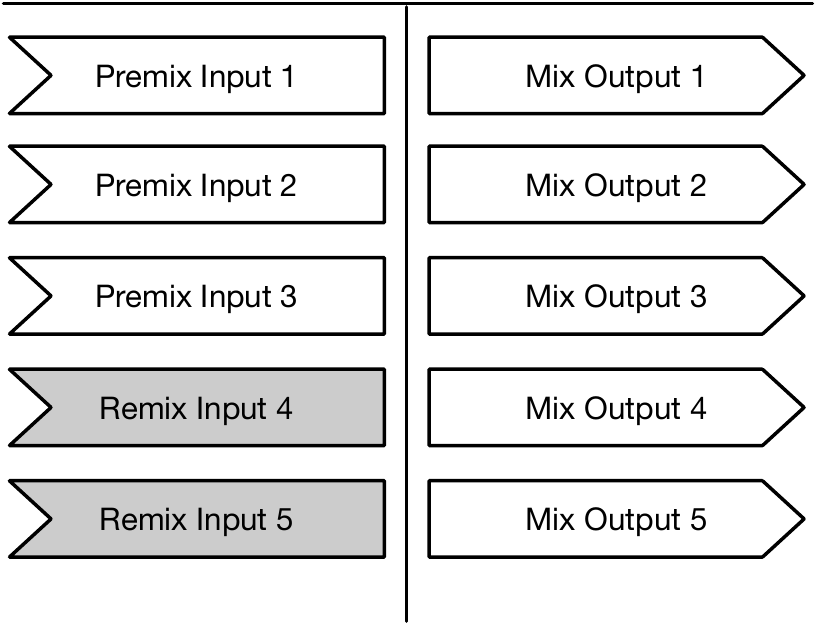}
      }
      \caption{Samourai CoinJoin transaction}
      \label{fig:samourai_coinjoin}
    \end{subfigure}
    \caption{The basic structure of a Wasabi CoinJoin (a) and a Samourai Whirlpool mix (b) transaction. Note that the number of inputs and outputs and the denomination of the outputs of a Samourai transaction are fixed. In contrast, the number of inputs, outputs, and denominations of mixed outputs may differ for different Wasabi transactions.}
    \label{fig:samourai_wasabi}
    \Description{TODO}
\end{figure*}

Wasabi Wallet is an implementation of the Chaumian CoinJoin and supports a mix of multiple denominations in the same CoinJoin transaction, with $0.01$ BTC being the lowest possible denomination. A coordinator fee of $0.003 \times a$, where $a$ is the target anonymity set, is charged for each transaction. Figure~\ref{fig:wasabi_coinjoin} illustrates the basic structure of a Wasabi CoinJoin transaction, which can contain $n$ possible inputs, $n$ mix outputs, and $n$ change outputs.

Samourai's implementation of the Chauminan CoinJoin, Samourai Whirlpool, does not support multi-denomination transactions but instead features four distinct \emph{pools} with denominations of $0.001$ BTC, $0.01$ BTC, $0.05$ BTC, and $0.5$ BTC. As shown in the illustrative example depicted in Figure~\ref{fig:samourai_coinjoin}, each Samourai Whirlpool transaction features exactly five inputs and five outputs. The mixing fees in Samourai are collected in so-called Tx0 transactions, which split selected UTXOs into other UTXOs with the appropriate sizes for the selected pool.

\subsection{Known Wasabi and Samourai Transaction Detection Methods}\label{sec:existing_approaches}

We are aware of two works focusing on detecting transactions generated by Wasabi and Samourai wallets.

% Most important existing work
First, and most importantly, our approach builds on the work of \Ficsor, the creator of Wasabi, who also implemented heuristics to compute various metrics such as transaction volume. He published them in two GitHub repositories~\cite{Ficsor2019,Ficsor2020} intending to compare various statistics for both Wasabi and Samourai, such as the volume of CoinJoin transactions, the number of fresh CoinJoin inputs, the average count of remix inputs, fees paid by users, and estimation of coordinator income.

% Known Wasabi heuristics
Detecting Wasabi transactions was straightforward in the early days because Wasabi wallets initially used two static coordinator addresses to collect all fees. This design made their detection relatively simple: a transaction is a Wasabi wallet CoinJoin if one of the static coordinator addresses is in the list of output addresses and if there are at least three indistinguishable outputs with the same value. 
However, since January 2020, Wasabi wallet has generated new coordinator addresses for every CoinJoin, rendering the original heuristic obsolete. \Ficsor, therefore, published a new heuristic that categorizes a transaction $t$ as Wasabi wallet CoinJoins if
it as at least ten equal value outputs, with $0.1 \pm 0.02$ BTC being the most
frequent one, and if it has at least two distinct output values that are
unique, and if it features at least as many inputs as occurrences of
the most frequent output\footnote{\url{https://github.com/nopara73/Dumplings/blob/84ae3747aa52349ab02aacf56493bdf0c19d9961/Dumplings/Scanning/Scanner.cs\#L129}}.

% Known Samourai heuristics
\Ficsor also proposed a heuristic for detecting Samourai Whirlpool transactions: if the number of inputs and outputs is equal to 5 and all outputs have the same value, which must equal one of the Samourai Whirlpool pool sizes (0.01, 0.05, or 0.5 BTC) $\pm 0.0011$ BTC for \cite{Ficsor2019} and $\pm 0.01$ BTC for \cite{Ficsor2020}, then a transaction can be categorized as Samourai wallet transaction.

% Other closely related work (1 para)
The second closely related work is by Wu et al.~\cite{Wu:2021a} who consider Wasabi CoinJoin transactions in their study. However, in contrast to analyzing historical transactions on the public ledger, they continuously queried the Wasabi coordinator API to retrieve ongoing CoinJoins. They used these sample transactions as a seed to detect subsequent Wasabi transactions by iterating the outputs of every transaction and checking whether a specific output is referenced by a transaction that features multiple indistinguishable outputs, which is an indicator of being a CoinJoin transaction. However, this approach based on forward-reasoning fails to consider historical transactions, an essential requirement for our study. Moreover, 
Wu et al.~\cite{Wu:2021a} do not detect Samourai CoinJoin transactions. 
 
% What is missing and how does our work go beyond state of the art (1 para)
In summary, we are aware of \Ficsor's heuristic methods for detecting Wasabi and Samourai CoinJoin transactions, and we build on these methods in our study. However, heuristic methods usually rely on manually defined thresholds that are often set without systematic tuning and evaluation. Furthermore, since we aim to study the adoption and actual privacy of Wasabi and Samourai transactions, we must first ensure that these wallets indeed create considered transactions. Therefore, we first understand the accuracy of possible detection methods and then choose the ones yielding the lowest false-positive rates.

Furthermore, there has been little research on the embedding of CoinJoin transactions in the larger Bitcoin ecosystem, especially their relation to cryptoasset exchanges, which are the focus of current regulatory efforts. In this work, we try to identify which entities conduct CoinJoin transactions either directly or are in the near vicinity of participants. For instance, if entity \emph{A} sends coins to entity \emph{B}, which then participates in a CoinJoin. If possible, we also try to label these entities as, e.g., exchanges or services. Finally, we also look at the practical anonymity offered by the ZeroLink framework and usage patterns from users of Wasabi and Samourai wallets and investigate whether there are any fundamental weaknesses. To the best of our knowledge, this has not been done before.

%% file: sections/3_coinjoin_detection.tex
% !TeX root = ../main.tex

\section{CoinJoin Detection Methods}
\label{sec:detection}

We now investigate algorithmic methods to detect Wasabi and Samourai CoinJoin
transactions and evaluate their accuracy.

\subsection{Wasabi}

We can detect Wasabi CoinJoins with two different approaches: we can apply a simple
threshold heuristics as proposed by \Ficsor (see Section~\ref{sec:existing_approaches}),
which yields varying accuracy depending on the tuning. Alternatively, we can train a statistical
machine learning model, which learns to separate Wasabi from non-Wasabi transactions
based on historical ground-truth data. Here, we consider and compare both approaches.

\subsubsection{Ground truth dataset}

We received a dataset of in total \num{28890} transactions in the date range
2018-07-19 to 2021-12-21 from the Wasabi coordinator service, which is naturally a
trust-worthy source for known Wasabi transactions. To establish a balanced ground truth
of true and false positives, we considered the corresponding block
range \numrange{532600}{715100} and randomly sampled
of \num{29202} negative instances from this block range.

\subsubsection{Feature engineering}

Next, we inspected random samples of true and false positive Wasabi transactions
and identified the following set of mainly transaction-level features, which may
allow the classifier to discriminate between Wasabi and non-Wasabi transactions:

\begin{description}
    \item[\texttt{num\_uniq\_output\_val}] number of unique output values.
    \item[\texttt{ratio\_num\_input\_num\_output}] ratio between the number
        of inputs and outputs.
    \item[\texttt{min\_output\_val}] minimum output value.
    \item[\texttt{rng\_output\_val}] the range of the output values, i.e.\
        the difference between the maximum and minimum output value.
    \item[\texttt{mean\_dec\_places}] average number of decimal places of
        the output values (value in BTC).
    \item[\texttt{num\_input\_reuse}] number of input addresses reused as
        outputs in the same transaction; reuse of addresses is strongly
        discouraged by the ZeroLink protocol and, therefore, should not occur
        in normal Wasabi Wallet transactions.
    \item[\texttt{mean\_output\_cluster\_size}] mean size (number of
        addresses) of the transaction output address clusters. The rationale
        behind this feature is that Wasabi CoinJoin outputs should not fall
        into larger clusters.
    \item[\texttt{is\_native\_segwit}] indicator if all inputs and outputs
        are using P2WPKH (Pay to Witness Public Key Hash)%
        \footnote{\url{https://github.com/bitcoin/bips/blob/master/bip-0173.mediawiki}}.
\end{description}

Correlation analysis shows that all features are not strongly correlated
(Pearson $\rho < 0.85$).

\subsubsection{Machine learning model}

Predicting the transaction type (Wasabi vs. non-Wasabi) is considered
a supervised learning problem. We use a random forest (RF) classifier,
an ensemble method that fits several decision trees on various
subsamples of the dataset (bagging). The fitted individual trees
are aggregated (majority vote for classification tasks) to improve the
predictive accuracy and control for over-fitting. RF can account
for correlation as well as interactions among predictors. They do not
require extensive hyper-parameter tuning and tend to perform usually very
well in a default setting~\cite{Liaw2002}. Furthermore, RF provides intrinsic
variable importance measures to rank predictors according to their predictive
power. We use the RF implementation from the \textsf{R}~package \emph{ranger}~%
\cite{Wright2017} within the \emph{mlr3}~\cite{Lang2019} machine learning
framework.

For an initial RF model, the dataset was randomly split into 70\% training and
30\% test set. A completely untuned RF model (500 trees, $\text{mtry} = 2$)
immediately yields an out-of-bag prediction error of only
\SI[mode=text]{0.029}{\percent} and a classification error of
\SI[mode=text]{0.017}{\percent} on the test set.
On the test set, the model achieves a false positive rate
$\text{FPR} = \SI{0.011}{\percent}$, and a false negative rate
$\text{FNR} = \SI{0.023}{\percent}$.

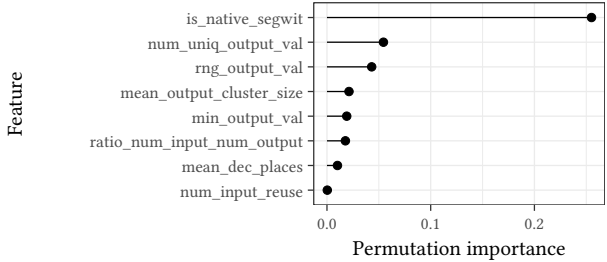
\begin{figure}
    \centering
    \resizebox{\columnwidth}{!}{
        \input{graphics/rf_varimp.tex}
    }
    \caption{Permutation importance of RF model.}
    \Description{TODO}
    \label{fig:rf-varimp}
\end{figure}

The feature importance scores are depicted in Figure~\ref{fig:rf-varimp}
in decreasing order. By far the most important predictor is the indicator
for native SegWit transactions (\texttt{is\_native\_segwit}),
followed by the number of unique output values (\texttt{num\_uniq\_output\_val})
and the range of the output values (\texttt{rng\_output\_val}), respectively.
Whereas the feature (\texttt{num\_input\_reuse}) does not seem to have any
predictive power. Only by using the feature \texttt{is\_native\_segwit}
as a single predictor on the ground-truth dataset, all positives instances
are recognized and only \num{663}~negatives instances are misclassified
($\text{FNR} = \SI{2.3}{\percent}$).

The robustness and stability of the model were assessed using different
resampling techniques. We applied subsampling (5 and 10 repeats) and
cross-validation (5 and 10-folds) on the full dataset.
Small standard errors (between \num{e-5} and \num{e-4}) indicate relatively
stable performance metrics.

\subsubsection{Evaluation and Validation}

For validation of the RF model outside of our ground-truth dataset, which
we used for training and tuning our machine learning model, we sampled
additional \num{100443} non-Wasabi transactions in the considered
block range (\num{532600} -- \num{715100}). Out of these \num{100443}~%
transactions, \num{10} instances were classified as Wasabi transactions
(false positives).

Furthermore, using the Wasabi wallet, we manually executed ten Wasabi transactions within the date
range 2022-01-14 -- 2022-02-02, which is also outside the range of our ground
truth dataset (see Appendix~\ref{sec:appendix_wasabi_manual} for a list of
transaction hashes). We applied our model and found that all transactions
were classified correctly.

Next, we compared the model predictions with \Ficsor's heuristic described
in Section~\ref{sec:existing_approaches}. This heuristic, which we denote
as Wasabi CoinJoin Detection Heuristics (WCDH), correctly classifies
all \num{29202}~non-Wasabi transactions
in the ground-truth dataset and all \num{100443}~non-Wasabi transactions
in the validation dataset. From the \num{28890}~Wasabi transactions,
\num{389} are missed ($\text{FNR} = \SI{1.3}{\percent}$).

A comparison of several performance metrics, which were evaluated on the
test dataset for both approaches, is shown in Table~\ref{tab:performance-metrics}.
It clearly shows that both the heuristics and the statistical machine
learning model are highly accurate, and the choice of procedure depends on
the application area. For example, to avoid transactions falsely identified
as Wasabi, one should choose the heuristic, which yields no false-positive predictions. On the other hand, if one needs even higher accuracy and can live with a
few false positives, one should choose the RF model with a better FNR
and accuracy.

\begin{table}%[tbp]
    \centering
    \begin{tabular*}{\columnwidth}{l@{\extracolsep{\fill}}rrr}
    \toprule
    Method & Accuracy & FPR & FNR \\
    \midrule
    WCDH & \SI{99.31}{\percent} &\SI{0.0000}{\percent} & \SI{1.3819}{\percent} \\
      RF & \SI{99.98}{\percent} &\SI{0.0114}{\percent} & \SI{0.0230}{\percent} \\
    \bottomrule
  \end{tabular*}
  \caption{Performance metrics on test data.}
  \label{tab:performance-metrics}
\end{table}

% Interpretation + Implications
We want to measure Wasabi transactions and avoid false positives in our
subsequent analysis. Therefore, we chose WCDH and, in total, identified
\nWasabiTotal Wasabi CoinJoin transactions within the block range
\numrange{530500}{725348} (2018-07-05 to 2022-02-28).

\subsection{Samourai}

To identify Samourai transactions, we can refer to a specific property dictated by Samourai's architecture. Each pool starts from a so-called \emph{genesis mix}, and each subsequent transaction features at least one remix address. This means that Samourai transactions form chains that we can traverse deterministically to identify individual transactions.

\subsubsection{Ground-Truth Dataset}

We rely on data provided by OXT Research\footnote{\url{https://oxt.me}}, which provides a blockchain explorer connected to Samourai Wallet and publishes snapshots of Samourai transaction graphs\footnote{\url{https://oxt.me/static/share/whirlpool/whirlpool_mix_txs_001.csv} - replace 001 with the denomination of choice}. However, these snapshots only include the first 17 characters of Whirlpool CoinJoins transaction hashes, and they are not guaranteed to be published continually. However, we can use these snapshots to validate further our results as the first 17 chars of a transaction hash are typically unique.

\subsubsection{Detection Algorithm}

We used a slightly modified version of \Ficsor's heuristic, which
can handle different \emph{premix} input denominations, to identify
the genesis mix transactions of a particular pool. Interestingly, we identified three distinct genesis mixes for the $0.01$ and $0.05$ BTC pools, although the vast majority of transactions can be traced to a
single one. For the $0.001$ and $0.5$ BTC pools, we only identified a single genesis mix. The genesis mixes have also been checked and verified against the snapshot data published by OXT Research.

Starting from a set of known, pool-specific genesis mix transactions, our Samourai transaction detection algorithm iterates the Bitcoin blockchain in chronological order and
considers a transaction to be a Samourai Whirlpool mix if:

\begin{itemize}
    \item it follows the structure of a Samourai Whirlpool transaction
    (5 inputs, 5 outputs), with each output value being equal to the
    pool denomination.
    \item at least one of the input transaction hashes has been identified
    before as a known Samourai Whirlpool transaction.
\end{itemize}

Conceptually, this detection algorithm can also be thought of as a breadth-first search starting at the genesis mixes. Every child transaction is added to the result set if it matches the structure of Samourai Whirlpool transactions.

\subsubsection{Evaluation and Algorithm Validation}

We also validated our identified transactions against the data published by OXT Research and found a near-perfect match. The single exception is a transaction listed by OXT Research for the $0.05$ BTC pool that is not included in our results with the hash \seqsplit{e04e5a5932e8d42e4ef641c836c6d08d9f0fff58ab4527ca788485a3fceb2416}. This transaction has the same structure as a Genesis mix (there are only \emph{premix} and no \emph{remix} inputs). However, it does not actually start a new chain as only three of the five outputs are used as \emph{remix} inputs for other Whirlpool transactions, all of which feature a second \emph{remix} input that can be traced back to an older Genesis mix.

Finally, we manually conducted five CoinJoin transactions ourselves using Samourai Wallet to verify our results further. We conducted five Tx0 transactions, one for each of the 0.001, 0.01, and 0.05 BTC pools and one for both the high and low 0.001 BTC pools. These Tx0 transactions resulted in 23 premix outputs which were then used in 23 distinct Whirlpool mixing transactions. See Appendix~\ref{sec:appendix_samourai_manual} for the complete list of transaction hashes. In total, we identified \nSamouraiTotal Samourai Whirlpool transactions
within the same block range as for Wasabi.

%% file: graphics/rf_varimp.tex
% !TEX encoding = UTF-8 Unicode
\begin{tikzpicture}[x=1pt,y=1pt]
\definecolor{fillColor}{RGB}{255,255,255}
\path[use as bounding box,fill=fillColor,fill opacity=0.00] (0,0) rectangle (289.08,130.09);
\begin{scope}
\path[clip] (  0.00,  0.00) rectangle (289.08,130.09);
\definecolor{drawColor}{RGB}{255,255,255}
\definecolor{fillColor}{RGB}{255,255,255}

\path[draw=drawColor,line width= 0.6pt,line join=round,line cap=round,fill=fillColor] (  0.00,  0.00) rectangle (289.08,130.09);
\end{scope}
\begin{scope}
\path[clip] (148.55, 30.69) rectangle (283.58,124.59);
\definecolor{fillColor}{RGB}{255,255,255}

\path[fill=fillColor] (148.55, 30.69) rectangle (283.58,124.59);
\definecolor{drawColor}{gray}{0.92}

\path[draw=drawColor,line width= 0.3pt,line join=round] (178.76, 30.69) --
	(178.76,124.59);

\path[draw=drawColor,line width= 0.3pt,line join=round] (226.89, 30.69) --
	(226.89,124.59);

\path[draw=drawColor,line width= 0.3pt,line join=round] (275.02, 30.69) --
	(275.02,124.59);

\path[draw=drawColor,line width= 0.6pt,line join=round] (148.55, 37.56) --
	(283.58, 37.56);

\path[draw=drawColor,line width= 0.6pt,line join=round] (148.55, 49.01) --
	(283.58, 49.01);

\path[draw=drawColor,line width= 0.6pt,line join=round] (148.55, 60.46) --
	(283.58, 60.46);

\path[draw=drawColor,line width= 0.6pt,line join=round] (148.55, 71.91) --
	(283.58, 71.91);

\path[draw=drawColor,line width= 0.6pt,line join=round] (148.55, 83.36) --
	(283.58, 83.36);

\path[draw=drawColor,line width= 0.6pt,line join=round] (148.55, 94.81) --
	(283.58, 94.81);

\path[draw=drawColor,line width= 0.6pt,line join=round] (148.55,106.26) --
	(283.58,106.26);

\path[draw=drawColor,line width= 0.6pt,line join=round] (148.55,117.72) --
	(283.58,117.72);

\path[draw=drawColor,line width= 0.6pt,line join=round] (154.69, 30.69) --
	(154.69,124.59);

\path[draw=drawColor,line width= 0.6pt,line join=round] (202.82, 30.69) --
	(202.82,124.59);

\path[draw=drawColor,line width= 0.6pt,line join=round] (250.95, 30.69) --
	(250.95,124.59);
\definecolor{drawColor}{RGB}{0,0,0}

\path[draw=drawColor,line width= 0.6pt,line join=round] (154.69,117.72) -- (277.44,117.72);

\path[draw=drawColor,line width= 0.6pt,line join=round] (154.69,106.26) -- (180.89,106.26);

\path[draw=drawColor,line width= 0.6pt,line join=round] (154.69, 94.81) -- (175.44, 94.81);

\path[draw=drawColor,line width= 0.6pt,line join=round] (154.69, 83.36) -- (164.91, 83.36);

\path[draw=drawColor,line width= 0.6pt,line join=round] (154.69, 71.91) -- (163.86, 71.91);

\path[draw=drawColor,line width= 0.6pt,line join=round] (154.69, 60.46) -- (163.24, 60.46);

\path[draw=drawColor,line width= 0.6pt,line join=round] (154.69, 49.01) -- (159.56, 49.01);

\path[draw=drawColor,line width= 0.6pt,line join=round] (154.69, 37.56) -- (154.84, 37.56);
\definecolor{fillColor}{RGB}{0,0,0}

\path[draw=drawColor,line width= 0.4pt,line join=round,line cap=round,fill=fillColor] (277.44,117.72) circle (  1.96);

\path[draw=drawColor,line width= 0.4pt,line join=round,line cap=round,fill=fillColor] (180.89,106.26) circle (  1.96);

\path[draw=drawColor,line width= 0.4pt,line join=round,line cap=round,fill=fillColor] (175.44, 94.81) circle (  1.96);

\path[draw=drawColor,line width= 0.4pt,line join=round,line cap=round,fill=fillColor] (164.91, 83.36) circle (  1.96);

\path[draw=drawColor,line width= 0.4pt,line join=round,line cap=round,fill=fillColor] (163.86, 71.91) circle (  1.96);

\path[draw=drawColor,line width= 0.4pt,line join=round,line cap=round,fill=fillColor] (163.24, 60.46) circle (  1.96);

\path[draw=drawColor,line width= 0.4pt,line join=round,line cap=round,fill=fillColor] (159.56, 49.01) circle (  1.96);

\path[draw=drawColor,line width= 0.4pt,line join=round,line cap=round,fill=fillColor] (154.84, 37.56) circle (  1.96);
\definecolor{drawColor}{gray}{0.20}

\path[draw=drawColor,line width= 0.6pt,line join=round,line cap=round] (148.55, 30.69) rectangle (283.58,124.59);
\end{scope}
\begin{scope}
\path[clip] (  0.00,  0.00) rectangle (289.08,130.09);
\definecolor{drawColor}{gray}{0.30}

\node[text=drawColor,anchor=base east,inner sep=0pt, outer sep=0pt, scale=  0.88] at (143.60, 34.53) {num\_input\_reuse};

\node[text=drawColor,anchor=base east,inner sep=0pt, outer sep=0pt, scale=  0.88] at (143.60, 45.98) {mean\_dec\_places};

\node[text=drawColor,anchor=base east,inner sep=0pt, outer sep=0pt, scale=  0.88] at (143.60, 57.43) {ratio\_num\_input\_num\_output};

\node[text=drawColor,anchor=base east,inner sep=0pt, outer sep=0pt, scale=  0.88] at (143.60, 68.88) {min\_output\_val};

\node[text=drawColor,anchor=base east,inner sep=0pt, outer sep=0pt, scale=  0.88] at (143.60, 80.33) {mean\_output\_cluster\_size};

\node[text=drawColor,anchor=base east,inner sep=0pt, outer sep=0pt, scale=  0.88] at (143.60, 91.78) {rng\_output\_val};

\node[text=drawColor,anchor=base east,inner sep=0pt, outer sep=0pt, scale=  0.88] at (143.60,103.23) {num\_uniq\_output\_val};

\node[text=drawColor,anchor=base east,inner sep=0pt, outer sep=0pt, scale=  0.88] at (143.60,114.68) {is\_native\_segwit};
\end{scope}
\begin{scope}
\path[clip] (  0.00,  0.00) rectangle (289.08,130.09);
\definecolor{drawColor}{gray}{0.20}

\path[draw=drawColor,line width= 0.6pt,line join=round] (145.80, 37.56) --
	(148.55, 37.56);

\path[draw=drawColor,line width= 0.6pt,line join=round] (145.80, 49.01) --
	(148.55, 49.01);

\path[draw=drawColor,line width= 0.6pt,line join=round] (145.80, 60.46) --
	(148.55, 60.46);

\path[draw=drawColor,line width= 0.6pt,line join=round] (145.80, 71.91) --
	(148.55, 71.91);

\path[draw=drawColor,line width= 0.6pt,line join=round] (145.80, 83.36) --
	(148.55, 83.36);

\path[draw=drawColor,line width= 0.6pt,line join=round] (145.80, 94.81) --
	(148.55, 94.81);

\path[draw=drawColor,line width= 0.6pt,line join=round] (145.80,106.26) --
	(148.55,106.26);

\path[draw=drawColor,line width= 0.6pt,line join=round] (145.80,117.72) --
	(148.55,117.72);
\end{scope}
\begin{scope}
\path[clip] (  0.00,  0.00) rectangle (289.08,130.09);
\definecolor{drawColor}{gray}{0.20}

\path[draw=drawColor,line width= 0.6pt,line join=round] (154.69, 27.94) --
	(154.69, 30.69);

\path[draw=drawColor,line width= 0.6pt,line join=round] (202.82, 27.94) --
	(202.82, 30.69);

\path[draw=drawColor,line width= 0.6pt,line join=round] (250.95, 27.94) --
	(250.95, 30.69);
\end{scope}
\begin{scope}
\path[clip] (  0.00,  0.00) rectangle (289.08,130.09);
\definecolor{drawColor}{gray}{0.30}

\node[text=drawColor,anchor=base,inner sep=0pt, outer sep=0pt, scale=  0.88] at (154.69, 19.68) {0.0};

\node[text=drawColor,anchor=base,inner sep=0pt, outer sep=0pt, scale=  0.88] at (202.82, 19.68) {0.1};

\node[text=drawColor,anchor=base,inner sep=0pt, outer sep=0pt, scale=  0.88] at (250.95, 19.68) {0.2};
\end{scope}
\begin{scope}
\path[clip] (  0.00,  0.00) rectangle (289.08,130.09);
\definecolor{drawColor}{RGB}{0,0,0}

\node[text=drawColor,anchor=base,inner sep=0pt, outer sep=0pt, scale=  1.10] at (216.07,  7.64) {Permutation importance};
\end{scope}
\begin{scope}
\path[clip] (  0.00,  0.00) rectangle (289.08,130.09);
\definecolor{drawColor}{RGB}{0,0,0}

\node[text=drawColor,rotate= 90.00,anchor=base,inner sep=0pt, outer sep=0pt, scale=  1.10] at ( 13.08, 77.64) {Feature};
\end{scope}
\end{tikzpicture}

%% file: sections/4_ecosystem.tex
% !TeX root = ../main.tex

\section{Empirical Analysis}
\label{sec:ecosystem}

Building on the CoinJoin detection mechanisms presented in the previous section, we now explore the role of both Wasabi and Samourai in the Bitcoin ecosystem. First, in Section~\ref{subsec:analysis_longitudional}, we investigate how usage of these services evolved. Next, in Section~\ref{subsec:network_analysis}, we then analyze the flow of generated CoinJoin transactions to cryptoasset exchanges.

\subsection{Longitudinal Analysis} % 1pg
\label{subsec:analysis_longitudional}

\begin{figure}
    \centering
    \resizebox{\columnwidth}{!}{
      \input{graphics/cj_number_of_txs.tex}
    }
    \caption{Number of transactions for Wasabi and Samourai (top)
             from October 2018 to February 2022, as well as the number
             of transactions by Samourai pool (bottom).}
    \label{fig:cj-no-txs}
    \Description{TODO}
\end{figure}
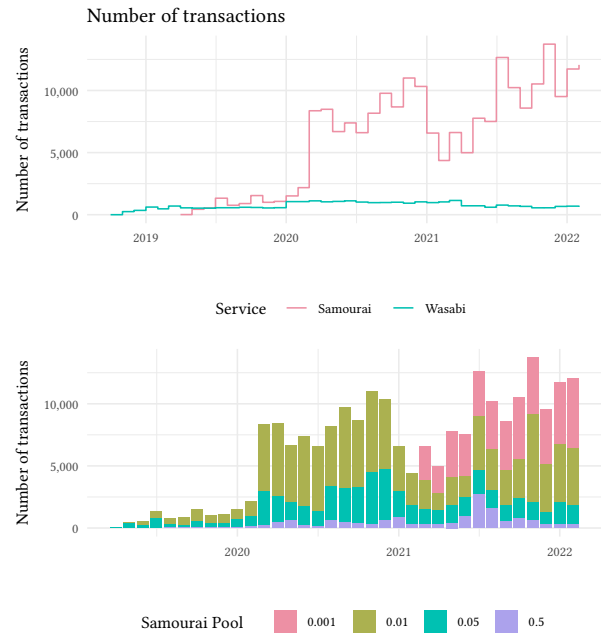

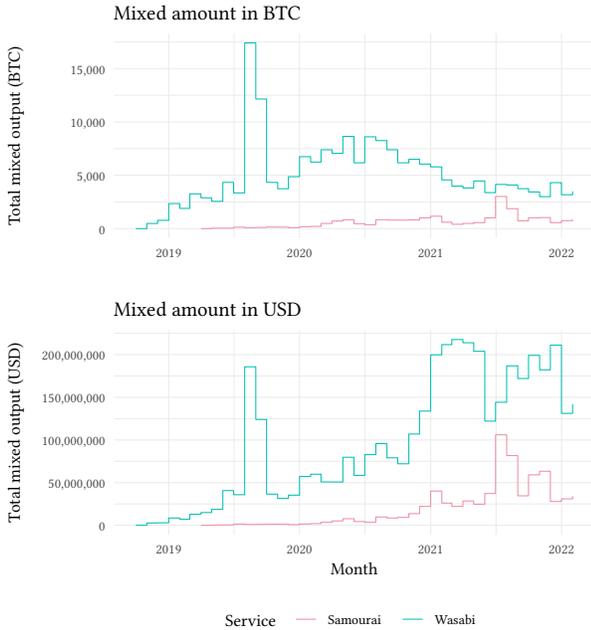
\begin{figure}
    \centering
    \resizebox{\columnwidth}{!}{
      \input{graphics/cj_mixed_amount.tex}
    }
    \caption{Amount of mixed coins for Wasabi and Samourai
             from October 2018 to February 2022.}
    \label{fig:cj-mixed-amount}
    \Description{TODO}
\end{figure}

Using our detection mechanisms, we identified \nWasabiTotal Wasabi and
\nSamouraiTotal Samourai CoinJoin transactions. For Wasabi, the average number
of inputs per transaction is \nWasabiInputs, with the average number of outputs
being \nWasabiOutputs. For Samourai, every transaction naturally (see Figure~\ref{fig:samourai_coinjoin}) has exactly five inputs and
outputs.

\textbf{Number of transactions.}
Figure~\ref{fig:cj-no-txs} shows how the number of transactions for both wallets has evolved in our observation period. We can observe that the number of Wasabi transactions has remained relatively stable since Wasabi's inception in late 2018, with minor changes around new wallet software releases (v1.1.10 in December 2019; v1.1.12 hotfixes around March 2021). This behavior reflects the design of Wasabi's CoinJoin coordination process, which starts a new round and thus creates a new CoinJoin transaction as soon as 100 peers have registered their coins or after one hour has elapsed since the last round. For Samourai, we can observe adoption since Jan 2020 and a drop in the number of transactions at the beginning of 2021 when the Bitcoin exchange rate exceeded 50K USD. Back then, the minimal possible mixing amount (0.01 pool) suddenly became 500 USD. After introducing the 0.001 pool around March 2021, which made Samourai CoinJoins more affordable, the number of transactions has again risen sharply. In general, as shown in Figure~\ref{fig:cj-no-txs} (bottom panel), we can observe higher use of smaller denomination (0.01, 0.001) pools, which is understandable because Samourai wallet users require more small than high denominations for assembling some target value. Interestingly, while the
0.5~BTC Samourai pool only makes up for \SI{\nSamouraiZeroFiveTxPct}{\percent} of Samourai
transactions, it is responsible for \SI{\nSamouraiZeroFiveBtcPct}{\percent} of the total Samourai
output. \emph{Overall, we can observe a steady adoption of Wasabi since Nov. 2018 and Samourai since Jan 2020. We can also see that software releases and design choices such as coordination interval and pool sizes affect the number of CoinJoin transactions generated by the users of these wallets.}

\textbf{Mixed coins.}
Next, as shown in Figure~\ref{fig:cj-mixed-amount}, we are interested in the value of mixed Wasabi and Samourai coins in the Bitcoin ecosystem; i.e., coins that have been mixed and are not used as remix inputs in subsequent CoinJoins. Figure~\ref{fig:cj-mixed-amount} (top panel) shows the amount of mixed BTC leaving the Wasabi and Samourai ecosystems. For Wasabi, the mixed amount in BTC is rising steadily until mid-2020, with a large spike in the second half of 2019. This period coincides with several successful law enforcement actions, including the closure of Bestmixer.io, a money-laundering machine that processed several millions of dollars worth of cryptocurrency~\cite{Abramova:2021a}. That suggests that some users of Bestmixer.io might have turned into Wasabi wallets around that period. From mid-2020 on, we can observe a decline in the mixed amount of BTC and an increase in the amount of mixed USD, which correlates strongly with (monthly aggregated) USD exchange rates (Pearson correlation $\rho=0.89$). For Samourai, we can also observe a spike in the 0.5 BTC transaction pool, which we cannot explain but increased the amount of mixed BTC and USD. \emph{In total, we found that the total value of mixed coins is \nTotalMixedBTC BTC (\nTotalMixedUSD USD\footnote{We converted BTC to USD using historic daily closing exchange rates retrieved from \url{https://api.coindesk.com/v1/bpi/historical/close.json}.}). Wasabi was used to mix \nWasabiMixedBTC BTC (\nWasabiMixedUSD USD) and Samourai for \nSamouraiMixedBTC BTC (\nSamouraiMixedUSD USD). Within recent months (Sep 2021 -- Feb 2022), Wasabi users mixed on average \nWasabiMixedRecentBTC BTC (\nWasabiMixedRecentUSD USD), Samourai users \nSamouraiMixedRecentBTC BTC (\nSamouraiMixedRecentUSD USD) per month.}

\subsection{Relations in the Bitcoin ecosystem} % 1pg
\label{subsec:network_analysis}

\begin{figure}
  \centering
  \includegraphics[width=\columnwidth]{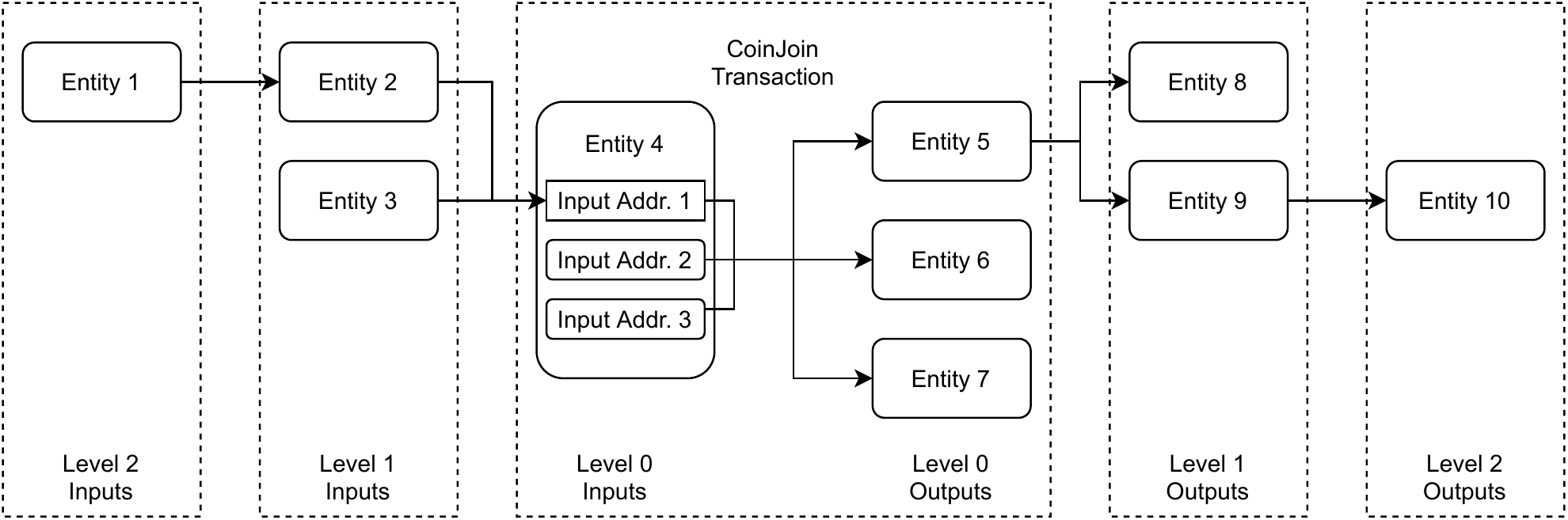}
  \caption{Entity levels considered in our analysis.}
  \label{fig:entityanalysis}
  \Description{TODO}
\end{figure}

% Method
To understand the role of Wasabi and Samourai CoinJoins in the greater Bitcoin ecosystem, we analyzed their connections to cryptoasset exchanges via direct or indirect transaction relationships, as shown in Figure~\ref{fig:entityanalysis}. First, we used \emph{GraphSense} (version 0.5.2)~\cite{Haslhofer2021}, which computes entities using the co-spent heuristic~\cite{Meiklejohn2013}, to map all input and output addresses (Level 0) of CoinJoin transactions to \emph{entities}. Since that heuristics joins addresses based on common input ownership, it merges all input addresses of a CoinJoin transaction into a single entity, which does not represent a single actor but the participants of a CoinJoin. We denote entities that directly forwarded or received values from Wasabi or Samourai CoinJoins as \emph{Level 1} entities and all entities that are sending or receiving coins via two hops as \emph{Level 2} entities. Under the assumption that CoinJoins are filtered out, these entities typically represent clusters of addresses (wallets) controlled by some real-world actor (e.g., an exchange service).

Since large address clusters typically represent services~\cite{Harrigan:2016a}
and service-to-service relations are not relevant for our analysis, we
introduced the rule-of-thumb assumption that human users typically do not
interact manually with more than 100 entities. Technically, we
stop traversal at entities with in- or out-degrees of more than $t$ other
entities. We considered different threshold values $t$,
in particular $t\in\{50, 75, 100\}$.
For attributing
addresses and categorizing entities, we rely on openly available attribution
tags retrieved from \url{walletexplorer.com} and manually curated attribution tags, which were collected by interacting with exchanges. In total, we consider 140 distinct cryptoasset exchanges. Since we are aware that our dataset might miss certain exchanges or clusters associated with exchanges, we consider the following findings as lower boundaries and point out that our analysis
method is reproducible and easy to repeat with a more comprehensive attribution tag dataset.

\textbf{Related exchange entities.}
On the output side of CoinJoin
transactions, we identified \nEntitiesReceivingBTCLOne distinct entities that
forward coins via one hop and \nEntitiesReceivingBTCLTwo entities that
forwarded coins via two hops ($t=100$). Interestingly, we found 251 entities associated with cryptoasset
exchange services that accepted Wasabi transactions directly and 307 that
accepted them via two hops ($t=100$). For Samourai, we found 187 direct and 253 indirect relations to exchange entities that accepted CoinJoin transactions via two hops
($t=100$). On the input side, we found that users associated with 155 cryptocurrency
exchange entities forwarded their funds from exchange-controlled hot wallets directly
to CoinJoin wallets (155 for Wasabi and two for Samourai) or indirectly via
two hops from 322 exchange entities (317 for Wasabi and 121 for Samourai; $t=100$).
\emph{In summary, we could identify a lower bound of \nExchangesReceivingBTC
cryptoasset exchange entities that accepted CoinJoin transactions from either Wasabi or
Samourai wallets either directly or indirectly via two hops.}

\begin{figure*}[tb]
  \centering
  \begin{subfigure}{\columnwidth}
    \centering
    \resizebox{\columnwidth}{!}{
      \input{graphics/exchange_txs_no_txs.tex}
    }
    \caption{Cumulative number of outgoing transactions to exchanges.}
    \label{fig:exchanges_out}
  \end{subfigure}
  \hfill
  \begin{subfigure}{\columnwidth}
    \centering
    \resizebox{\columnwidth}{!}{
      \input{graphics/exchange_txs_amount}
    }
    \caption{Cumulative amount of transferred mixed BTC to exchanges.}
    \label{fig:exchanges_out_ratio}
  \end{subfigure}
  \caption{Monthly aggregated lower-bounds of CoinJoin transactions and amount
           of mixed coins received by exchanges.}
  \label{fig:exchange_coinjoins}
  \Description{TODO}
\end{figure*}
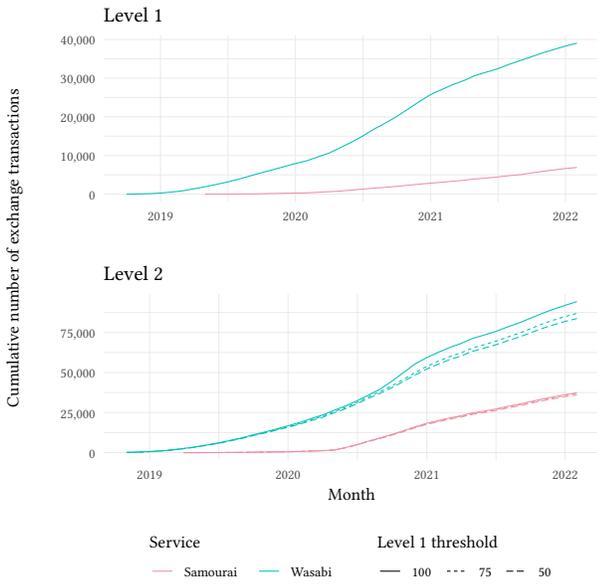
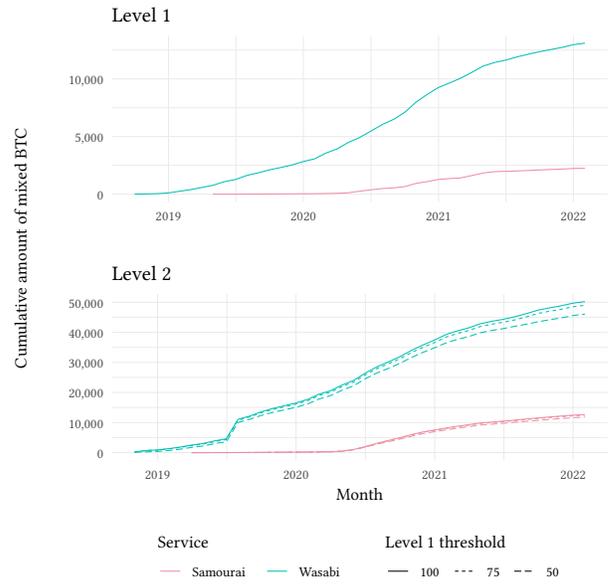

\textbf{CoinJoin transactions received by exchanges}
To understand whether
the direct and indirect relations between CoinJoin transactions and entities controlled by cryptoasset
exchanges reflect the past or the present, we analyzed their evolution in terms
of numbers and amounts of mixed BTC. For each CoinJoin output, we checked the shortest path between the output entity (level 0) to an entity categorized as
exchange via one or two hops. The cumulative number of transactions, as well as
the cumulative mixed amount of BTC is illustrated in
Figure~\ref{fig:exchange_coinjoins}. The evolution of the transactions numbers
to exchanges shows a nonlinear increase until 2021 for both services and entity
levels, respectively, which then turns into almost linear growth.
\emph{In total, we could pinpoint \nWasabiTXsToExchangeLevelOne Wasabi and
\nSamouraiTXsToExchangeLevelOne Samourai transactions that were directly
accepted by exchange entities and \nWasabiTXsToExchangeLevelTwo and
\nSamouraiTXsToExchangeLevelTwo transactions that were accepted indirectly via
two hops. In terms of mixed coins, this sums up to
\nExchangesReceivedBTCLevelOne directly received BTC
(\nExchangesReceivedUSDLevelOne USD) and \nExchangesReceivedBTCLevelTwo
indirectly received BTC (\nExchangesReceivedUSDLevelTwo USD).}

%% file: graphics/cj_number_of_txs.tex
% !TEX encoding = UTF-8 Unicode
\begin{tikzpicture}[x=1pt,y=1pt]
\definecolor{fillColor}{RGB}{255,255,255}
\path[use as bounding box,fill=fillColor,fill opacity=0.00] (0,0) rectangle (361.35,390.26);
\begin{scope}
\path[clip] ( 52.85,253.69) rectangle (350.35,362.10);
\definecolor{drawColor}{gray}{0.92}

\path[draw=drawColor,line width= 0.4pt,line join=round] ( 52.85,276.56) --
	(350.35,276.56);

\path[draw=drawColor,line width= 0.4pt,line join=round] ( 52.85,312.45) --
	(350.35,312.45);

\path[draw=drawColor,line width= 0.4pt,line join=round] ( 52.85,348.35) --
	(350.35,348.35);

\path[draw=drawColor,line width= 0.4pt,line join=round] (127.28,253.69) --
	(127.28,362.10);

\path[draw=drawColor,line width= 0.4pt,line join=round] (208.37,253.69) --
	(208.37,362.10);

\path[draw=drawColor,line width= 0.4pt,line join=round] (289.46,253.69) --
	(289.46,362.10);

\path[draw=drawColor,line width= 0.9pt,line join=round] ( 52.85,258.61) --
	(350.35,258.61);

\path[draw=drawColor,line width= 0.9pt,line join=round] ( 52.85,294.51) --
	(350.35,294.51);

\path[draw=drawColor,line width= 0.9pt,line join=round] ( 52.85,330.40) --
	(350.35,330.40);

\path[draw=drawColor,line width= 0.9pt,line join=round] ( 86.79,253.69) --
	( 86.79,362.10);

\path[draw=drawColor,line width= 0.9pt,line join=round] (167.77,253.69) --
	(167.77,362.10);

\path[draw=drawColor,line width= 0.9pt,line join=round] (248.97,253.69) --
	(248.97,362.10);

\path[draw=drawColor,line width= 0.9pt,line join=round] (329.95,253.69) --
	(329.95,362.10);
\definecolor{drawColor}{RGB}{237,144,164}

\path[draw=drawColor,line width= 0.9pt,line join=round] (106.75,258.69) --
	(113.41,258.69) --
	(113.41,261.74) --
	(120.29,261.74) --
	(120.29,262.51) --
	(126.94,262.51) --
	(126.94,268.15) --
	(133.82,268.15) --
	(133.82,264.07) --
	(140.70,264.07) --
	(140.70,265.06) --
	(147.36,265.06) --
	(147.36,269.68) --
	(154.23,269.68) --
	(154.23,265.78) --
	(160.89,265.78) --
	(160.89,266.31) --
	(167.77,266.31) --
	(167.77,269.47) --
	(174.65,269.47) --
	(174.65,274.25) --
	(181.08,274.25) --
	(181.08,318.68) --
	(187.96,318.68) --
	(187.96,319.53) --
	(194.61,319.53) --
	(194.61,306.65) --
	(201.49,306.65) --
	(201.49,311.65) --
	(208.15,311.65) --
	(208.15,306.02) --
	(215.02,306.02) --
	(215.02,317.26) --
	(221.90,317.26) --
	(221.90,328.78) --
	(228.56,328.78) --
	(228.56,320.90) --
	(235.44,320.90) --
	(235.44,337.60) --
	(242.09,337.60) --
	(242.09,332.73) --
	(248.97,332.73) --
	(248.97,305.79) --
	(255.85,305.79) --
	(255.85,289.93) --
	(262.06,289.93) --
	(262.06,306.10) --
	(268.94,306.10) --
	(268.94,294.45) --
	(275.59,294.45) --
	(275.59,314.36) --
	(282.47,314.36) --
	(282.47,312.50) --
	(289.13,312.50) --
	(289.13,349.46) --
	(296.00,349.46) --
	(296.00,332.05) --
	(302.88,332.05) --
	(302.88,320.22) --
	(309.54,320.22) --
	(309.54,334.17) --
	(316.42,334.17) --
	(316.42,357.17) --
	(323.07,357.17) --
	(323.07,326.89) --
	(329.95,326.89) --
	(329.95,342.81) --
	(336.83,342.81) --
	(336.83,345.19);
\definecolor{drawColor}{RGB}{0,193,178}

\path[draw=drawColor,line width= 0.9pt,line join=round] ( 66.38,258.62) --
	( 73.25,258.62) --
	( 73.25,260.40) --
	( 79.91,260.40) --
	( 79.91,261.11) --
	( 86.79,261.11) --
	( 86.79,263.03) --
	( 93.66,263.03) --
	( 93.66,262.01) --
	( 99.88,262.01) --
	( 99.88,263.62) --
	(106.75,263.62) --
	(106.75,262.58) --
	(113.41,262.58) --
	(113.41,262.41) --
	(120.29,262.41) --
	(120.29,262.40) --
	(126.94,262.40) --
	(126.94,262.65) --
	(133.82,262.65) --
	(133.82,262.65) --
	(140.70,262.65) --
	(140.70,262.91) --
	(147.36,262.91) --
	(147.36,262.80) --
	(154.23,262.80) --
	(154.23,262.52) --
	(160.89,262.52) --
	(160.89,262.67) --
	(167.77,262.67) --
	(167.77,266.18) --
	(174.65,266.18) --
	(174.65,266.18) --
	(181.08,266.18) --
	(181.08,266.62) --
	(187.96,266.62) --
	(187.96,266.06) --
	(194.61,266.06) --
	(194.61,266.30) --
	(201.49,266.30) --
	(201.49,266.63) --
	(208.15,266.63) --
	(208.15,265.92) --
	(215.02,265.92) --
	(215.02,265.59) --
	(221.90,265.59) --
	(221.90,265.65) --
	(228.56,265.65) --
	(228.56,265.83) --
	(235.44,265.83) --
	(235.44,265.27) --
	(242.09,265.27) --
	(242.09,266.04) --
	(248.97,266.04) --
	(248.97,265.61) --
	(255.85,265.61) --
	(255.85,266.01) --
	(262.06,266.01) --
	(262.06,266.85) --
	(268.94,266.85) --
	(268.94,263.77) --
	(275.59,263.77) --
	(275.59,263.77) --
	(282.47,263.77) --
	(282.47,262.93) --
	(289.13,262.93) --
	(289.13,264.17) --
	(296.00,264.17) --
	(296.00,263.72) --
	(302.88,263.72) --
	(302.88,263.43) --
	(309.54,263.43) --
	(309.54,262.60) --
	(316.42,262.60) --
	(316.42,262.62) --
	(323.07,262.62) --
	(323.07,263.41) --
	(329.95,263.41) --
	(329.95,263.56) --
	(336.83,263.56) --
	(336.83,263.14);
\end{scope}
\begin{scope}
\path[clip] (  0.00,  0.00) rectangle (361.35,390.26);
\definecolor{drawColor}{gray}{0.30}

\node[text=drawColor,anchor=base east,inner sep=0pt, outer sep=0pt, scale=  0.88] at ( 47.90,255.58) {0};

\node[text=drawColor,anchor=base east,inner sep=0pt, outer sep=0pt, scale=  0.88] at ( 47.90,291.48) {5,000};

\node[text=drawColor,anchor=base east,inner sep=0pt, outer sep=0pt, scale=  0.88] at ( 47.90,327.37) {10,000};
\end{scope}
\begin{scope}
\path[clip] (  0.00,  0.00) rectangle (361.35,390.26);
\definecolor{drawColor}{gray}{0.30}

\node[text=drawColor,anchor=base,inner sep=0pt, outer sep=0pt, scale=  0.88] at ( 86.79,242.68) {2019};

\node[text=drawColor,anchor=base,inner sep=0pt, outer sep=0pt, scale=  0.88] at (167.77,242.68) {2020};

\node[text=drawColor,anchor=base,inner sep=0pt, outer sep=0pt, scale=  0.88] at (248.97,242.68) {2021};

\node[text=drawColor,anchor=base,inner sep=0pt, outer sep=0pt, scale=  0.88] at (329.95,242.68) {2022};
\end{scope}
\begin{scope}
\path[clip] (  0.00,  0.00) rectangle (361.35,390.26);
\definecolor{drawColor}{RGB}{0,0,0}

\node[text=drawColor,rotate= 90.00,anchor=base,inner sep=0pt, outer sep=0pt, scale=  1.10] at ( 18.58,307.90) {Number of transactions};
\end{scope}
\begin{scope}
\path[clip] (  0.00,  0.00) rectangle (361.35,390.26);
\definecolor{drawColor}{RGB}{0,0,0}

\node[text=drawColor,anchor=base west,inner sep=0pt, outer sep=0pt, scale=  1.10] at (127.13,200.99) {Service};
\end{scope}
\begin{scope}
\path[clip] (  0.00,  0.00) rectangle (361.35,390.26);
\definecolor{drawColor}{RGB}{237,144,164}

\path[draw=drawColor,line width= 0.9pt,line join=round] (168.01,204.78) -- (179.57,204.78);
\end{scope}
\begin{scope}
\path[clip] (  0.00,  0.00) rectangle (361.35,390.26);
\definecolor{drawColor}{RGB}{0,193,178}

\path[draw=drawColor,line width= 0.9pt,line join=round] (229.66,204.78) -- (241.22,204.78);
\end{scope}
\begin{scope}
\path[clip] (  0.00,  0.00) rectangle (361.35,390.26);
\definecolor{drawColor}{RGB}{0,0,0}

\node[text=drawColor,anchor=base west,inner sep=0pt, outer sep=0pt, scale=  0.88] at (186.52,201.75) {Samourai};
\end{scope}
\begin{scope}
\path[clip] (  0.00,  0.00) rectangle (361.35,390.26);
\definecolor{drawColor}{RGB}{0,0,0}

\node[text=drawColor,anchor=base west,inner sep=0pt, outer sep=0pt, scale=  0.88] at (248.17,201.75) {Wasabi};
\end{scope}
\begin{scope}
\path[clip] (  0.00,  0.00) rectangle (361.35,390.26);
\definecolor{drawColor}{RGB}{0,0,0}

\node[text=drawColor,anchor=base west,inner sep=0pt, outer sep=0pt, scale=  1.32] at ( 52.85,370.17) {Number of transactions};
\end{scope}
\begin{scope}
\path[clip] ( 52.85, 72.64) rectangle (350.35,181.05);
\definecolor{drawColor}{gray}{0.92}

\path[draw=drawColor,line width= 0.4pt,line join=round] ( 52.85, 95.51) --
	(350.35, 95.51);

\path[draw=drawColor,line width= 0.4pt,line join=round] ( 52.85,131.41) --
	(350.35,131.41);

\path[draw=drawColor,line width= 0.4pt,line join=round] ( 52.85,167.30) --
	(350.35,167.30);

\path[draw=drawColor,line width= 0.4pt,line join=round] ( 93.01, 72.64) --
	( 93.01,181.05);

\path[draw=drawColor,line width= 0.4pt,line join=round] (186.20, 72.64) --
	(186.20,181.05);

\path[draw=drawColor,line width= 0.4pt,line join=round] (279.26, 72.64) --
	(279.26,181.05);

\path[draw=drawColor,line width= 0.9pt,line join=round] ( 52.85, 77.57) --
	(350.35, 77.57);

\path[draw=drawColor,line width= 0.9pt,line join=round] ( 52.85,113.46) --
	(350.35,113.46);

\path[draw=drawColor,line width= 0.9pt,line join=round] ( 52.85,149.35) --
	(350.35,149.35);

\path[draw=drawColor,line width= 0.9pt,line join=round] (139.60, 72.64) --
	(139.60,181.05);

\path[draw=drawColor,line width= 0.9pt,line join=round] (232.79, 72.64) --
	(232.79,181.05);

\path[draw=drawColor,line width= 0.9pt,line join=round] (325.73, 72.64) --
	(325.73,181.05);
\definecolor{fillColor}{RGB}{0,193,178}

\path[fill=fillColor] ( 66.38, 77.57) rectangle ( 72.79, 77.65);
\definecolor{fillColor}{RGB}{171,177,80}

\path[fill=fillColor] ( 74.01, 79.94) rectangle ( 80.43, 80.70);
\definecolor{fillColor}{RGB}{0,193,178}

\path[fill=fillColor] ( 74.01, 77.57) rectangle ( 80.43, 79.94);
\definecolor{fillColor}{RGB}{171,177,80}

\path[fill=fillColor] ( 81.91, 79.30) rectangle ( 88.32, 81.47);
\definecolor{fillColor}{RGB}{0,193,178}

\path[fill=fillColor] ( 81.91, 77.57) rectangle ( 88.32, 79.30);
\definecolor{fillColor}{RGB}{171,177,80}

\path[fill=fillColor] ( 89.55, 83.04) rectangle ( 95.96, 87.11);
\definecolor{fillColor}{RGB}{0,193,178}

\path[fill=fillColor] ( 89.55, 77.57) rectangle ( 95.96, 83.04);
\definecolor{fillColor}{RGB}{171,177,80}

\path[fill=fillColor] ( 97.44, 79.76) rectangle (103.85, 83.03);
\definecolor{fillColor}{RGB}{0,193,178}

\path[fill=fillColor] ( 97.44, 77.78) rectangle (103.85, 79.76);
\definecolor{fillColor}{RGB}{172,162,236}

\path[fill=fillColor] ( 97.44, 77.57) rectangle (103.85, 77.78);
\definecolor{fillColor}{RGB}{171,177,80}

\path[fill=fillColor] (105.33, 79.58) rectangle (111.75, 84.01);
\definecolor{fillColor}{RGB}{0,193,178}

\path[fill=fillColor] (105.33, 77.88) rectangle (111.75, 79.58);
\definecolor{fillColor}{RGB}{172,162,236}

\path[fill=fillColor] (105.33, 77.57) rectangle (111.75, 77.88);
\definecolor{fillColor}{RGB}{171,177,80}

\path[fill=fillColor] (112.97, 81.34) rectangle (119.39, 88.64);
\definecolor{fillColor}{RGB}{0,193,178}

\path[fill=fillColor] (112.97, 77.95) rectangle (119.39, 81.34);
\definecolor{fillColor}{RGB}{172,162,236}

\path[fill=fillColor] (112.97, 77.57) rectangle (119.39, 77.95);
\definecolor{fillColor}{RGB}{171,177,80}

\path[fill=fillColor] (120.86, 80.68) rectangle (127.28, 84.74);
\definecolor{fillColor}{RGB}{0,193,178}

\path[fill=fillColor] (120.86, 77.89) rectangle (127.28, 80.68);
\definecolor{fillColor}{RGB}{172,162,236}

\path[fill=fillColor] (120.86, 77.57) rectangle (127.28, 77.89);
\definecolor{fillColor}{RGB}{171,177,80}

\path[fill=fillColor] (128.50, 80.40) rectangle (134.92, 85.26);
\definecolor{fillColor}{RGB}{0,193,178}

\path[fill=fillColor] (128.50, 77.70) rectangle (134.92, 80.40);
\definecolor{fillColor}{RGB}{172,162,236}

\path[fill=fillColor] (128.50, 77.57) rectangle (134.92, 77.70);
\definecolor{fillColor}{RGB}{171,177,80}

\path[fill=fillColor] (136.39, 82.48) rectangle (142.81, 88.43);
\definecolor{fillColor}{RGB}{0,193,178}

\path[fill=fillColor] (136.39, 78.03) rectangle (142.81, 82.48);
\definecolor{fillColor}{RGB}{172,162,236}

\path[fill=fillColor] (136.39, 77.57) rectangle (142.81, 78.03);
\definecolor{fillColor}{RGB}{171,177,80}

\path[fill=fillColor] (144.29, 84.55) rectangle (150.70, 93.21);
\definecolor{fillColor}{RGB}{0,193,178}

\path[fill=fillColor] (144.29, 78.29) rectangle (150.70, 84.55);
\definecolor{fillColor}{RGB}{172,162,236}

\path[fill=fillColor] (144.29, 77.57) rectangle (150.70, 78.29);
\definecolor{fillColor}{RGB}{171,177,80}

\path[fill=fillColor] (151.67, 99.00) rectangle (158.09,137.63);
\definecolor{fillColor}{RGB}{0,193,178}

\path[fill=fillColor] (151.67, 79.19) rectangle (158.09, 99.00);
\definecolor{fillColor}{RGB}{172,162,236}

\path[fill=fillColor] (151.67, 77.57) rectangle (158.09, 79.19);
\definecolor{fillColor}{RGB}{171,177,80}

\path[fill=fillColor] (159.56, 95.78) rectangle (165.98,138.49);
\definecolor{fillColor}{RGB}{0,193,178}

\path[fill=fillColor] (159.56, 80.94) rectangle (165.98, 95.78);
\definecolor{fillColor}{RGB}{172,162,236}

\path[fill=fillColor] (159.56, 77.57) rectangle (165.98, 80.94);
\definecolor{fillColor}{RGB}{171,177,80}

\path[fill=fillColor] (167.20, 92.81) rectangle (173.62,125.61);
\definecolor{fillColor}{RGB}{0,193,178}

\path[fill=fillColor] (167.20, 81.90) rectangle (173.62, 92.81);
\definecolor{fillColor}{RGB}{172,162,236}

\path[fill=fillColor] (167.20, 77.57) rectangle (173.62, 81.90);
\definecolor{fillColor}{RGB}{171,177,80}

\path[fill=fillColor] (175.10, 90.10) rectangle (181.51,130.60);
\definecolor{fillColor}{RGB}{0,193,178}

\path[fill=fillColor] (175.10, 78.99) rectangle (181.51, 90.10);
\definecolor{fillColor}{RGB}{172,162,236}

\path[fill=fillColor] (175.10, 77.57) rectangle (181.51, 78.99);
\definecolor{fillColor}{RGB}{171,177,80}

\path[fill=fillColor] (182.73, 87.36) rectangle (189.15,124.98);
\definecolor{fillColor}{RGB}{0,193,178}

\path[fill=fillColor] (182.73, 78.38) rectangle (189.15, 87.36);
\definecolor{fillColor}{RGB}{172,162,236}

\path[fill=fillColor] (182.73, 77.57) rectangle (189.15, 78.38);
\definecolor{fillColor}{RGB}{171,177,80}

\path[fill=fillColor] (190.63,101.46) rectangle (197.04,136.21);
\definecolor{fillColor}{RGB}{0,193,178}

\path[fill=fillColor] (190.63, 81.78) rectangle (197.04,101.46);
\definecolor{fillColor}{RGB}{172,162,236}

\path[fill=fillColor] (190.63, 77.57) rectangle (197.04, 81.78);
\definecolor{fillColor}{RGB}{171,177,80}

\path[fill=fillColor] (198.52,100.42) rectangle (204.94,147.73);
\definecolor{fillColor}{RGB}{0,193,178}

\path[fill=fillColor] (198.52, 80.79) rectangle (204.94,100.42);
\definecolor{fillColor}{RGB}{172,162,236}

\path[fill=fillColor] (198.52, 77.57) rectangle (204.94, 80.79);
\definecolor{fillColor}{RGB}{171,177,80}

\path[fill=fillColor] (206.16,101.45) rectangle (212.58,139.86);
\definecolor{fillColor}{RGB}{0,193,178}

\path[fill=fillColor] (206.16, 80.11) rectangle (212.58,101.45);
\definecolor{fillColor}{RGB}{172,162,236}

\path[fill=fillColor] (206.16, 77.57) rectangle (212.58, 80.11);
\definecolor{fillColor}{RGB}{171,177,80}

\path[fill=fillColor] (214.05,109.56) rectangle (220.47,156.55);
\definecolor{fillColor}{RGB}{0,193,178}

\path[fill=fillColor] (214.05, 79.78) rectangle (220.47,109.56);
\definecolor{fillColor}{RGB}{172,162,236}

\path[fill=fillColor] (214.05, 77.57) rectangle (220.47, 79.78);
\definecolor{fillColor}{RGB}{171,177,80}

\path[fill=fillColor] (221.69,111.20) rectangle (228.11,151.68);
\definecolor{fillColor}{RGB}{0,193,178}

\path[fill=fillColor] (221.69, 81.65) rectangle (228.11,111.20);
\definecolor{fillColor}{RGB}{172,162,236}

\path[fill=fillColor] (221.69, 77.57) rectangle (228.11, 81.65);
\definecolor{fillColor}{RGB}{171,177,80}

\path[fill=fillColor] (229.58, 98.97) rectangle (236.00,124.75);
\definecolor{fillColor}{RGB}{0,193,178}

\path[fill=fillColor] (229.58, 83.43) rectangle (236.00, 98.97);
\definecolor{fillColor}{RGB}{172,162,236}

\path[fill=fillColor] (229.58, 77.57) rectangle (236.00, 83.43);
\definecolor{fillColor}{RGB}{171,177,80}

\path[fill=fillColor] (237.48, 90.34) rectangle (243.89,108.88);
\definecolor{fillColor}{RGB}{0,193,178}

\path[fill=fillColor] (237.48, 79.88) rectangle (243.89, 90.34);
\definecolor{fillColor}{RGB}{172,162,236}

\path[fill=fillColor] (237.48, 77.57) rectangle (243.89, 79.88);
\definecolor{fillColor}{RGB}{237,144,164}

\path[fill=fillColor] (244.61,104.87) rectangle (251.02,125.05);
\definecolor{fillColor}{RGB}{171,177,80}

\path[fill=fillColor] (244.61, 88.52) rectangle (251.02,104.87);
\definecolor{fillColor}{RGB}{0,193,178}

\path[fill=fillColor] (244.61, 79.67) rectangle (251.02, 88.52);
\definecolor{fillColor}{RGB}{172,162,236}

\path[fill=fillColor] (244.61, 77.57) rectangle (251.02, 79.67);
\definecolor{fillColor}{RGB}{237,144,164}

\path[fill=fillColor] (252.50, 97.52) rectangle (258.92,113.40);
\definecolor{fillColor}{RGB}{171,177,80}

\path[fill=fillColor] (252.50, 88.08) rectangle (258.92, 97.52);
\definecolor{fillColor}{RGB}{0,193,178}

\path[fill=fillColor] (252.50, 79.62) rectangle (258.92, 88.08);
\definecolor{fillColor}{RGB}{172,162,236}

\path[fill=fillColor] (252.50, 77.57) rectangle (258.92, 79.62);
\definecolor{fillColor}{RGB}{237,144,164}

\path[fill=fillColor] (260.14,107.04) rectangle (266.55,133.31);
\definecolor{fillColor}{RGB}{171,177,80}

\path[fill=fillColor] (260.14, 90.60) rectangle (266.55,107.04);
\definecolor{fillColor}{RGB}{0,193,178}

\path[fill=fillColor] (260.14, 80.48) rectangle (266.55, 90.60);
\definecolor{fillColor}{RGB}{172,162,236}

\path[fill=fillColor] (260.14, 77.57) rectangle (266.55, 80.48);
\definecolor{fillColor}{RGB}{237,144,164}

\path[fill=fillColor] (268.03,107.64) rectangle (274.45,131.45);
\definecolor{fillColor}{RGB}{171,177,80}

\path[fill=fillColor] (268.03, 95.54) rectangle (274.45,107.64);
\definecolor{fillColor}{RGB}{0,193,178}

\path[fill=fillColor] (268.03, 84.37) rectangle (274.45, 95.54);
\definecolor{fillColor}{RGB}{172,162,236}

\path[fill=fillColor] (268.03, 77.57) rectangle (274.45, 84.37);
\definecolor{fillColor}{RGB}{237,144,164}

\path[fill=fillColor] (275.67,141.95) rectangle (282.09,168.41);
\definecolor{fillColor}{RGB}{171,177,80}

\path[fill=fillColor] (275.67,111.29) rectangle (282.09,141.95);
\definecolor{fillColor}{RGB}{0,193,178}

\path[fill=fillColor] (275.67, 96.94) rectangle (282.09,111.29);
\definecolor{fillColor}{RGB}{172,162,236}

\path[fill=fillColor] (275.67, 77.57) rectangle (282.09, 96.94);
\definecolor{fillColor}{RGB}{237,144,164}

\path[fill=fillColor] (283.56,122.99) rectangle (289.98,151.00);
\definecolor{fillColor}{RGB}{171,177,80}

\path[fill=fillColor] (283.56, 99.22) rectangle (289.98,122.99);
\definecolor{fillColor}{RGB}{0,193,178}

\path[fill=fillColor] (283.56, 88.72) rectangle (289.98, 99.22);
\definecolor{fillColor}{RGB}{172,162,236}

\path[fill=fillColor] (283.56, 77.57) rectangle (289.98, 88.72);
\definecolor{fillColor}{RGB}{237,144,164}

\path[fill=fillColor] (291.46,111.16) rectangle (297.87,139.17);
\definecolor{fillColor}{RGB}{171,177,80}

\path[fill=fillColor] (291.46, 90.70) rectangle (297.87,111.16);
\definecolor{fillColor}{RGB}{0,193,178}

\path[fill=fillColor] (291.46, 81.19) rectangle (297.87, 90.70);
\definecolor{fillColor}{RGB}{172,162,236}

\path[fill=fillColor] (291.46, 77.57) rectangle (297.87, 81.19);
\definecolor{fillColor}{RGB}{237,144,164}

\path[fill=fillColor] (299.09,117.37) rectangle (305.51,153.12);
\definecolor{fillColor}{RGB}{171,177,80}

\path[fill=fillColor] (299.09, 94.70) rectangle (305.51,117.37);
\definecolor{fillColor}{RGB}{0,193,178}

\path[fill=fillColor] (299.09, 83.01) rectangle (305.51, 94.70);
\definecolor{fillColor}{RGB}{172,162,236}

\path[fill=fillColor] (299.09, 77.57) rectangle (305.51, 83.01);
\definecolor{fillColor}{RGB}{237,144,164}

\path[fill=fillColor] (306.99,143.37) rectangle (313.40,176.12);
\definecolor{fillColor}{RGB}{171,177,80}

\path[fill=fillColor] (306.99, 92.12) rectangle (313.40,143.37);
\definecolor{fillColor}{RGB}{0,193,178}

\path[fill=fillColor] (306.99, 81.65) rectangle (313.40, 92.12);
\definecolor{fillColor}{RGB}{172,162,236}

\path[fill=fillColor] (306.99, 77.57) rectangle (313.40, 81.65);
\definecolor{fillColor}{RGB}{237,144,164}

\path[fill=fillColor] (314.62,114.52) rectangle (321.04,145.84);
\definecolor{fillColor}{RGB}{171,177,80}

\path[fill=fillColor] (314.62, 86.45) rectangle (321.04,114.52);
\definecolor{fillColor}{RGB}{0,193,178}

\path[fill=fillColor] (314.62, 79.43) rectangle (321.04, 86.45);
\definecolor{fillColor}{RGB}{172,162,236}

\path[fill=fillColor] (314.62, 77.57) rectangle (321.04, 79.43);
\definecolor{fillColor}{RGB}{237,144,164}

\path[fill=fillColor] (322.52,126.04) rectangle (328.93,161.77);
\definecolor{fillColor}{RGB}{171,177,80}

\path[fill=fillColor] (322.52, 92.76) rectangle (328.93,126.04);
\definecolor{fillColor}{RGB}{0,193,178}

\path[fill=fillColor] (322.52, 79.64) rectangle (328.93, 92.76);
\definecolor{fillColor}{RGB}{172,162,236}

\path[fill=fillColor] (322.52, 77.57) rectangle (328.93, 79.64);
\definecolor{fillColor}{RGB}{237,144,164}

\path[fill=fillColor] (330.41,123.57) rectangle (336.83,164.14);
\definecolor{fillColor}{RGB}{171,177,80}

\path[fill=fillColor] (330.41, 90.95) rectangle (336.83,123.57);
\definecolor{fillColor}{RGB}{0,193,178}

\path[fill=fillColor] (330.41, 79.41) rectangle (336.83, 90.95);
\definecolor{fillColor}{RGB}{172,162,236}

\path[fill=fillColor] (330.41, 77.57) rectangle (336.83, 79.41);
\end{scope}
\begin{scope}
\path[clip] (  0.00,  0.00) rectangle (361.35,390.26);
\definecolor{drawColor}{gray}{0.30}

\node[text=drawColor,anchor=base east,inner sep=0pt, outer sep=0pt, scale=  0.88] at ( 47.90, 74.54) {0};

\node[text=drawColor,anchor=base east,inner sep=0pt, outer sep=0pt, scale=  0.88] at ( 47.90,110.43) {5,000};

\node[text=drawColor,anchor=base east,inner sep=0pt, outer sep=0pt, scale=  0.88] at ( 47.90,146.32) {10,000};
\end{scope}
\begin{scope}
\path[clip] (  0.00,  0.00) rectangle (361.35,390.26);
\definecolor{drawColor}{gray}{0.30}

\node[text=drawColor,anchor=base,inner sep=0pt, outer sep=0pt, scale=  0.88] at (139.60, 61.63) {2020};

\node[text=drawColor,anchor=base,inner sep=0pt, outer sep=0pt, scale=  0.88] at (232.79, 61.63) {2021};

\node[text=drawColor,anchor=base,inner sep=0pt, outer sep=0pt, scale=  0.88] at (325.73, 61.63) {2022};
\end{scope}
\begin{scope}
\path[clip] (  0.00,  0.00) rectangle (361.35,390.26);
\definecolor{drawColor}{RGB}{0,0,0}

\node[text=drawColor,rotate= 90.00,anchor=base,inner sep=0pt, outer sep=0pt, scale=  1.10] at ( 18.58,126.85) {Number of transactions};
\end{scope}
\begin{scope}
\path[clip] (  0.00,  0.00) rectangle (361.35,390.26);
\definecolor{drawColor}{RGB}{0,0,0}

\node[text=drawColor,anchor=base west,inner sep=0pt, outer sep=0pt, scale=  1.10] at ( 84.19, 19.94) {Samourai Pool};
\end{scope}
\begin{scope}
\path[clip] (  0.00,  0.00) rectangle (361.35,390.26);
\definecolor{fillColor}{RGB}{237,144,164}

\path[fill=fillColor] (160.85, 17.21) rectangle (173.88, 30.24);
\end{scope}
\begin{scope}
\path[clip] (  0.00,  0.00) rectangle (361.35,390.26);
\definecolor{fillColor}{RGB}{171,177,80}

\path[fill=fillColor] (206.34, 17.21) rectangle (219.37, 30.24);
\end{scope}
\begin{scope}
\path[clip] (  0.00,  0.00) rectangle (361.35,390.26);
\definecolor{fillColor}{RGB}{0,193,178}

\path[fill=fillColor] (247.43, 17.21) rectangle (260.47, 30.24);
\end{scope}
\begin{scope}
\path[clip] (  0.00,  0.00) rectangle (361.35,390.26);
\definecolor{fillColor}{RGB}{172,162,236}

\path[fill=fillColor] (288.53, 17.21) rectangle (301.56, 30.24);
\end{scope}
\begin{scope}
\path[clip] (  0.00,  0.00) rectangle (361.35,390.26);
\definecolor{drawColor}{RGB}{0,0,0}

\node[text=drawColor,anchor=base west,inner sep=0pt, outer sep=0pt, scale=  0.88] at (180.09, 20.70) {0.001};
\end{scope}
\begin{scope}
\path[clip] (  0.00,  0.00) rectangle (361.35,390.26);
\definecolor{drawColor}{RGB}{0,0,0}

\node[text=drawColor,anchor=base west,inner sep=0pt, outer sep=0pt, scale=  0.88] at (225.58, 20.70) {0.01};
\end{scope}
\begin{scope}
\path[clip] (  0.00,  0.00) rectangle (361.35,390.26);
\definecolor{drawColor}{RGB}{0,0,0}

\node[text=drawColor,anchor=base west,inner sep=0pt, outer sep=0pt, scale=  0.88] at (266.68, 20.70) {0.05};
\end{scope}
\begin{scope}
\path[clip] (  0.00,  0.00) rectangle (361.35,390.26);
\definecolor{drawColor}{RGB}{0,0,0}

\node[text=drawColor,anchor=base west,inner sep=0pt, outer sep=0pt, scale=  0.88] at (307.77, 20.70) {0.5};
\end{scope}
\end{tikzpicture}

%% file: graphics/cj_mixed_amount.tex
% !TEX encoding = UTF-8 Unicode
\begin{tikzpicture}[x=1pt,y=1pt]
\definecolor{fillColor}{RGB}{255,255,255}
\path[use as bounding box,fill=fillColor,fill opacity=0.00] (0,0) rectangle (361.35,390.26);
\begin{scope}
\path[clip] ( 72.89,244.04) rectangle (350.35,362.10);
\definecolor{drawColor}{gray}{0.92}

\path[draw=drawColor,line width= 0.3pt,line join=round] ( 72.89,264.82) --
	(350.35,264.82);

\path[draw=drawColor,line width= 0.3pt,line join=round] ( 72.89,295.65) --
	(350.35,295.65);

\path[draw=drawColor,line width= 0.3pt,line join=round] ( 72.89,326.49) --
	(350.35,326.49);

\path[draw=drawColor,line width= 0.3pt,line join=round] ( 72.89,357.32) --
	(350.35,357.32);

\path[draw=drawColor,line width= 0.3pt,line join=round] (142.30,244.04) --
	(142.30,362.10);

\path[draw=drawColor,line width= 0.3pt,line join=round] (217.93,244.04) --
	(217.93,362.10);

\path[draw=drawColor,line width= 0.3pt,line join=round] (293.56,244.04) --
	(293.56,362.10);

\path[draw=drawColor,line width= 0.6pt,line join=round] ( 72.89,249.40) --
	(350.35,249.40);

\path[draw=drawColor,line width= 0.6pt,line join=round] ( 72.89,280.23) --
	(350.35,280.23);

\path[draw=drawColor,line width= 0.6pt,line join=round] ( 72.89,311.07) --
	(350.35,311.07);

\path[draw=drawColor,line width= 0.6pt,line join=round] ( 72.89,341.90) --
	(350.35,341.90);

\path[draw=drawColor,line width= 0.6pt,line join=round] (104.54,244.04) --
	(104.54,362.10);

\path[draw=drawColor,line width= 0.6pt,line join=round] (180.07,244.04) --
	(180.07,362.10);

\path[draw=drawColor,line width= 0.6pt,line join=round] (255.80,244.04) --
	(255.80,362.10);

\path[draw=drawColor,line width= 0.6pt,line join=round] (331.32,244.04) --
	(331.32,362.10);
\definecolor{drawColor}{RGB}{237,144,164}

\path[draw=drawColor,line width= 0.6pt,line join=round] (123.16,249.41) --
	(129.37,249.41) --
	(129.37,249.67) --
	(135.79,249.67) --
	(135.79,249.67) --
	(141.99,249.67) --
	(141.99,250.24) --
	(148.41,250.24) --
	(148.41,249.97) --
	(154.82,249.97) --
	(154.82,250.10) --
	(161.03,250.10) --
	(161.03,250.31) --
	(167.44,250.31) --
	(167.44,250.27) --
	(173.65,250.27) --
	(173.65,250.00) --
	(180.07,250.00) --
	(180.07,250.50) --
	(186.48,250.50) --
	(186.48,250.70) --
	(192.48,250.70) --
	(192.48,252.43) --
	(198.90,252.43) --
	(198.90,253.81) --
	(205.10,253.81) --
	(205.10,254.56) --
	(211.52,254.56) --
	(211.52,252.25) --
	(217.73,252.25) --
	(217.73,251.68) --
	(224.14,251.68) --
	(224.14,254.58) --
	(230.55,254.58) --
	(230.55,254.35) --
	(236.76,254.35) --
	(236.76,254.32) --
	(243.18,254.32) --
	(243.18,254.45) --
	(249.38,254.45) --
	(249.38,255.63) --
	(255.80,255.63) --
	(255.80,256.58) --
	(262.21,256.58) --
	(262.21,253.12) --
	(268.01,253.12) --
	(268.01,251.92) --
	(274.42,251.92) --
	(274.42,252.41) --
	(280.63,252.41) --
	(280.63,252.88) --
	(287.04,252.88) --
	(287.04,255.62) --
	(293.25,255.62) --
	(293.25,268.00) --
	(299.67,268.00) --
	(299.67,260.94) --
	(306.08,260.94) --
	(306.08,253.95) --
	(312.29,253.95) --
	(312.29,255.69) --
	(318.70,255.69) --
	(318.70,255.81) --
	(324.91,255.81) --
	(324.91,252.88) --
	(331.32,252.88) --
	(331.32,254.05) --
	(337.74,254.05) --
	(337.74,254.70);
\definecolor{drawColor}{RGB}{0,193,178}

\path[draw=drawColor,line width= 0.6pt,line join=round] ( 85.50,249.41) --
	( 91.92,249.41) --
	( 91.92,252.40) --
	( 98.13,252.40) --
	( 98.13,254.21) --
	(104.54,254.21) --
	(104.54,263.88) --
	(110.96,263.88) --
	(110.96,261.10) --
	(116.75,261.10) --
	(116.75,269.51) --
	(123.16,269.51) --
	(123.16,267.23) --
	(129.37,267.23) --
	(129.37,265.18) --
	(135.79,265.18) --
	(135.79,276.13) --
	(141.99,276.13) --
	(141.99,270.06) --
	(148.41,270.06) --
	(148.41,356.73) --
	(154.82,356.73) --
	(154.82,324.30) --
	(161.03,324.30) --
	(161.03,276.13) --
	(167.44,276.13) --
	(167.44,272.49) --
	(173.65,272.49) --
	(173.65,279.44) --
	(180.07,279.44) --
	(180.07,291.00) --
	(186.48,291.00) --
	(186.48,287.87) --
	(192.48,287.87) --
	(192.48,294.99) --
	(198.90,294.99) --
	(198.90,292.91) --
	(205.10,292.91) --
	(205.10,302.72) --
	(211.52,302.72) --
	(211.52,287.45) --
	(217.73,287.45) --
	(217.73,302.48) --
	(224.14,302.48) --
	(224.14,300.29) --
	(230.55,300.29) --
	(230.55,295.00) --
	(236.76,295.00) --
	(236.76,287.45) --
	(243.18,287.45) --
	(243.18,289.48) --
	(249.38,289.48) --
	(249.38,286.72) --
	(255.80,286.72) --
	(255.80,285.10) --
	(262.21,285.10) --
	(262.21,277.57) --
	(268.01,277.57) --
	(268.01,273.99) --
	(274.42,273.99) --
	(274.42,272.91) --
	(280.63,272.91) --
	(280.63,276.95) --
	(287.04,276.95) --
	(287.04,270.21) --
	(293.25,270.21) --
	(293.25,274.95) --
	(299.67,274.95) --
	(299.67,274.63) --
	(306.08,274.63) --
	(306.08,272.51) --
	(312.29,272.51) --
	(312.29,270.61) --
	(318.70,270.61) --
	(318.70,267.89) --
	(324.91,267.89) --
	(324.91,275.99) --
	(331.32,275.99) --
	(331.32,269.01) --
	(337.74,269.01) --
	(337.74,270.91);
\end{scope}
\begin{scope}
\path[clip] (  0.00,  0.00) rectangle (361.35,390.26);
\definecolor{drawColor}{gray}{0.30}

\node[text=drawColor,anchor=base east,inner sep=0pt, outer sep=0pt, scale=  0.88] at ( 67.94,246.37) {0};

\node[text=drawColor,anchor=base east,inner sep=0pt, outer sep=0pt, scale=  0.88] at ( 67.94,277.20) {5,000};

\node[text=drawColor,anchor=base east,inner sep=0pt, outer sep=0pt, scale=  0.88] at ( 67.94,308.04) {10,000};

\node[text=drawColor,anchor=base east,inner sep=0pt, outer sep=0pt, scale=  0.88] at ( 67.94,338.87) {15,000};
\end{scope}
\begin{scope}
\path[clip] (  0.00,  0.00) rectangle (361.35,390.26);
\definecolor{drawColor}{gray}{0.30}

\node[text=drawColor,anchor=base,inner sep=0pt, outer sep=0pt, scale=  0.88] at (104.54,233.03) {2019};

\node[text=drawColor,anchor=base,inner sep=0pt, outer sep=0pt, scale=  0.88] at (180.07,233.03) {2020};

\node[text=drawColor,anchor=base,inner sep=0pt, outer sep=0pt, scale=  0.88] at (255.80,233.03) {2021};

\node[text=drawColor,anchor=base,inner sep=0pt, outer sep=0pt, scale=  0.88] at (331.32,233.03) {2022};
\end{scope}
\begin{scope}
\path[clip] (  0.00,  0.00) rectangle (361.35,390.26);
\definecolor{drawColor}{RGB}{0,0,0}

\node[text=drawColor,rotate= 90.00,anchor=base,inner sep=0pt, outer sep=0pt, scale=  1.10] at ( 18.58,303.07) {Total mixed output (BTC)};
\end{scope}
\begin{scope}
\path[clip] (  0.00,  0.00) rectangle (361.35,390.26);
\definecolor{drawColor}{RGB}{0,0,0}

\node[text=drawColor,anchor=base west,inner sep=0pt, outer sep=0pt, scale=  1.32] at ( 72.89,370.17) {Mixed amount in BTC};
\end{scope}
\begin{scope}
\path[clip] ( 72.89, 72.64) rectangle (350.35,190.70);
\definecolor{drawColor}{gray}{0.92}

\path[draw=drawColor,line width= 0.3pt,line join=round] ( 72.89, 90.33) --
	(350.35, 90.33);

\path[draw=drawColor,line width= 0.3pt,line join=round] ( 72.89,114.98) --
	(350.35,114.98);

\path[draw=drawColor,line width= 0.3pt,line join=round] ( 72.89,139.62) --
	(350.35,139.62);

\path[draw=drawColor,line width= 0.3pt,line join=round] ( 72.89,164.27) --
	(350.35,164.27);

\path[draw=drawColor,line width= 0.3pt,line join=round] ( 72.89,188.92) --
	(350.35,188.92);

\path[draw=drawColor,line width= 0.3pt,line join=round] (142.30, 72.64) --
	(142.30,190.70);

\path[draw=drawColor,line width= 0.3pt,line join=round] (217.93, 72.64) --
	(217.93,190.70);

\path[draw=drawColor,line width= 0.3pt,line join=round] (293.56, 72.64) --
	(293.56,190.70);

\path[draw=drawColor,line width= 0.6pt,line join=round] ( 72.89, 78.00) --
	(350.35, 78.00);

\path[draw=drawColor,line width= 0.6pt,line join=round] ( 72.89,102.65) --
	(350.35,102.65);

\path[draw=drawColor,line width= 0.6pt,line join=round] ( 72.89,127.30) --
	(350.35,127.30);

\path[draw=drawColor,line width= 0.6pt,line join=round] ( 72.89,151.95) --
	(350.35,151.95);

\path[draw=drawColor,line width= 0.6pt,line join=round] ( 72.89,176.60) --
	(350.35,176.60);

\path[draw=drawColor,line width= 0.6pt,line join=round] (104.54, 72.64) --
	(104.54,190.70);

\path[draw=drawColor,line width= 0.6pt,line join=round] (180.07, 72.64) --
	(180.07,190.70);

\path[draw=drawColor,line width= 0.6pt,line join=round] (255.80, 72.64) --
	(255.80,190.70);

\path[draw=drawColor,line width= 0.6pt,line join=round] (331.32, 72.64) --
	(331.32,190.70);
\definecolor{drawColor}{RGB}{237,144,164}

\path[draw=drawColor,line width= 0.6pt,line join=round] (123.16, 78.01) --
	(129.37, 78.01) --
	(129.37, 78.14) --
	(135.79, 78.14) --
	(135.79, 78.22) --
	(141.99, 78.22) --
	(141.99, 78.70) --
	(148.41, 78.70) --
	(148.41, 78.49) --
	(154.82, 78.49) --
	(154.82, 78.54) --
	(161.03, 78.54) --
	(161.03, 78.61) --
	(167.44, 78.61) --
	(167.44, 78.59) --
	(173.65, 78.59) --
	(173.65, 78.35) --
	(180.07, 78.35) --
	(180.07, 78.76) --
	(186.48, 78.76) --
	(186.48, 78.99) --
	(192.48, 78.99) --
	(192.48, 79.70) --
	(198.90, 79.70) --
	(198.90, 80.53) --
	(205.10, 80.53) --
	(205.10, 81.80) --
	(211.52, 81.80) --
	(211.52, 80.15) --
	(217.73, 80.15) --
	(217.73, 79.74) --
	(224.14, 79.74) --
	(224.14, 82.82) --
	(230.55, 82.82) --
	(230.55, 82.23) --
	(236.76, 82.23) --
	(236.76, 82.56) --
	(243.18, 82.56) --
	(243.18, 84.73) --
	(249.38, 84.73) --
	(249.38, 88.91) --
	(255.80, 88.91) --
	(255.80, 97.78) --
	(262.21, 97.78) --
	(262.21, 90.83) --
	(268.01, 90.83) --
	(268.01, 88.93) --
	(274.42, 88.93) --
	(274.42, 92.03) --
	(280.63, 92.03) --
	(280.63, 90.16) --
	(287.04, 90.16) --
	(287.04, 96.37) --
	(293.25, 96.37) --
	(293.25,130.27) --
	(299.67,130.27) --
	(299.67,118.27) --
	(306.08,118.27) --
	(306.08, 95.03) --
	(312.29, 95.03) --
	(312.29,107.17) --
	(318.70,107.17) --
	(318.70,109.23) --
	(324.91,109.23) --
	(324.91, 91.78) --
	(331.32, 91.78) --
	(331.32, 93.28) --
	(337.74, 93.28) --
	(337.74, 94.92);
\definecolor{drawColor}{RGB}{0,193,178}

\path[draw=drawColor,line width= 0.6pt,line join=round] ( 85.50, 78.01) --
	( 91.92, 78.01) --
	( 91.92, 79.32) --
	( 98.13, 79.32) --
	( 98.13, 79.43) --
	(104.54, 79.43) --
	(104.54, 82.23) --
	(110.96, 82.23) --
	(110.96, 81.48) --
	(116.75, 81.48) --
	(116.75, 84.36) --
	(123.16, 84.36) --
	(123.16, 85.35) --
	(129.37, 85.35) --
	(129.37, 87.23) --
	(135.79, 87.23) --
	(135.79, 98.06) --
	(141.99, 98.06) --
	(141.99, 95.68) --
	(148.41, 95.68) --
	(148.41,169.59) --
	(154.82,169.59) --
	(154.82,139.14) --
	(161.03,139.14) --
	(161.03, 96.01) --
	(167.44, 96.01) --
	(167.44, 93.52) --
	(173.65, 93.52) --
	(173.65, 95.41) --
	(180.07, 95.41) --
	(180.07,106.21) --
	(186.48,106.21) --
	(186.48,107.53) --
	(192.48,107.53) --
	(192.48,103.04) --
	(198.90,103.04) --
	(198.90,103.00) --
	(205.10,103.00) --
	(205.10,117.31) --
	(211.52,117.31) --
	(211.52,106.84) --
	(217.73,106.84) --
	(217.73,118.80) --
	(224.14,118.80) --
	(224.14,125.17) --
	(230.55,125.17) --
	(230.55,116.98) --
	(236.76,116.98) --
	(236.76,113.53) --
	(243.18,113.53) --
	(243.18,130.73) --
	(249.38,130.73) --
	(249.38,143.97) --
	(255.80,143.97) --
	(255.80,176.50) --
	(262.21,176.50) --
	(262.21,182.39) --
	(268.01,182.39) --
	(268.01,185.33) --
	(274.42,185.33) --
	(274.42,183.48) --
	(280.63,183.48) --
	(280.63,178.54) --
	(287.04,178.54) --
	(287.04,138.19) --
	(293.25,138.19) --
	(293.25,149.07) --
	(299.67,149.07) --
	(299.67,170.16) --
	(306.08,170.16) --
	(306.08,162.80) --
	(312.29,162.80) --
	(312.29,176.19) --
	(318.70,176.19) --
	(318.70,167.82) --
	(324.91,167.82) --
	(324.91,182.11) --
	(331.32,182.11) --
	(331.32,142.65) --
	(337.74,142.65) --
	(337.74,147.95);
\end{scope}
\begin{scope}
\path[clip] (  0.00,  0.00) rectangle (361.35,390.26);
\definecolor{drawColor}{gray}{0.30}

\node[text=drawColor,anchor=base east,inner sep=0pt, outer sep=0pt, scale=  0.88] at ( 67.94, 74.97) {0};

\node[text=drawColor,anchor=base east,inner sep=0pt, outer sep=0pt, scale=  0.88] at ( 67.94, 99.62) {50,000,000};

\node[text=drawColor,anchor=base east,inner sep=0pt, outer sep=0pt, scale=  0.88] at ( 67.94,124.27) {100,000,000};

\node[text=drawColor,anchor=base east,inner sep=0pt, outer sep=0pt, scale=  0.88] at ( 67.94,148.92) {150,000,000};

\node[text=drawColor,anchor=base east,inner sep=0pt, outer sep=0pt, scale=  0.88] at ( 67.94,173.57) {200,000,000};
\end{scope}
\begin{scope}
\path[clip] (  0.00,  0.00) rectangle (361.35,390.26);
\definecolor{drawColor}{gray}{0.30}

\node[text=drawColor,anchor=base,inner sep=0pt, outer sep=0pt, scale=  0.88] at (104.54, 61.63) {2019};

\node[text=drawColor,anchor=base,inner sep=0pt, outer sep=0pt, scale=  0.88] at (180.07, 61.63) {2020};

\node[text=drawColor,anchor=base,inner sep=0pt, outer sep=0pt, scale=  0.88] at (255.80, 61.63) {2021};

\node[text=drawColor,anchor=base,inner sep=0pt, outer sep=0pt, scale=  0.88] at (331.32, 61.63) {2022};
\end{scope}
\begin{scope}
\path[clip] (  0.00,  0.00) rectangle (361.35,390.26);
\definecolor{drawColor}{RGB}{0,0,0}

\node[text=drawColor,anchor=base,inner sep=0pt, outer sep=0pt, scale=  1.10] at (211.62, 49.59) {Month};
\end{scope}
\begin{scope}
\path[clip] (  0.00,  0.00) rectangle (361.35,390.26);
\definecolor{drawColor}{RGB}{0,0,0}

\node[text=drawColor,rotate= 90.00,anchor=base,inner sep=0pt, outer sep=0pt, scale=  1.10] at ( 18.58,131.67) {Total mixed output (USD)};
\end{scope}
\begin{scope}
\path[clip] (  0.00,  0.00) rectangle (361.35,390.26);
\definecolor{drawColor}{RGB}{0,0,0}

\node[text=drawColor,anchor=base west,inner sep=0pt, outer sep=0pt, scale=  1.32] at ( 72.89,198.76) {Mixed amount in USD};
\end{scope}
\begin{scope}
\path[clip] (  0.00,  0.00) rectangle (361.35,390.26);
\definecolor{drawColor}{RGB}{0,0,0}

\node[text=drawColor,anchor=base west,inner sep=0pt, outer sep=0pt, scale=  1.10] at (137.15, 19.94) {Service};
\end{scope}
\begin{scope}
\path[clip] (  0.00,  0.00) rectangle (361.35,390.26);
\definecolor{drawColor}{RGB}{237,144,164}

\path[draw=drawColor,line width= 0.6pt,line join=round] (178.03, 23.73) -- (189.59, 23.73);
\end{scope}
\begin{scope}
\path[clip] (  0.00,  0.00) rectangle (361.35,390.26);
\definecolor{drawColor}{RGB}{0,193,178}

\path[draw=drawColor,line width= 0.6pt,line join=round] (239.68, 23.73) -- (251.24, 23.73);
\end{scope}
\begin{scope}
\path[clip] (  0.00,  0.00) rectangle (361.35,390.26);
\definecolor{drawColor}{RGB}{0,0,0}

\node[text=drawColor,anchor=base west,inner sep=0pt, outer sep=0pt, scale=  0.88] at (196.54, 20.70) {Samourai};
\end{scope}
\begin{scope}
\path[clip] (  0.00,  0.00) rectangle (361.35,390.26);
\definecolor{drawColor}{RGB}{0,0,0}

\node[text=drawColor,anchor=base west,inner sep=0pt, outer sep=0pt, scale=  0.88] at (258.19, 20.70) {Wasabi};
\end{scope}
\end{tikzpicture}

%% file: graphics/exchange_txs_no_txs.tex
% !TEX encoding = UTF-8 Unicode
\begin{tikzpicture}[x=1pt,y=1pt]
\definecolor{fillColor}{RGB}{255,255,255}
\path[use as bounding box,fill=fillColor,fill opacity=0.00] (0,0) rectangle (361.35,361.35);
\begin{scope}
\path[clip] (  0.00,  0.00) rectangle (361.35,361.35);
\definecolor{drawColor}{RGB}{0,0,0}

\node[text=drawColor,rotate= 90.00,anchor=base,inner sep=0pt, outer sep=0pt, scale=  1.10] at ( 15.02,210.52) {Cumulative number of exchange transactions};
\end{scope}
\begin{scope}
\path[clip] ( 64.30,237.19) rectangle (350.35,333.19);
\definecolor{drawColor}{gray}{0.92}

\path[draw=drawColor,line width= 0.3pt,line join=round] ( 64.30,252.73) --
	(350.35,252.73);

\path[draw=drawColor,line width= 0.3pt,line join=round] ( 64.30,275.09) --
	(350.35,275.09);

\path[draw=drawColor,line width= 0.3pt,line join=round] ( 64.30,297.44) --
	(350.35,297.44);

\path[draw=drawColor,line width= 0.3pt,line join=round] ( 64.30,319.79) --
	(350.35,319.79);

\path[draw=drawColor,line width= 0.3pt,line join=round] (135.86,237.19) --
	(135.86,333.19);

\path[draw=drawColor,line width= 0.3pt,line join=round] (213.83,237.19) --
	(213.83,333.19);

\path[draw=drawColor,line width= 0.3pt,line join=round] (291.80,237.19) --
	(291.80,333.19);

\path[draw=drawColor,line width= 0.6pt,line join=round] ( 64.30,241.56) --
	(350.35,241.56);

\path[draw=drawColor,line width= 0.6pt,line join=round] ( 64.30,263.91) --
	(350.35,263.91);

\path[draw=drawColor,line width= 0.6pt,line join=round] ( 64.30,286.26) --
	(350.35,286.26);

\path[draw=drawColor,line width= 0.6pt,line join=round] ( 64.30,308.62) --
	(350.35,308.62);

\path[draw=drawColor,line width= 0.6pt,line join=round] ( 64.30,330.97) --
	(350.35,330.97);

\path[draw=drawColor,line width= 0.6pt,line join=round] ( 96.92,237.19) --
	( 96.92,333.19);

\path[draw=drawColor,line width= 0.6pt,line join=round] (174.79,237.19) --
	(174.79,333.19);

\path[draw=drawColor,line width= 0.6pt,line join=round] (252.87,237.19) --
	(252.87,333.19);

\path[draw=drawColor,line width= 0.6pt,line join=round] (330.73,237.19) --
	(330.73,333.19);
\definecolor{drawColor}{RGB}{237,144,164}

\path[draw=drawColor,line width= 0.6pt,line join=round] (122.52,241.56) --
	(129.14,241.58) --
	(135.54,241.59) --
	(142.15,241.66) --
	(148.76,241.73) --
	(155.16,241.80) --
	(161.78,241.99) --
	(168.18,242.09) --
	(174.79,242.20) --
	(181.40,242.35) --
	(187.59,242.65) --
	(194.20,242.98) --
	(200.60,243.36) --
	(207.22,243.94) --
	(213.62,244.48) --
	(220.23,245.09) --
	(226.84,245.55) --
	(233.24,246.09) --
	(239.86,246.73) --
	(246.26,247.35) --
	(252.87,247.94) --
	(259.48,248.46) --
	(265.46,249.04) --
	(272.07,249.63) --
	(278.47,250.35) --
	(285.08,250.89) --
	(291.48,251.51) --
	(298.09,252.26) --
	(304.71,252.86) --
	(311.11,253.78) --
	(317.72,254.81) --
	(324.12,255.60) --
	(330.73,256.37) --
	(337.35,257.02);
\definecolor{drawColor}{RGB}{0,193,178}

\path[draw=drawColor,line width= 0.6pt,line join=round] ( 77.30,241.56) --
	( 83.91,241.66) --
	( 90.31,241.87) --
	( 96.92,242.20) --
	(103.54,242.83) --
	(109.51,243.49) --
	(116.12,244.65) --
	(122.52,245.84) --
	(129.14,247.20) --
	(135.54,248.58) --
	(142.15,250.25) --
	(148.76,252.15) --
	(155.16,253.95) --
	(161.78,255.71) --
	(168.18,257.45) --
	(174.79,259.27) --
	(181.40,260.88) --
	(187.59,263.05) --
	(194.20,265.31) --
	(200.60,268.38) --
	(207.22,271.63) --
	(213.62,275.16) --
	(220.23,279.18) --
	(226.84,282.73) --
	(233.24,286.40) --
	(239.86,290.75) --
	(246.26,295.02) --
	(252.87,299.11) --
	(259.48,302.09) --
	(265.46,304.79) --
	(272.07,307.20) --
	(278.47,310.12) --
	(285.08,312.08) --
	(291.48,313.98) --
	(298.09,316.55) --
	(304.71,318.82) --
	(311.11,321.00) --
	(317.72,323.25) --
	(324.12,325.20) --
	(330.73,327.17) --
	(337.35,328.83);
\end{scope}
\begin{scope}
\path[clip] (  0.00,  0.00) rectangle (361.35,361.35);
\definecolor{drawColor}{gray}{0.30}

\node[text=drawColor,anchor=base east,inner sep=0pt, outer sep=0pt, scale=  0.88] at ( 59.35,238.53) {0};

\node[text=drawColor,anchor=base east,inner sep=0pt, outer sep=0pt, scale=  0.88] at ( 59.35,260.88) {10,000};

\node[text=drawColor,anchor=base east,inner sep=0pt, outer sep=0pt, scale=  0.88] at ( 59.35,283.23) {20,000};

\node[text=drawColor,anchor=base east,inner sep=0pt, outer sep=0pt, scale=  0.88] at ( 59.35,305.59) {30,000};

\node[text=drawColor,anchor=base east,inner sep=0pt, outer sep=0pt, scale=  0.88] at ( 59.35,327.94) {40,000};
\end{scope}
\begin{scope}
\path[clip] (  0.00,  0.00) rectangle (361.35,361.35);
\definecolor{drawColor}{gray}{0.30}

\node[text=drawColor,anchor=base,inner sep=0pt, outer sep=0pt, scale=  0.88] at ( 96.92,226.18) {2019};

\node[text=drawColor,anchor=base,inner sep=0pt, outer sep=0pt, scale=  0.88] at (174.79,226.18) {2020};

\node[text=drawColor,anchor=base,inner sep=0pt, outer sep=0pt, scale=  0.88] at (252.87,226.18) {2021};

\node[text=drawColor,anchor=base,inner sep=0pt, outer sep=0pt, scale=  0.88] at (330.73,226.18) {2022};
\end{scope}
\begin{scope}
\path[clip] (  0.00,  0.00) rectangle (361.35,361.35);
\definecolor{drawColor}{RGB}{0,0,0}

\node[text=drawColor,rotate= 90.00,anchor=base,inner sep=0pt, outer sep=0pt, scale=  1.10] at ( 30.02,285.19) { };
\end{scope}
\begin{scope}
\path[clip] (  0.00,  0.00) rectangle (361.35,361.35);
\definecolor{drawColor}{RGB}{0,0,0}

\node[text=drawColor,anchor=base west,inner sep=0pt, outer sep=0pt, scale=  1.32] at ( 64.30,341.26) {Level 1};
\end{scope}
\begin{scope}
\path[clip] ( 64.30, 87.85) rectangle (350.35,183.85);
\definecolor{drawColor}{gray}{0.92}

\path[draw=drawColor,line width= 0.3pt,line join=round] ( 64.30,103.79) --
	(350.35,103.79);

\path[draw=drawColor,line width= 0.3pt,line join=round] ( 64.30,126.94) --
	(350.35,126.94);

\path[draw=drawColor,line width= 0.3pt,line join=round] ( 64.30,150.08) --
	(350.35,150.08);

\path[draw=drawColor,line width= 0.3pt,line join=round] ( 64.30,173.23) --
	(350.35,173.23);

\path[draw=drawColor,line width= 0.3pt,line join=round] (130.60, 87.85) --
	(130.60,183.85);

\path[draw=drawColor,line width= 0.3pt,line join=round] (210.61, 87.85) --
	(210.61,183.85);

\path[draw=drawColor,line width= 0.3pt,line join=round] (290.61, 87.85) --
	(290.61,183.85);

\path[draw=drawColor,line width= 0.6pt,line join=round] ( 64.30, 92.22) --
	(350.35, 92.22);

\path[draw=drawColor,line width= 0.6pt,line join=round] ( 64.30,115.36) --
	(350.35,115.36);

\path[draw=drawColor,line width= 0.6pt,line join=round] ( 64.30,138.51) --
	(350.35,138.51);

\path[draw=drawColor,line width= 0.6pt,line join=round] ( 64.30,161.66) --
	(350.35,161.66);

\path[draw=drawColor,line width= 0.6pt,line join=round] ( 90.65, 87.85) --
	( 90.65,183.85);

\path[draw=drawColor,line width= 0.6pt,line join=round] (170.55, 87.85) --
	(170.55,183.85);

\path[draw=drawColor,line width= 0.6pt,line join=round] (250.66, 87.85) --
	(250.66,183.85);

\path[draw=drawColor,line width= 0.6pt,line join=round] (330.56, 87.85) --
	(330.56,183.85);
\definecolor{drawColor}{RGB}{237,144,164}

\path[draw=drawColor,line width= 0.6pt,line join=round] (110.35, 92.22) --
	(116.92, 92.22) --
	(123.70, 92.24) --
	(130.27, 92.27) --
	(137.06, 92.32) --
	(143.84, 92.37) --
	(150.41, 92.46) --
	(157.20, 92.59) --
	(163.76, 92.68) --
	(170.55, 92.77) --
	(177.33, 92.90) --
	(183.68, 93.13) --
	(190.47, 93.38) --
	(197.03, 93.73) --
	(203.82, 95.14) --
	(210.39, 96.86) --
	(217.17, 98.89) --
	(223.96,100.81) --
	(230.53,102.72) --
	(237.31,104.90) --
	(243.88,107.10) --
	(250.66,109.18) --
	(257.45,110.77) --
	(263.58,112.23) --
	(270.36,113.53) --
	(276.93,115.12) --
	(283.72,116.21) --
	(290.28,117.38) --
	(297.07,118.90) --
	(303.86,120.18) --
	(310.42,121.66) --
	(317.21,123.15) --
	(323.78,124.42) --
	(330.56,125.70) --
	(337.35,126.78);

\path[draw=drawColor,line width= 0.6pt,dash pattern=on 2pt off 2pt ,line join=round] (110.35, 92.22) --
	(116.92, 92.22) --
	(123.70, 92.24) --
	(130.27, 92.27) --
	(137.06, 92.31) --
	(143.84, 92.37) --
	(150.41, 92.45) --
	(157.20, 92.58) --
	(163.76, 92.68) --
	(170.55, 92.77) --
	(177.33, 92.90) --
	(183.68, 93.12) --
	(190.47, 93.37) --
	(197.03, 93.70) --
	(203.82, 95.10) --
	(210.39, 96.81) --
	(217.17, 98.82) --
	(223.96,100.72) --
	(230.53,102.62) --
	(237.31,104.78) --
	(243.88,106.96) --
	(250.66,109.01) --
	(257.45,110.57) --
	(263.58,112.00) --
	(270.36,113.28) --
	(276.93,114.84) --
	(283.72,115.92) --
	(290.28,117.07) --
	(297.07,118.57) --
	(303.86,119.85) --
	(310.42,121.31) --
	(317.21,122.77) --
	(323.78,124.02) --
	(330.56,125.28) --
	(337.35,126.35);

\path[draw=drawColor,line width= 0.6pt,dash pattern=on 4pt off 2pt ,line join=round] (110.35, 92.22) --
	(116.92, 92.22) --
	(123.70, 92.24) --
	(130.27, 92.27) --
	(137.06, 92.31) --
	(143.84, 92.37) --
	(150.41, 92.45) --
	(157.20, 92.57) --
	(163.76, 92.67) --
	(170.55, 92.76) --
	(177.33, 92.88) --
	(183.68, 93.11) --
	(190.47, 93.35) --
	(197.03, 93.66) --
	(203.82, 95.06) --
	(210.39, 96.74) --
	(217.17, 98.70) --
	(223.96,100.45) --
	(230.53,102.31) --
	(237.31,104.43) --
	(243.88,106.57) --
	(250.66,108.57) --
	(257.45,110.10) --
	(263.58,111.50) --
	(270.36,112.76) --
	(276.93,114.31) --
	(283.72,115.37) --
	(290.28,116.50) --
	(297.07,117.97) --
	(303.86,119.22) --
	(310.42,120.66) --
	(317.21,122.10) --
	(323.78,123.33) --
	(330.56,124.58) --
	(337.35,125.61);
\definecolor{drawColor}{RGB}{0,193,178}

\path[draw=drawColor,line width= 0.6pt,line join=round] ( 77.30, 92.33) --
	( 83.86, 92.56) --
	( 90.65, 92.83) --
	( 97.44, 93.29) --
	(103.57, 93.81) --
	(110.35, 94.62) --
	(116.92, 95.52) --
	(123.70, 96.65) --
	(130.27, 97.80) --
	(137.06, 99.26) --
	(143.84,100.74) --
	(150.41,102.54) --
	(157.20,104.15) --
	(163.76,105.86) --
	(170.55,107.76) --
	(177.33,109.63) --
	(183.68,111.83) --
	(190.47,113.95) --
	(197.03,116.64) --
	(203.82,119.23) --
	(210.39,122.18) --
	(217.17,125.59) --
	(223.96,129.03) --
	(230.53,133.53) --
	(237.31,138.57) --
	(243.88,143.46) --
	(250.66,147.15) --
	(257.45,150.26) --
	(263.58,152.87) --
	(270.36,155.28) --
	(276.93,158.13) --
	(283.72,160.17) --
	(290.28,162.17) --
	(297.07,164.76) --
	(303.86,167.15) --
	(310.42,169.83) --
	(317.21,172.66) --
	(323.78,175.10) --
	(330.56,177.37) --
	(337.35,179.49);

\path[draw=drawColor,line width= 0.6pt,dash pattern=on 2pt off 2pt ,line join=round] ( 77.30, 92.33) --
	( 83.86, 92.55) --
	( 90.65, 92.82) --
	( 97.44, 93.28) --
	(103.57, 93.80) --
	(110.35, 94.59) --
	(116.92, 95.48) --
	(123.70, 96.60) --
	(130.27, 97.72) --
	(137.06, 99.16) --
	(143.84,100.61) --
	(150.41,102.34) --
	(157.20,103.87) --
	(163.76,105.50) --
	(170.55,107.27) --
	(177.33,109.03) --
	(183.68,111.16) --
	(190.47,113.23) --
	(197.03,115.87) --
	(203.82,118.37) --
	(210.39,121.27) --
	(217.17,124.60) --
	(223.96,127.65) --
	(230.53,130.85) --
	(237.31,134.60) --
	(243.88,138.45) --
	(250.66,142.03) --
	(257.45,145.04) --
	(263.58,147.58) --
	(270.36,149.87) --
	(276.93,152.62) --
	(283.72,154.58) --
	(290.28,156.50) --
	(297.07,158.93) --
	(303.86,161.21) --
	(310.42,163.79) --
	(317.21,166.48) --
	(323.78,168.79) --
	(330.56,170.89) --
	(337.35,172.68);

\path[draw=drawColor,line width= 0.6pt,dash pattern=on 4pt off 2pt ,line join=round] ( 77.30, 92.31) --
	( 83.86, 92.50) --
	( 90.65, 92.77) --
	( 97.44, 93.21) --
	(103.57, 93.72) --
	(110.35, 94.49) --
	(116.92, 95.36) --
	(123.70, 96.47) --
	(130.27, 97.56) --
	(137.06, 98.97) --
	(143.84,100.40) --
	(150.41,102.08) --
	(157.20,103.57) --
	(163.76,105.18) --
	(170.55,106.91) --
	(177.33,108.64) --
	(183.68,110.71) --
	(190.47,112.63) --
	(197.03,115.18) --
	(203.82,117.62) --
	(210.39,120.47) --
	(217.17,123.69) --
	(223.96,126.63) --
	(230.53,129.76) --
	(237.31,133.39) --
	(243.88,137.09) --
	(250.66,140.54) --
	(257.45,143.42) --
	(263.58,145.88) --
	(270.36,148.07) --
	(276.93,150.72) --
	(283.72,152.61) --
	(290.28,154.45) --
	(297.07,156.77) --
	(303.86,158.95) --
	(310.42,161.41) --
	(317.21,163.96) --
	(323.78,166.08) --
	(330.56,168.06) --
	(337.35,169.71);
\end{scope}
\begin{scope}
\path[clip] (  0.00,  0.00) rectangle (361.35,361.35);
\definecolor{drawColor}{gray}{0.30}

\node[text=drawColor,anchor=base east,inner sep=0pt, outer sep=0pt, scale=  0.88] at ( 59.35, 89.19) {0};

\node[text=drawColor,anchor=base east,inner sep=0pt, outer sep=0pt, scale=  0.88] at ( 59.35,112.33) {25,000};

\node[text=drawColor,anchor=base east,inner sep=0pt, outer sep=0pt, scale=  0.88] at ( 59.35,135.48) {50,000};

\node[text=drawColor,anchor=base east,inner sep=0pt, outer sep=0pt, scale=  0.88] at ( 59.35,158.63) {75,000};
\end{scope}
\begin{scope}
\path[clip] (  0.00,  0.00) rectangle (361.35,361.35);
\definecolor{drawColor}{gray}{0.30}

\node[text=drawColor,anchor=base,inner sep=0pt, outer sep=0pt, scale=  0.88] at ( 90.65, 76.84) {2019};

\node[text=drawColor,anchor=base,inner sep=0pt, outer sep=0pt, scale=  0.88] at (170.55, 76.84) {2020};

\node[text=drawColor,anchor=base,inner sep=0pt, outer sep=0pt, scale=  0.88] at (250.66, 76.84) {2021};

\node[text=drawColor,anchor=base,inner sep=0pt, outer sep=0pt, scale=  0.88] at (330.56, 76.84) {2022};
\end{scope}
\begin{scope}
\path[clip] (  0.00,  0.00) rectangle (361.35,361.35);
\definecolor{drawColor}{RGB}{0,0,0}

\node[text=drawColor,anchor=base,inner sep=0pt, outer sep=0pt, scale=  1.10] at (207.32, 64.81) {Month};
\end{scope}
\begin{scope}
\path[clip] (  0.00,  0.00) rectangle (361.35,361.35);
\definecolor{drawColor}{RGB}{0,0,0}

\node[text=drawColor,rotate= 90.00,anchor=base,inner sep=0pt, outer sep=0pt, scale=  1.10] at ( 30.02,135.85) { };
\end{scope}
\begin{scope}
\path[clip] (  0.00,  0.00) rectangle (361.35,361.35);
\definecolor{drawColor}{RGB}{0,0,0}

\node[text=drawColor,anchor=base west,inner sep=0pt, outer sep=0pt, scale=  1.10] at ( 90.74, 37.52) {Service};
\end{scope}
\begin{scope}
\path[clip] (  0.00,  0.00) rectangle (361.35,361.35);
\definecolor{drawColor}{RGB}{237,144,164}

\path[draw=drawColor,line width= 0.6pt,line join=round] ( 92.19, 23.73) -- (103.75, 23.73);
\end{scope}
\begin{scope}
\path[clip] (  0.00,  0.00) rectangle (361.35,361.35);
\definecolor{drawColor}{RGB}{0,193,178}

\path[draw=drawColor,line width= 0.6pt,line join=round] (153.83, 23.73) -- (165.40, 23.73);
\end{scope}
\begin{scope}
\path[clip] (  0.00,  0.00) rectangle (361.35,361.35);
\definecolor{drawColor}{RGB}{0,0,0}

\node[text=drawColor,anchor=base west,inner sep=0pt, outer sep=0pt, scale=  0.88] at (110.69, 20.70) {Samourai};
\end{scope}
\begin{scope}
\path[clip] (  0.00,  0.00) rectangle (361.35,361.35);
\definecolor{drawColor}{RGB}{0,0,0}

\node[text=drawColor,anchor=base west,inner sep=0pt, outer sep=0pt, scale=  0.88] at (172.34, 20.70) {Wasabi};
\end{scope}
\begin{scope}
\path[clip] (  0.00,  0.00) rectangle (361.35,361.35);
\definecolor{drawColor}{RGB}{0,0,0}

\node[text=drawColor,anchor=base west,inner sep=0pt, outer sep=0pt, scale=  1.10] at (222.25, 37.52) {Level 1 threshold};
\end{scope}
\begin{scope}
\path[clip] (  0.00,  0.00) rectangle (361.35,361.35);
\definecolor{drawColor}{RGB}{0,0,0}

\path[draw=drawColor,line width= 0.6pt,line join=round] (223.70, 23.73) -- (235.26, 23.73);
\end{scope}
\begin{scope}
\path[clip] (  0.00,  0.00) rectangle (361.35,361.35);
\definecolor{drawColor}{RGB}{0,0,0}

\path[draw=drawColor,line width= 0.6pt,dash pattern=on 2pt off 2pt ,line join=round] (262.35, 23.73) -- (273.91, 23.73);
\end{scope}
\begin{scope}
\path[clip] (  0.00,  0.00) rectangle (361.35,361.35);
\definecolor{drawColor}{RGB}{0,0,0}

\path[draw=drawColor,line width= 0.6pt,dash pattern=on 4pt off 2pt ,line join=round] (296.60, 23.73) -- (308.16, 23.73);
\end{scope}
\begin{scope}
\path[clip] (  0.00,  0.00) rectangle (361.35,361.35);
\definecolor{drawColor}{RGB}{0,0,0}

\node[text=drawColor,anchor=base west,inner sep=0pt, outer sep=0pt, scale=  0.88] at (242.20, 20.70) {100};
\end{scope}
\begin{scope}
\path[clip] (  0.00,  0.00) rectangle (361.35,361.35);
\definecolor{drawColor}{RGB}{0,0,0}

\node[text=drawColor,anchor=base west,inner sep=0pt, outer sep=0pt, scale=  0.88] at (280.86, 20.70) {75};
\end{scope}
\begin{scope}
\path[clip] (  0.00,  0.00) rectangle (361.35,361.35);
\definecolor{drawColor}{RGB}{0,0,0}

\node[text=drawColor,anchor=base west,inner sep=0pt, outer sep=0pt, scale=  0.88] at (315.11, 20.70) {50};
\end{scope}
\begin{scope}
\path[clip] (  0.00,  0.00) rectangle (361.35,361.35);
\definecolor{drawColor}{RGB}{0,0,0}

\node[text=drawColor,anchor=base west,inner sep=0pt, outer sep=0pt, scale=  1.32] at ( 64.30,191.92) {Level 2};
\end{scope}
\end{tikzpicture}

%% file: graphics/exchange_txs_amount.tex
% !TEX encoding = UTF-8 Unicode
\begin{tikzpicture}[x=1pt,y=1pt]
\definecolor{fillColor}{RGB}{255,255,255}
\path[use as bounding box,fill=fillColor,fill opacity=0.00] (0,0) rectangle (361.35,361.35);
\begin{scope}
\path[clip] (  0.00,  0.00) rectangle (361.35,361.35);
\definecolor{drawColor}{RGB}{0,0,0}

\node[text=drawColor,rotate= 90.00,anchor=base,inner sep=0pt, outer sep=0pt, scale=  1.10] at ( 15.02,210.52) {Cumulative amount of mixed BTC};
\end{scope}
\begin{scope}
\path[clip] ( 64.30,237.19) rectangle (350.35,333.19);
\definecolor{drawColor}{gray}{0.92}

\path[draw=drawColor,line width= 0.3pt,line join=round] ( 64.30,258.21) --
	(350.35,258.21);

\path[draw=drawColor,line width= 0.3pt,line join=round] ( 64.30,291.52) --
	(350.35,291.52);

\path[draw=drawColor,line width= 0.3pt,line join=round] ( 64.30,324.84) --
	(350.35,324.84);

\path[draw=drawColor,line width= 0.3pt,line join=round] (135.86,237.19) --
	(135.86,333.19);

\path[draw=drawColor,line width= 0.3pt,line join=round] (213.83,237.19) --
	(213.83,333.19);

\path[draw=drawColor,line width= 0.3pt,line join=round] (291.80,237.19) --
	(291.80,333.19);

\path[draw=drawColor,line width= 0.6pt,line join=round] ( 64.30,241.56) --
	(350.35,241.56);

\path[draw=drawColor,line width= 0.6pt,line join=round] ( 64.30,274.87) --
	(350.35,274.87);

\path[draw=drawColor,line width= 0.6pt,line join=round] ( 64.30,308.18) --
	(350.35,308.18);

\path[draw=drawColor,line width= 0.6pt,line join=round] ( 96.92,237.19) --
	( 96.92,333.19);

\path[draw=drawColor,line width= 0.6pt,line join=round] (174.79,237.19) --
	(174.79,333.19);

\path[draw=drawColor,line width= 0.6pt,line join=round] (252.87,237.19) --
	(252.87,333.19);

\path[draw=drawColor,line width= 0.6pt,line join=round] (330.73,237.19) --
	(330.73,333.19);
\definecolor{drawColor}{RGB}{237,144,164}

\path[draw=drawColor,line width= 0.6pt,line join=round] (122.52,241.56) --
	(129.14,241.57) --
	(135.54,241.58) --
	(142.15,241.62) --
	(148.76,241.64) --
	(155.16,241.67) --
	(161.78,241.74) --
	(168.18,241.76) --
	(174.79,241.81) --
	(181.40,241.83) --
	(187.59,241.95) --
	(194.20,242.07) --
	(200.60,242.36) --
	(207.22,243.26) --
	(213.62,244.07) --
	(220.23,244.81) --
	(226.84,245.21) --
	(233.24,245.92) --
	(239.86,247.81) --
	(246.26,248.85) --
	(252.87,250.10) --
	(259.48,250.61) --
	(265.46,250.92) --
	(272.07,252.34) --
	(278.47,253.77) --
	(285.08,254.58) --
	(291.48,254.80) --
	(298.09,255.00) --
	(304.71,255.27) --
	(311.11,255.56) --
	(317.72,255.84) --
	(324.12,256.07) --
	(330.73,256.44) --
	(337.35,256.55);
\definecolor{drawColor}{RGB}{0,193,178}

\path[draw=drawColor,line width= 0.6pt,line join=round] ( 77.30,241.56) --
	( 83.91,241.71) --
	( 90.31,241.87) --
	( 96.92,242.30) --
	(103.54,243.29) --
	(109.51,244.17) --
	(116.12,245.43) --
	(122.52,246.69) --
	(129.14,248.79) --
	(135.54,250.04) --
	(142.15,252.40) --
	(148.76,253.92) --
	(155.16,255.60) --
	(161.78,256.99) --
	(168.18,258.42) --
	(174.79,260.44) --
	(181.40,262.02) --
	(187.59,265.21) --
	(194.20,267.67) --
	(200.60,271.38) --
	(207.22,274.31) --
	(213.62,277.92) --
	(220.23,281.74) --
	(226.84,284.77) --
	(233.24,288.79) --
	(239.86,294.76) --
	(246.26,299.20) --
	(252.87,303.24) --
	(259.48,305.94) --
	(265.46,308.52) --
	(272.07,311.95) --
	(278.47,315.53) --
	(285.08,317.62) --
	(291.48,319.00) --
	(298.09,320.86) --
	(304.71,322.48) --
	(311.11,323.89) --
	(317.72,325.14) --
	(324.12,326.37) --
	(330.73,327.95) --
	(337.35,328.83);
\end{scope}
\begin{scope}
\path[clip] (  0.00,  0.00) rectangle (361.35,361.35);
\definecolor{drawColor}{gray}{0.30}

\node[text=drawColor,anchor=base east,inner sep=0pt, outer sep=0pt, scale=  0.88] at ( 59.35,238.53) {0};

\node[text=drawColor,anchor=base east,inner sep=0pt, outer sep=0pt, scale=  0.88] at ( 59.35,271.84) {5,000};

\node[text=drawColor,anchor=base east,inner sep=0pt, outer sep=0pt, scale=  0.88] at ( 59.35,305.15) {10,000};
\end{scope}
\begin{scope}
\path[clip] (  0.00,  0.00) rectangle (361.35,361.35);
\definecolor{drawColor}{gray}{0.30}

\node[text=drawColor,anchor=base,inner sep=0pt, outer sep=0pt, scale=  0.88] at ( 96.92,226.18) {2019};

\node[text=drawColor,anchor=base,inner sep=0pt, outer sep=0pt, scale=  0.88] at (174.79,226.18) {2020};

\node[text=drawColor,anchor=base,inner sep=0pt, outer sep=0pt, scale=  0.88] at (252.87,226.18) {2021};

\node[text=drawColor,anchor=base,inner sep=0pt, outer sep=0pt, scale=  0.88] at (330.73,226.18) {2022};
\end{scope}
\begin{scope}
\path[clip] (  0.00,  0.00) rectangle (361.35,361.35);
\definecolor{drawColor}{RGB}{0,0,0}

\node[text=drawColor,rotate= 90.00,anchor=base,inner sep=0pt, outer sep=0pt, scale=  1.10] at ( 30.02,285.19) { };
\end{scope}
\begin{scope}
\path[clip] (  0.00,  0.00) rectangle (361.35,361.35);
\definecolor{drawColor}{RGB}{0,0,0}

\node[text=drawColor,anchor=base west,inner sep=0pt, outer sep=0pt, scale=  1.32] at ( 64.30,341.26) {Level 1};
\end{scope}
\begin{scope}
\path[clip] ( 64.30, 87.85) rectangle (350.35,183.85);
\definecolor{drawColor}{gray}{0.92}

\path[draw=drawColor,line width= 0.3pt,line join=round] ( 64.30,100.91) --
	(350.35,100.91);

\path[draw=drawColor,line width= 0.3pt,line join=round] ( 64.30,118.29) --
	(350.35,118.29);

\path[draw=drawColor,line width= 0.3pt,line join=round] ( 64.30,135.68) --
	(350.35,135.68);

\path[draw=drawColor,line width= 0.3pt,line join=round] ( 64.30,153.06) --
	(350.35,153.06);

\path[draw=drawColor,line width= 0.3pt,line join=round] ( 64.30,170.45) --
	(350.35,170.45);

\path[draw=drawColor,line width= 0.3pt,line join=round] (130.60, 87.85) --
	(130.60,183.85);

\path[draw=drawColor,line width= 0.3pt,line join=round] (210.61, 87.85) --
	(210.61,183.85);

\path[draw=drawColor,line width= 0.3pt,line join=round] (290.61, 87.85) --
	(290.61,183.85);

\path[draw=drawColor,line width= 0.6pt,line join=round] ( 64.30, 92.22) --
	(350.35, 92.22);

\path[draw=drawColor,line width= 0.6pt,line join=round] ( 64.30,109.60) --
	(350.35,109.60);

\path[draw=drawColor,line width= 0.6pt,line join=round] ( 64.30,126.99) --
	(350.35,126.99);

\path[draw=drawColor,line width= 0.6pt,line join=round] ( 64.30,144.37) --
	(350.35,144.37);

\path[draw=drawColor,line width= 0.6pt,line join=round] ( 64.30,161.75) --
	(350.35,161.75);

\path[draw=drawColor,line width= 0.6pt,line join=round] ( 64.30,179.14) --
	(350.35,179.14);

\path[draw=drawColor,line width= 0.6pt,line join=round] ( 90.65, 87.85) --
	( 90.65,183.85);

\path[draw=drawColor,line width= 0.6pt,line join=round] (170.55, 87.85) --
	(170.55,183.85);

\path[draw=drawColor,line width= 0.6pt,line join=round] (250.66, 87.85) --
	(250.66,183.85);

\path[draw=drawColor,line width= 0.6pt,line join=round] (330.56, 87.85) --
	(330.56,183.85);
\definecolor{drawColor}{RGB}{237,144,164}

\path[draw=drawColor,line width= 0.6pt,line join=round] (110.35, 92.22) --
	(116.92, 92.22) --
	(123.70, 92.24) --
	(130.27, 92.24) --
	(137.06, 92.29) --
	(143.84, 92.32) --
	(150.41, 92.38) --
	(157.20, 92.46) --
	(163.76, 92.50) --
	(170.55, 92.54) --
	(177.33, 92.58) --
	(183.68, 92.70) --
	(190.47, 92.77) --
	(197.03, 93.10) --
	(203.82, 94.04) --
	(210.39, 95.64) --
	(217.17, 97.57) --
	(223.96, 99.27) --
	(230.53,100.86) --
	(237.31,102.63) --
	(243.88,104.18) --
	(250.66,105.34) --
	(257.45,106.48) --
	(263.58,107.47) --
	(270.36,108.40) --
	(276.93,109.35) --
	(283.72,109.89) --
	(290.28,110.48) --
	(297.07,111.03) --
	(303.86,111.72) --
	(310.42,112.26) --
	(317.21,112.88) --
	(323.78,113.36) --
	(330.56,113.90) --
	(337.35,114.23);

\path[draw=drawColor,line width= 0.6pt,dash pattern=on 2pt off 2pt ,line join=round] (110.35, 92.22) --
	(116.92, 92.22) --
	(123.70, 92.24) --
	(130.27, 92.24) --
	(137.06, 92.28) --
	(143.84, 92.30) --
	(150.41, 92.34) --
	(157.20, 92.42) --
	(163.76, 92.47) --
	(170.55, 92.51) --
	(177.33, 92.54) --
	(183.68, 92.66) --
	(190.47, 92.73) --
	(197.03, 93.05) --
	(203.82, 93.99) --
	(210.39, 95.57) --
	(217.17, 97.47) --
	(223.96, 99.15) --
	(230.53,100.73) --
	(237.31,102.47) --
	(243.88,103.96) --
	(250.66,105.11) --
	(257.45,106.23) --
	(263.58,107.21) --
	(270.36,108.14) --
	(276.93,109.08) --
	(283.72,109.59) --
	(290.28,110.17) --
	(297.07,110.73) --
	(303.86,111.37) --
	(310.42,111.91) --
	(317.21,112.52) --
	(323.78,112.96) --
	(330.56,113.49) --
	(337.35,113.82);

\path[draw=drawColor,line width= 0.6pt,dash pattern=on 4pt off 2pt ,line join=round] (110.35, 92.22) --
	(116.92, 92.22) --
	(123.70, 92.22) --
	(130.27, 92.23) --
	(137.06, 92.26) --
	(143.84, 92.27) --
	(150.41, 92.29) --
	(157.20, 92.37) --
	(163.76, 92.42) --
	(170.55, 92.45) --
	(177.33, 92.49) --
	(183.68, 92.61) --
	(190.47, 92.67) --
	(197.03, 92.98) --
	(203.82, 93.88) --
	(210.39, 95.38) --
	(217.17, 97.09) --
	(223.96, 98.59) --
	(230.53,100.04) --
	(237.31,101.74) --
	(243.88,103.21) --
	(250.66,104.33) --
	(257.45,105.41) --
	(263.58,106.37) --
	(270.36,107.25) --
	(276.93,108.16) --
	(283.72,108.66) --
	(290.28,109.23) --
	(297.07,109.78) --
	(303.86,110.38) --
	(310.42,110.87) --
	(317.21,111.46) --
	(323.78,111.90) --
	(330.56,112.42) --
	(337.35,112.75);
\definecolor{drawColor}{RGB}{0,193,178}

\path[draw=drawColor,line width= 0.6pt,line join=round] ( 77.30, 92.61) --
	( 83.86, 93.46) --
	( 90.65, 93.76) --
	( 97.44, 94.55) --
	(103.57, 95.41) --
	(110.35, 96.61) --
	(116.92, 97.51) --
	(123.70, 99.05) --
	(130.27,100.21) --
	(137.06,111.46) --
	(143.84,113.35) --
	(150.41,115.78) --
	(157.20,117.77) --
	(163.76,119.29) --
	(170.55,120.92) --
	(177.33,123.00) --
	(183.68,125.93) --
	(190.47,127.98) --
	(197.03,131.23) --
	(203.82,133.94) --
	(210.39,138.07) --
	(217.17,141.77) --
	(223.96,144.67) --
	(230.53,147.68) --
	(237.31,151.36) --
	(243.88,154.45) --
	(250.66,157.35) --
	(257.45,160.58) --
	(263.58,162.53) --
	(270.36,164.36) --
	(276.93,166.69) --
	(283.72,168.13) --
	(290.28,169.26) --
	(297.07,170.77) --
	(303.86,172.62) --
	(310.42,174.59) --
	(317.21,175.93) --
	(323.78,177.12) --
	(330.56,178.60) --
	(337.35,179.49);

\path[draw=drawColor,line width= 0.6pt,dash pattern=on 2pt off 2pt ,line join=round] ( 77.30, 92.61) --
	( 83.86, 93.44) --
	( 90.65, 93.72) --
	( 97.44, 94.44) --
	(103.57, 95.19) --
	(110.35, 96.28) --
	(116.92, 97.17) --
	(123.70, 98.65) --
	(130.27, 99.77) --
	(137.06,110.96) --
	(143.84,112.85) --
	(150.41,115.25) --
	(157.20,117.18) --
	(163.76,118.67) --
	(170.55,120.26) --
	(177.33,122.28) --
	(183.68,125.17) --
	(190.47,127.18) --
	(197.03,130.36) --
	(203.82,133.03) --
	(210.39,137.10) --
	(217.17,140.73) --
	(223.96,143.54) --
	(230.53,146.51) --
	(237.31,150.08) --
	(243.88,153.10) --
	(250.66,155.93) --
	(257.45,159.09) --
	(263.58,161.02) --
	(270.36,162.83) --
	(276.93,165.07) --
	(283.72,166.46) --
	(290.28,167.57) --
	(297.07,168.97) --
	(303.86,170.73) --
	(310.42,172.68) --
	(317.21,173.97) --
	(323.78,175.09) --
	(330.56,176.53) --
	(337.35,177.39);

\path[draw=drawColor,line width= 0.6pt,dash pattern=on 4pt off 2pt ,line join=round] ( 77.30, 92.45) --
	( 83.86, 92.64) --
	( 90.65, 92.92) --
	( 97.44, 93.50) --
	(103.57, 94.19) --
	(110.35, 95.27) --
	(116.92, 96.06) --
	(123.70, 97.44) --
	(130.27, 98.48) --
	(137.06,109.65) --
	(143.84,111.49) --
	(150.41,113.70) --
	(157.20,115.49) --
	(163.76,116.91) --
	(170.55,118.45) --
	(177.33,120.39) --
	(183.68,123.17) --
	(190.47,125.16) --
	(197.03,128.23) --
	(203.82,130.84) --
	(210.39,134.79) --
	(217.17,138.22) --
	(223.96,140.92) --
	(230.53,143.81) --
	(237.31,147.25) --
	(243.88,150.17) --
	(250.66,152.87) --
	(257.45,155.74) --
	(263.58,157.64) --
	(270.36,159.34) --
	(276.93,161.48) --
	(283.72,162.84) --
	(290.28,163.90) --
	(297.07,165.28) --
	(303.86,166.51) --
	(310.42,167.86) --
	(317.21,169.06) --
	(323.78,170.05) --
	(330.56,171.46) --
	(337.35,172.27);
\end{scope}
\begin{scope}
\path[clip] (  0.00,  0.00) rectangle (361.35,361.35);
\definecolor{drawColor}{gray}{0.30}

\node[text=drawColor,anchor=base east,inner sep=0pt, outer sep=0pt, scale=  0.88] at ( 59.35, 89.19) {0};

\node[text=drawColor,anchor=base east,inner sep=0pt, outer sep=0pt, scale=  0.88] at ( 59.35,106.57) {10,000};

\node[text=drawColor,anchor=base east,inner sep=0pt, outer sep=0pt, scale=  0.88] at ( 59.35,123.96) {20,000};

\node[text=drawColor,anchor=base east,inner sep=0pt, outer sep=0pt, scale=  0.88] at ( 59.35,141.34) {30,000};

\node[text=drawColor,anchor=base east,inner sep=0pt, outer sep=0pt, scale=  0.88] at ( 59.35,158.72) {40,000};

\node[text=drawColor,anchor=base east,inner sep=0pt, outer sep=0pt, scale=  0.88] at ( 59.35,176.11) {50,000};
\end{scope}
\begin{scope}
\path[clip] (  0.00,  0.00) rectangle (361.35,361.35);
\definecolor{drawColor}{gray}{0.30}

\node[text=drawColor,anchor=base,inner sep=0pt, outer sep=0pt, scale=  0.88] at ( 90.65, 76.84) {2019};

\node[text=drawColor,anchor=base,inner sep=0pt, outer sep=0pt, scale=  0.88] at (170.55, 76.84) {2020};

\node[text=drawColor,anchor=base,inner sep=0pt, outer sep=0pt, scale=  0.88] at (250.66, 76.84) {2021};

\node[text=drawColor,anchor=base,inner sep=0pt, outer sep=0pt, scale=  0.88] at (330.56, 76.84) {2022};
\end{scope}
\begin{scope}
\path[clip] (  0.00,  0.00) rectangle (361.35,361.35);
\definecolor{drawColor}{RGB}{0,0,0}

\node[text=drawColor,anchor=base,inner sep=0pt, outer sep=0pt, scale=  1.10] at (207.32, 64.81) {Month};
\end{scope}
\begin{scope}
\path[clip] (  0.00,  0.00) rectangle (361.35,361.35);
\definecolor{drawColor}{RGB}{0,0,0}

\node[text=drawColor,rotate= 90.00,anchor=base,inner sep=0pt, outer sep=0pt, scale=  1.10] at ( 30.02,135.85) { };
\end{scope}
\begin{scope}
\path[clip] (  0.00,  0.00) rectangle (361.35,361.35);
\definecolor{drawColor}{RGB}{0,0,0}

\node[text=drawColor,anchor=base west,inner sep=0pt, outer sep=0pt, scale=  1.10] at ( 90.74, 37.52) {Service};
\end{scope}
\begin{scope}
\path[clip] (  0.00,  0.00) rectangle (361.35,361.35);
\definecolor{drawColor}{RGB}{237,144,164}

\path[draw=drawColor,line width= 0.6pt,line join=round] ( 92.19, 23.73) -- (103.75, 23.73);
\end{scope}
\begin{scope}
\path[clip] (  0.00,  0.00) rectangle (361.35,361.35);
\definecolor{drawColor}{RGB}{0,193,178}

\path[draw=drawColor,line width= 0.6pt,line join=round] (153.83, 23.73) -- (165.40, 23.73);
\end{scope}
\begin{scope}
\path[clip] (  0.00,  0.00) rectangle (361.35,361.35);
\definecolor{drawColor}{RGB}{0,0,0}

\node[text=drawColor,anchor=base west,inner sep=0pt, outer sep=0pt, scale=  0.88] at (110.69, 20.70) {Samourai};
\end{scope}
\begin{scope}
\path[clip] (  0.00,  0.00) rectangle (361.35,361.35);
\definecolor{drawColor}{RGB}{0,0,0}

\node[text=drawColor,anchor=base west,inner sep=0pt, outer sep=0pt, scale=  0.88] at (172.34, 20.70) {Wasabi};
\end{scope}
\begin{scope}
\path[clip] (  0.00,  0.00) rectangle (361.35,361.35);
\definecolor{drawColor}{RGB}{0,0,0}

\node[text=drawColor,anchor=base west,inner sep=0pt, outer sep=0pt, scale=  1.10] at (222.25, 37.52) {Level 1 threshold};
\end{scope}
\begin{scope}
\path[clip] (  0.00,  0.00) rectangle (361.35,361.35);
\definecolor{drawColor}{RGB}{0,0,0}

\path[draw=drawColor,line width= 0.6pt,line join=round] (223.70, 23.73) -- (235.26, 23.73);
\end{scope}
\begin{scope}
\path[clip] (  0.00,  0.00) rectangle (361.35,361.35);
\definecolor{drawColor}{RGB}{0,0,0}

\path[draw=drawColor,line width= 0.6pt,dash pattern=on 2pt off 2pt ,line join=round] (262.35, 23.73) -- (273.91, 23.73);
\end{scope}
\begin{scope}
\path[clip] (  0.00,  0.00) rectangle (361.35,361.35);
\definecolor{drawColor}{RGB}{0,0,0}

\path[draw=drawColor,line width= 0.6pt,dash pattern=on 4pt off 2pt ,line join=round] (296.60, 23.73) -- (308.16, 23.73);
\end{scope}
\begin{scope}
\path[clip] (  0.00,  0.00) rectangle (361.35,361.35);
\definecolor{drawColor}{RGB}{0,0,0}

\node[text=drawColor,anchor=base west,inner sep=0pt, outer sep=0pt, scale=  0.88] at (242.20, 20.70) {100};
\end{scope}
\begin{scope}
\path[clip] (  0.00,  0.00) rectangle (361.35,361.35);
\definecolor{drawColor}{RGB}{0,0,0}

\node[text=drawColor,anchor=base west,inner sep=0pt, outer sep=0pt, scale=  0.88] at (280.86, 20.70) {75};
\end{scope}
\begin{scope}
\path[clip] (  0.00,  0.00) rectangle (361.35,361.35);
\definecolor{drawColor}{RGB}{0,0,0}

\node[text=drawColor,anchor=base west,inner sep=0pt, outer sep=0pt, scale=  0.88] at (315.11, 20.70) {50};
\end{scope}
\begin{scope}
\path[clip] (  0.00,  0.00) rectangle (361.35,361.35);
\definecolor{drawColor}{RGB}{0,0,0}

\node[text=drawColor,anchor=base west,inner sep=0pt, outer sep=0pt, scale=  1.32] at ( 64.30,191.92) {Level 2};
\end{scope}
\end{tikzpicture}

%% file: sections/5_security_privacy.tex
% !TeX root = ../main.tex

\section{Anonymity Analysis}
\label{sec:sp}

In this section, we quantify the anonymity perceived by Wasabi and Samourai users and propose methods to quantify how much the real anonymity differs from the perceived one due to the provenance of addresses before and after the mixing.
Unlike in an ideal world where pre- and post-mixed addresses are fresh and unlinkable to any other, 
the traceability of addresses during the pre-mixing and post-mixing allows us to narrow down the 
anonymity set participating in the mixing. We also observe usage patterns during 
the mixing itself that harms anonymity.

\subsection{Upper Bounding Perceived Anonymity Set} 
Here, we are interested in  the (upper) bound on the number of different transaction outputs of the same denomination that are 
mixed in Wasabi and Samourai over their lifetime. Intuitively, this study thereby sheds light on the 
maximum gain in the anonymity set provided by 
each of the wallets over time. 
Towards this goal, we first compute the 
number of bitcoins that are being actively mixed at each point of 
time (i.e., bitcoins added to the wallet until time $t$ minus the number 
of bitcoins withdrawn until time $t$). We denote this quantity by $\alpha_t$.
Figure~\ref{fig:btc-in-pool} (solid line)  shows the concrete results for both wallets, 
while Figure~\ref{fig:btc-in-samourai-by-pool} shows the results for each of the pools in Samourai.  

\begin{figure}[tb]
  \centering
  \resizebox{\columnwidth}{!}{\input{graphics/btc_in_pool.tex}}
  \caption{Monthly aggregated number of BTC in the mixing process,
           as well as the maximal number of outputs.}
  \label{fig:btc-in-pool}
\end{figure}
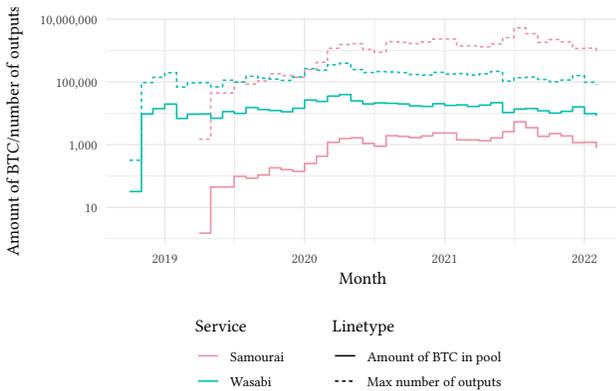
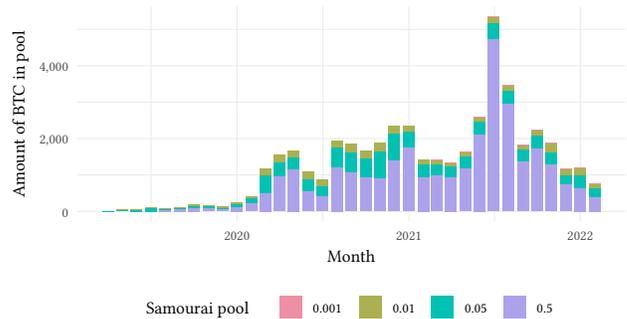
\begin{figure}[tb]
  \centering
  \resizebox{\columnwidth}{!}{\input{graphics/btc_in_samourai_by_pool.tex}}
  \caption{Monthly aggregated number of BTC in the mixing process
           by Samourai pool.}
  \label{fig:btc-in-samourai-by-pool}
\end{figure}

Second, since Wasabi and Samourai consider several mixing denominations, 
we consider the extreme case where all participants use the minimum denomination as we are interested in the upper bound  (i.e., the maximum possible number of different outputs, each of them with the minimum denomination, being mixed). In the concrete case of Wasabi, since it supports 0.01 BTC as the minimum mixing denomination $\beta$,  
we define the upper bound of the 
number of outputs being mixed at each point of time as $\alpha_t/\beta$. 
The result is shown in Figure~\ref{fig:btc-in-pool} (dashed line). 
Following the same reasoning as 
with Wasabi, in Samourai we establish the upper bound on the number of distinct outputs as  $\alpha_t/\beta_{\textit{min}}$, where  $\beta_{\textit{min}}$ is the denomination used in the pool with the minimum denomination. 
% Figure~\ref{fig:btc-in-samourai-by-pool}. 

As mentioned above, these results show the maximum number of outputs being mixed and thus the maximum anonymity set (in an ideal setting where each output belongs to a different user) that each of these wallets has provided over time. These are, however, loose bounds because, e.g., it is likely that a single user owns more than one output, or some of the outputs might be already tainted by an adversary with side information. Therefore, we next study tighter upper bounds on the anonymity set.

\begin{figure*}
  \centering
  \includegraphics[width=1\textwidth]{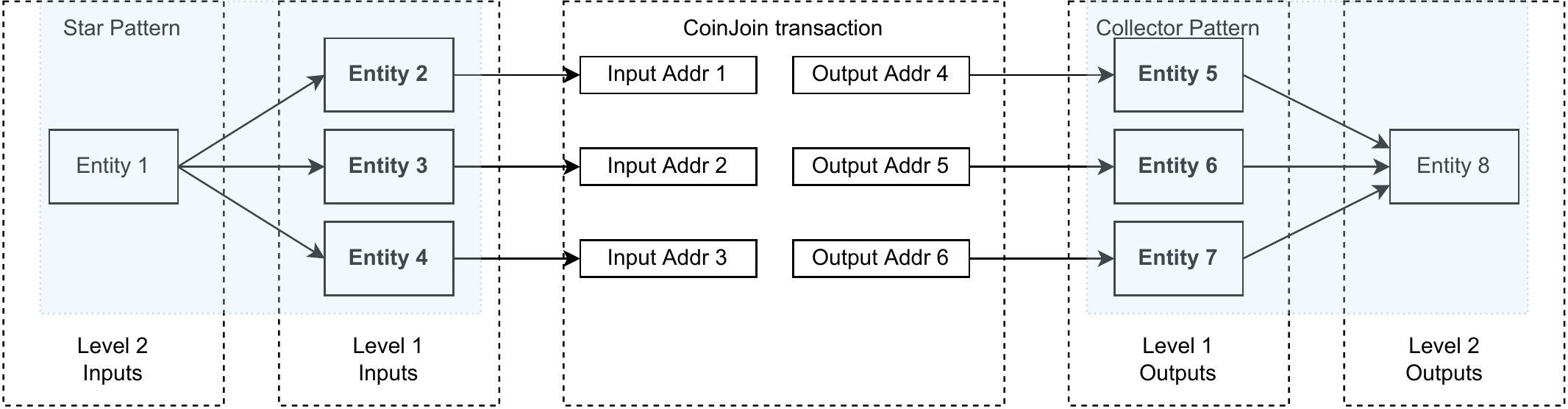}
  \caption{Reduction on the pre- and post-mixing anonymity set due to provenance.}
  %figure source: https://drive.google.com/file/d/1hzGUtHAoJoeeVIxDnK5UkpYcvhbaFGUh/view?usp=sharing
  \label{fig:entityprivacy}
\end{figure*}

\subsection{Pre- and post-mixing anonymity} 
The mixing wallets studied in this work improve the anonymity of the addresses \emph{within} the mixing itself. 
In other words, assume that a certain address has an anonymity set $s_\textit{init}$ when first used by the mixing wallet. 
Then, the mixing wallet's job is to boost such anonymity set by $s_\textit{boost}$ so that the address ends up with a bigger  anonymity set $s_\textit{final} := s_\textit{init} + s_\textit{boost}$ after the mixing is finished. 
%A mixing wallet's goal is to make $\beta \geq \alpha$. 
The ideal scenario would be that $s_\textit{init}$ is as big as possible so that a small effort (i.e., a small  $s_\textit{boost}$) is required to get a good enough final anonymity set $s_\textit{final}$. 
Similarly, whatever is the final anonymity set $s_\textit{final}$ achieved by an address, the ideal scenario is that such anonymity set stays afterward when the user uses such address outside the mixing (e.g., to pay a merchant).

Unfortunately, the pre- and post-mixing anonymity sets are far from the ideal situation perceived by the users. 
In the remaining of this section, we show the results of our study on how far are the anonymity sets of current addresses from the ideal scenarios described above. 

\paragraph{Pre-mixing anonymity loss} The addresses used in the mixing wallets come with 
a history that is publicly available in the blockchain itself and that effectively 
reduces their ideal pre-mixing anonymity set. 
In order to quantify that, we use the entities and their relations as provided by GraphSense 
(Section \ref{subsec:network_analysis}) to compute the provenance of such addresses and 
quantify the reduction of the pre-mixing anonymity. As an illustrative example, 
input address $1$ in~\autoref{fig:entityprivacy}  (respectively input addresses $2$ and $3$) ideally does not have 
any relation to the other two addresses, therefore 
having an initial anonymity set $s_\textit{init}$ of $3$. This would correspond to the best case of 
having three different unrelated entities, and this address could belong to any of the three. 
However, 
studying the provenance of these addresses and their corresponding entities,  
we can observe that all of them belong to the 
same entity (i.e., entity 1). This pattern that one entity is used to fund different addresses is studied as the star pattern in~\cite{Romiti2021}. In our setting, this appears because a user may use her common Bitcoin wallet to fund the pre-mix wallet of her Wasabi (or Samourai) wallet.  
In summary, the 
actual pre-mixing anonymity set of the addresses in this running example is reduced to a single entity, the owner of 
the entity 1. 

\begin{figure}[tb]
  \centering
  \resizebox{\columnwidth}{!}{\input{graphics/incoming_outgoing_entities.tex}}
  \caption{Comparison between number of addresses as well as their corresponding 
           pre-mixing and post-mixing Level 1 entities.}
  \label{fig:incoming-outgoing-entities}
\end{figure}
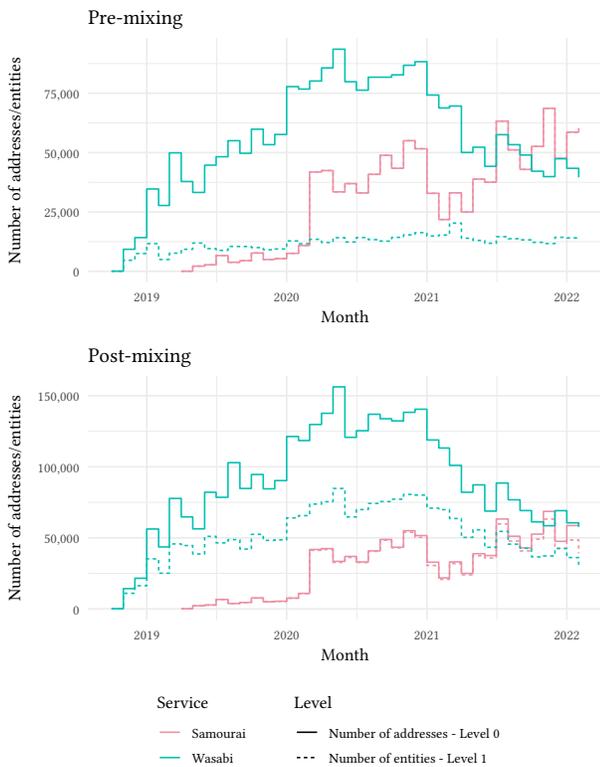

Following the logic in this example, 
we evaluate the actual loss of pre-mixing anonymity by systematically applying the previous 
approach to the complete set of addresses used as input in Wasabi and Samourai wallets 
along with the entities provided by GraphSense.
% Using the entities provided by GraphSense (Section \ref{subsec:network_analysis}),  we compute the number of Level-1 entities that transfer bitcoins into Samourai or 
%  Wasabi CoinJoins and we omit Level-0 entities since they do not represent distinct mixing participants (see Section \ref{subsec:network_analysis}). 
%As shown in Figure \ref{fig:entityanalysis}, GraphSense clusters all inputs of a CoinJoin as 
% a single entity following the multi-input heuristic. However, this does not represent a single participant and 
% thus we ignore Level-0 input entities. Instead, 
% we compute the number of Level-1 input entities that are using Wasabi and Samourai 
% wallets over time and aggregate all entities by month 
% in order to better observe the evolution. 
The results are shown in~\autoref{fig:incoming-outgoing-entities} (top panel). 
For Wasabi, we observe that the number of 
distinct addresses (i.e., ideal anonymity set $s_\textit{init}$)  
is significantly narrowed down when clustering them into entities. This implies that the 
ideally large anonymity set of addresses is significantly reduced if they are not freshly used as input to the Wasabi wallet. For instance, the possible anonymity set of almost $75K$ addresses in Jan'21 is 
largely reduced to less than $25K$ entities. On the other hand, we observe that such a  
reduction in the anonymity sets for addresses used in Samourai wallet does not occur. 

In summary, this exemplifies how the potential anonymity perceived by the users of Wasabi (possibly magnified by the fact that CoinJoin transactions in Wasabi have a larger number of inputs than those from Samourai) is effectively smaller, in fact, even smaller than that of Samourai. 

\paragraph{Post-mixing anonymity loss}
Similar to pre-mixing, the anonymity of the post-mixed addresses 
can be reduced depending on how they 
are used after they are mixed. For instance, if multiple post-mixed addresses can be linked to a single entity, the anonymity set gained during the mixing process is reduced. 
%The effect is even more severe because all participants of the mixing suffer since their anonymity set is 
%also reduced by the number of mixed outputs linked to this entity.
In the simple illustrative example of~\autoref{fig:entityprivacy} where only a single CoinJoin transaction is executed, output address 4 (respectively addresses 5 and 6) ideally has a post-mixing anonymity set $s_\textit{final}$ of $3$, meaning the best case of having three different unrelated entities and this address could belong to any of the three. However, as it was the case for pre-mixing, studying the provenance of addresses 4, 5, and 6, we can observe that they belong to the same entity (entity 8). This usage pattern can be quite common in practice because a user might be tempted to collect her mixed coins from the post-mixing wallet in Wasabi or Samourai into her unique Bitcoin wallet to use them further. This usage pattern has been studied as the collector pattern in~\cite{Romiti2021}.

In~\autoref{fig:incoming-outgoing-entities} (bottom panel) we show the relation
between post-mixed addresses in Wasabi and Samourai (i.e., ideal post-mixing
anonymity sets) and the number of entities computed by GraphSense. As with the
pre-mixing case, we observe that the anonymity sets of post-mixed addresses in Wasabi are
significantly narrowed down to a smaller number of entities, while it is maintained for Samourai.

\subsection{The Impact of Remix Inputs} 
The anonymity set of an output increases by having it being an input of 
other CoinJoin transactions (i.e., a remix input). Conversely, if a CoinJoin exclusively features fresh inputs, i.e., no remix inputs, its anonymity set is only as big as the number of its inputs. The importance of remix inputs is such that the design of Samourai CoinJoin mandates that every CoinJoin features at least one remix input (other than initial Genesis mixes).
%Indeed, as described in Section \ref{sec:detection}, it is possible to trace a path of CoinJoin transactions from every output of a CoinJoin transaction created by Samourai to a respective Genesis mix.

However,  Wasabi wallet does not require remix inputs.  
In fact, we detect that \nWasabiStandalone/\nWasabiTotal transactions do not feature any remix inputs. This adds to the misperception of anonymity 
given by Wasabi wallets because these transactions only provide the anonymity of a single mix (instead of the anonymity 
of several intertwined mixers). This should prompt users that find their addresses in such transactions to remix them as soon as possible.

%% file: graphics/btc_in_pool.tex
% !TEX encoding = UTF-8 Unicode
\begin{tikzpicture}[x=1pt,y=1pt]
\definecolor{fillColor}{RGB}{255,255,255}
\path[use as bounding box,fill=fillColor,fill opacity=0.00] (0,0) rectangle (361.35,231.26);
\begin{scope}
\path[clip] ( 62.99, 96.81) rectangle (355.85,225.76);
\definecolor{drawColor}{gray}{0.92}

\path[draw=drawColor,line width= 0.4pt,line join=round] ( 62.99, 99.52) --
	(355.85, 99.52);

\path[draw=drawColor,line width= 0.4pt,line join=round] ( 62.99,135.30) --
	(355.85,135.30);

\path[draw=drawColor,line width= 0.4pt,line join=round] ( 62.99,171.08) --
	(355.85,171.08);

\path[draw=drawColor,line width= 0.4pt,line join=round] ( 62.99,206.86) --
	(355.85,206.86);

\path[draw=drawColor,line width= 0.4pt,line join=round] (136.26, 96.81) --
	(136.26,225.76);

\path[draw=drawColor,line width= 0.4pt,line join=round] (216.08, 96.81) --
	(216.08,225.76);

\path[draw=drawColor,line width= 0.4pt,line join=round] (295.91, 96.81) --
	(295.91,225.76);

\path[draw=drawColor,line width= 0.9pt,line join=round] ( 62.99,117.41) --
	(355.85,117.41);

\path[draw=drawColor,line width= 0.9pt,line join=round] ( 62.99,153.19) --
	(355.85,153.19);

\path[draw=drawColor,line width= 0.9pt,line join=round] ( 62.99,188.97) --
	(355.85,188.97);

\path[draw=drawColor,line width= 0.9pt,line join=round] ( 62.99,224.75) --
	(355.85,224.75);

\path[draw=drawColor,line width= 0.9pt,line join=round] ( 96.40, 96.81) --
	( 96.40,225.76);

\path[draw=drawColor,line width= 0.9pt,line join=round] (176.12, 96.81) --
	(176.12,225.76);

\path[draw=drawColor,line width= 0.9pt,line join=round] (256.05, 96.81) --
	(256.05,225.76);

\path[draw=drawColor,line width= 0.9pt,line join=round] (335.77, 96.81) --
	(335.77,225.76);
\definecolor{drawColor}{RGB}{237,144,164}

\path[draw=drawColor,line width= 0.9pt,line join=round] (116.05,102.67) --
	(122.61,102.67) --
	(122.61,129.04) --
	(129.38,129.04) --
	(129.38,129.02) --
	(135.93,129.02) --
	(135.93,135.13) --
	(142.70,135.13) --
	(142.70,134.06) --
	(149.47,134.06) --
	(149.47,135.90) --
	(156.02,135.90) --
	(156.02,139.97) --
	(162.79,139.97) --
	(162.79,138.91) --
	(169.34,138.91) --
	(169.34,137.94) --
	(176.12,137.94) --
	(176.12,142.41) --
	(182.89,142.41) --
	(182.89,146.56) --
	(189.22,146.56) --
	(189.22,154.53) --
	(195.99,154.53) --
	(195.99,156.67) --
	(202.54,156.67) --
	(202.54,157.12) --
	(209.31,157.12) --
	(209.31,153.90) --
	(215.86,153.90) --
	(215.86,152.27) --
	(222.64,152.27) --
	(222.64,158.31) --
	(229.41,158.31) --
	(229.41,157.95) --
	(235.96,157.95) --
	(235.96,157.18) --
	(242.73,157.18) --
	(242.73,158.16) --
	(249.28,158.16) --
	(249.28,159.85) --
	(256.05,159.85) --
	(256.05,159.82) --
	(262.82,159.82) --
	(262.82,155.89) --
	(268.94,155.89) --
	(268.94,155.91) --
	(275.71,155.91) --
	(275.71,155.38) --
	(282.26,155.38) --
	(282.26,157.01) --
	(289.03,157.01) --
	(289.03,160.59) --
	(295.58,160.59) --
	(295.58,166.23) --
	(302.35,166.23) --
	(302.35,162.87) --
	(309.12,162.87) --
	(309.12,157.94) --
	(315.67,157.94) --
	(315.67,159.51) --
	(322.45,159.51) --
	(322.45,158.17) --
	(329.00,158.17) --
	(329.00,154.40) --
	(335.77,154.40) --
	(335.77,154.55) --
	(342.54,154.55) --
	(342.54,151.20);

\path[draw=drawColor,line width= 0.9pt,dash pattern=on 2pt off 2pt ,line join=round] (116.05,156.34) --
	(122.61,156.34) --
	(122.61,182.71) --
	(129.38,182.71) --
	(129.38,182.69) --
	(135.93,182.69) --
	(135.93,188.80) --
	(142.70,188.80) --
	(142.70,187.73) --
	(149.47,187.73) --
	(149.47,189.57) --
	(156.02,189.57) --
	(156.02,193.64) --
	(162.79,193.64) --
	(162.79,192.58) --
	(169.34,192.58) --
	(169.34,191.61) --
	(176.12,191.61) --
	(176.12,196.08) --
	(182.89,196.08) --
	(182.89,200.23) --
	(189.22,200.23) --
	(189.22,208.19) --
	(195.99,208.19) --
	(195.99,210.34) --
	(202.54,210.34) --
	(202.54,210.79) --
	(209.31,210.79) --
	(209.31,207.57) --
	(215.86,207.57) --
	(215.86,205.94) --
	(222.64,205.94) --
	(222.64,211.98) --
	(229.41,211.98) --
	(229.41,211.62) --
	(235.96,211.62) --
	(235.96,210.85) --
	(242.73,210.85) --
	(242.73,211.83) --
	(249.28,211.83) --
	(249.28,213.52) --
	(256.05,213.52) --
	(256.05,213.49) --
	(262.82,213.49) --
	(262.82,209.56) --
	(268.94,209.56) --
	(268.94,209.58) --
	(275.71,209.58) --
	(275.71,209.05) --
	(282.26,209.05) --
	(282.26,210.68) --
	(289.03,210.68) --
	(289.03,214.26) --
	(295.58,214.26) --
	(295.58,219.90) --
	(302.35,219.90) --
	(302.35,216.54) --
	(309.12,216.54) --
	(309.12,211.61) --
	(315.67,211.61) --
	(315.67,213.18) --
	(322.45,213.18) --
	(322.45,211.84) --
	(329.00,211.84) --
	(329.00,208.07) --
	(335.77,208.07) --
	(335.77,208.22) --
	(342.54,208.22) --
	(342.54,204.87);
\definecolor{drawColor}{RGB}{0,193,178}

\path[draw=drawColor,line width= 0.9pt,line join=round] ( 76.31,126.43) --
	( 83.08,126.43) --
	( 83.08,170.71) --
	( 89.63,170.71) --
	( 89.63,173.74) --
	( 96.40,173.74) --
	( 96.40,176.37) --
	(103.17,176.37) --
	(103.17,168.15) --
	(109.28,168.15) --
	(109.28,170.57) --
	(116.05,170.57) --
	(116.05,170.68) --
	(122.61,170.68) --
	(122.61,168.28) --
	(129.38,168.28) --
	(129.38,172.08) --
	(135.93,172.08) --
	(135.93,171.02) --
	(142.70,171.02) --
	(142.70,174.38) --
	(149.47,174.38) --
	(149.47,173.23) --
	(156.02,173.23) --
	(156.02,172.67) --
	(162.79,172.67) --
	(162.79,171.97) --
	(169.34,171.97) --
	(169.34,173.94) --
	(176.12,173.94) --
	(176.12,178.68) --
	(182.89,178.68) --
	(182.89,177.84) --
	(189.22,177.84) --
	(189.22,180.91) --
	(195.99,180.91) --
	(195.99,181.78) --
	(202.54,181.78) --
	(202.54,178.17) --
	(209.31,178.17) --
	(209.31,176.44) --
	(215.86,176.44) --
	(215.86,177.02) --
	(222.64,177.02) --
	(222.64,176.78) --
	(229.41,176.78) --
	(229.41,176.42) --
	(235.96,176.42) --
	(235.96,175.32) --
	(242.73,175.32) --
	(242.73,175.01) --
	(249.28,175.01) --
	(249.28,176.56) --
	(256.05,176.56) --
	(256.05,175.58) --
	(262.82,175.58) --
	(262.82,175.89) --
	(268.94,175.89) --
	(268.94,174.99) --
	(275.71,174.99) --
	(275.71,175.76) --
	(282.26,175.76) --
	(282.26,177.14) --
	(289.03,177.14) --
	(289.03,171.53) --
	(295.58,171.53) --
	(295.58,173.52) --
	(302.35,173.52) --
	(302.35,173.81) --
	(309.12,173.81) --
	(309.12,172.52) --
	(315.67,172.52) --
	(315.67,171.21) --
	(322.45,171.21) --
	(322.45,172.22) --
	(329.00,172.22) --
	(329.00,174.73) --
	(335.77,174.73) --
	(335.77,170.90) --
	(342.54,170.90) --
	(342.54,169.56);

\path[draw=drawColor,line width= 0.9pt,dash pattern=on 2pt off 2pt ,line join=round] ( 76.31,144.32) --
	( 83.08,144.32) --
	( 83.08,188.60) --
	( 89.63,188.60) --
	( 89.63,191.63) --
	( 96.40,191.63) --
	( 96.40,194.26) --
	(103.17,194.26) --
	(103.17,186.04) --
	(109.28,186.04) --
	(109.28,188.46) --
	(116.05,188.46) --
	(116.05,188.57) --
	(122.61,188.57) --
	(122.61,186.17) --
	(129.38,186.17) --
	(129.38,189.97) --
	(135.93,189.97) --
	(135.93,188.91) --
	(142.70,188.91) --
	(142.70,192.27) --
	(149.47,192.27) --
	(149.47,191.12) --
	(156.02,191.12) --
	(156.02,190.56) --
	(162.79,190.56) --
	(162.79,189.86) --
	(169.34,189.86) --
	(169.34,191.83) --
	(176.12,191.83) --
	(176.12,196.57) --
	(182.89,196.57) --
	(182.89,195.73) --
	(189.22,195.73) --
	(189.22,198.80) --
	(195.99,198.80) --
	(195.99,199.67) --
	(202.54,199.67) --
	(202.54,196.06) --
	(209.31,196.06) --
	(209.31,194.33) --
	(215.86,194.33) --
	(215.86,194.91) --
	(222.64,194.91) --
	(222.64,194.67) --
	(229.41,194.67) --
	(229.41,194.31) --
	(235.96,194.31) --
	(235.96,193.21) --
	(242.73,193.21) --
	(242.73,192.90) --
	(249.28,192.90) --
	(249.28,194.45) --
	(256.05,194.45) --
	(256.05,193.47) --
	(262.82,193.47) --
	(262.82,193.78) --
	(268.94,193.78) --
	(268.94,192.88) --
	(275.71,192.88) --
	(275.71,193.65) --
	(282.26,193.65) --
	(282.26,195.02) --
	(289.03,195.02) --
	(289.03,189.42) --
	(295.58,189.42) --
	(295.58,191.41) --
	(302.35,191.41) --
	(302.35,191.70) --
	(309.12,191.70) --
	(309.12,190.41) --
	(315.67,190.41) --
	(315.67,189.10) --
	(322.45,189.10) --
	(322.45,190.11) --
	(329.00,190.11) --
	(329.00,192.62) --
	(335.77,192.62) --
	(335.77,188.79) --
	(342.54,188.79) --
	(342.54,187.45);
\end{scope}
\begin{scope}
\path[clip] (  0.00,  0.00) rectangle (361.35,231.26);
\definecolor{drawColor}{gray}{0.30}

\node[text=drawColor,anchor=base east,inner sep=0pt, outer sep=0pt, scale=  0.88] at ( 58.04,114.38) {10};

\node[text=drawColor,anchor=base east,inner sep=0pt, outer sep=0pt, scale=  0.88] at ( 58.04,150.16) {1,000};

\node[text=drawColor,anchor=base east,inner sep=0pt, outer sep=0pt, scale=  0.88] at ( 58.04,185.94) {100,000};

\node[text=drawColor,anchor=base east,inner sep=0pt, outer sep=0pt, scale=  0.88] at ( 58.04,221.72) {10,000,000};
\end{scope}
\begin{scope}
\path[clip] (  0.00,  0.00) rectangle (361.35,231.26);
\definecolor{drawColor}{gray}{0.30}

\node[text=drawColor,anchor=base,inner sep=0pt, outer sep=0pt, scale=  0.88] at ( 96.40, 85.80) {2019};

\node[text=drawColor,anchor=base,inner sep=0pt, outer sep=0pt, scale=  0.88] at (176.12, 85.80) {2020};

\node[text=drawColor,anchor=base,inner sep=0pt, outer sep=0pt, scale=  0.88] at (256.05, 85.80) {2021};

\node[text=drawColor,anchor=base,inner sep=0pt, outer sep=0pt, scale=  0.88] at (335.77, 85.80) {2022};
\end{scope}
\begin{scope}
\path[clip] (  0.00,  0.00) rectangle (361.35,231.26);
\definecolor{drawColor}{RGB}{0,0,0}

\node[text=drawColor,anchor=base,inner sep=0pt, outer sep=0pt, scale=  1.10] at (209.42, 73.76) {Month};
\end{scope}
\begin{scope}
\path[clip] (  0.00,  0.00) rectangle (361.35,231.26);
\definecolor{drawColor}{RGB}{0,0,0}

\node[text=drawColor,rotate= 90.00,anchor=base,inner sep=0pt, outer sep=0pt, scale=  1.10] at ( 13.08,161.29) {Amount of BTC/number of outputs};
\end{scope}
\begin{scope}
\path[clip] (  0.00,  0.00) rectangle (361.35,231.26);
\definecolor{drawColor}{RGB}{0,0,0}

\node[text=drawColor,anchor=base west,inner sep=0pt, outer sep=0pt, scale=  1.10] at (113.88, 46.48) {Service};
\end{scope}
\begin{scope}
\path[clip] (  0.00,  0.00) rectangle (361.35,231.26);
\definecolor{drawColor}{RGB}{237,144,164}

\path[draw=drawColor,line width= 0.9pt,line join=round] (115.32, 32.68) -- (126.89, 32.68);
\end{scope}
\begin{scope}
\path[clip] (  0.00,  0.00) rectangle (361.35,231.26);
\definecolor{drawColor}{RGB}{0,193,178}

\path[draw=drawColor,line width= 0.9pt,line join=round] (115.32, 18.23) -- (126.89, 18.23);
\end{scope}
\begin{scope}
\path[clip] (  0.00,  0.00) rectangle (361.35,231.26);
\definecolor{drawColor}{RGB}{0,0,0}

\node[text=drawColor,anchor=base west,inner sep=0pt, outer sep=0pt, scale=  0.88] at (133.83, 29.65) {Samourai};
\end{scope}
\begin{scope}
\path[clip] (  0.00,  0.00) rectangle (361.35,231.26);
\definecolor{drawColor}{RGB}{0,0,0}

\node[text=drawColor,anchor=base west,inner sep=0pt, outer sep=0pt, scale=  0.88] at (133.83, 15.20) {Wasabi};
\end{scope}
\begin{scope}
\path[clip] (  0.00,  0.00) rectangle (361.35,231.26);
\definecolor{drawColor}{RGB}{0,0,0}

\node[text=drawColor,anchor=base west,inner sep=0pt, outer sep=0pt, scale=  1.10] at (192.02, 46.48) {Linetype};
\end{scope}
\begin{scope}
\path[clip] (  0.00,  0.00) rectangle (361.35,231.26);
\definecolor{drawColor}{RGB}{0,0,0}

\path[draw=drawColor,line width= 0.9pt,line join=round] (193.47, 32.68) -- (205.03, 32.68);
\end{scope}
\begin{scope}
\path[clip] (  0.00,  0.00) rectangle (361.35,231.26);
\definecolor{drawColor}{RGB}{0,0,0}

\path[draw=drawColor,line width= 0.9pt,dash pattern=on 2pt off 2pt ,line join=round] (193.47, 18.23) -- (205.03, 18.23);
\end{scope}
\begin{scope}
\path[clip] (  0.00,  0.00) rectangle (361.35,231.26);
\definecolor{drawColor}{RGB}{0,0,0}

\node[text=drawColor,anchor=base west,inner sep=0pt, outer sep=0pt, scale=  0.88] at (211.98, 29.65) {Amount of BTC in pool};
\end{scope}
\begin{scope}
\path[clip] (  0.00,  0.00) rectangle (361.35,231.26);
\definecolor{drawColor}{RGB}{0,0,0}

\node[text=drawColor,anchor=base west,inner sep=0pt, outer sep=0pt, scale=  0.88] at (211.98, 15.20) {Max number of outputs};
\end{scope}
\end{tikzpicture}

%% file: graphics/btc_in_samourai_by_pool.tex
% !TEX encoding = UTF-8 Unicode
\begin{tikzpicture}[x=1pt,y=1pt]
\definecolor{fillColor}{RGB}{255,255,255}
\path[use as bounding box,fill=fillColor,fill opacity=0.00] (0,0) rectangle (361.35,195.13);
\begin{scope}
\path[clip] ( 42.95, 67.14) rectangle (355.85,189.63);
\definecolor{drawColor}{gray}{0.92}

\path[draw=drawColor,line width= 0.3pt,line join=round] ( 42.95, 93.48) --
	(355.85, 93.48);

\path[draw=drawColor,line width= 0.3pt,line join=round] ( 42.95,135.04) --
	(355.85,135.04);

\path[draw=drawColor,line width= 0.3pt,line join=round] ( 42.95,176.59) --
	(355.85,176.59);

\path[draw=drawColor,line width= 0.3pt,line join=round] ( 85.19, 67.14) --
	( 85.19,189.63);

\path[draw=drawColor,line width= 0.3pt,line join=round] (183.20, 67.14) --
	(183.20,189.63);

\path[draw=drawColor,line width= 0.3pt,line join=round] (281.08, 67.14) --
	(281.08,189.63);

\path[draw=drawColor,line width= 0.6pt,line join=round] ( 42.95, 72.71) --
	(355.85, 72.71);

\path[draw=drawColor,line width= 0.6pt,line join=round] ( 42.95,114.26) --
	(355.85,114.26);

\path[draw=drawColor,line width= 0.6pt,line join=round] ( 42.95,155.81) --
	(355.85,155.81);

\path[draw=drawColor,line width= 0.6pt,line join=round] (134.19, 67.14) --
	(134.19,189.63);

\path[draw=drawColor,line width= 0.6pt,line join=round] (232.21, 67.14) --
	(232.21,189.63);

\path[draw=drawColor,line width= 0.6pt,line join=round] (329.95, 67.14) --
	(329.95,189.63);
\definecolor{fillColor}{RGB}{0,193,178}

\path[fill=fillColor] ( 57.18, 72.71) rectangle ( 63.92, 72.74);

\path[fill=fillColor] ( 65.21, 72.71) rectangle ( 71.96, 73.58);
\definecolor{fillColor}{RGB}{171,177,80}

\path[fill=fillColor] ( 65.21, 73.58) rectangle ( 71.96, 73.64);
\definecolor{fillColor}{RGB}{0,193,178}

\path[fill=fillColor] ( 73.51, 72.71) rectangle ( 80.26, 73.48);
\definecolor{fillColor}{RGB}{171,177,80}

\path[fill=fillColor] ( 73.51, 73.48) rectangle ( 80.26, 73.63);
\definecolor{fillColor}{RGB}{0,193,178}

\path[fill=fillColor] ( 81.55, 72.71) rectangle ( 88.29, 74.45);
\definecolor{fillColor}{RGB}{171,177,80}

\path[fill=fillColor] ( 81.55, 74.45) rectangle ( 88.29, 74.74);
\definecolor{fillColor}{RGB}{0,193,178}

\path[fill=fillColor] ( 89.85, 73.51) rectangle ( 96.60, 74.23);
\definecolor{fillColor}{RGB}{172,162,236}

\path[fill=fillColor] ( 89.85, 72.71) rectangle ( 96.60, 73.51);
\definecolor{fillColor}{RGB}{171,177,80}

\path[fill=fillColor] ( 89.85, 74.23) rectangle ( 96.60, 74.48);
\definecolor{fillColor}{RGB}{0,193,178}

\path[fill=fillColor] ( 98.15, 73.97) rectangle (104.90, 74.61);
\definecolor{fillColor}{RGB}{172,162,236}

\path[fill=fillColor] ( 98.15, 72.71) rectangle (104.90, 73.97);
\definecolor{fillColor}{RGB}{171,177,80}

\path[fill=fillColor] ( 98.15, 74.61) rectangle (104.90, 74.95);
\definecolor{fillColor}{RGB}{0,193,178}

\path[fill=fillColor] (106.18, 74.61) rectangle (112.93, 75.88);
\definecolor{fillColor}{RGB}{172,162,236}

\path[fill=fillColor] (106.18, 72.71) rectangle (112.93, 74.61);
\definecolor{fillColor}{RGB}{171,177,80}

\path[fill=fillColor] (106.18, 75.88) rectangle (112.93, 76.50);
\definecolor{fillColor}{RGB}{0,193,178}

\path[fill=fillColor] (114.48, 74.36) rectangle (121.23, 75.54);
\definecolor{fillColor}{RGB}{172,162,236}

\path[fill=fillColor] (114.48, 72.71) rectangle (121.23, 74.36);
\definecolor{fillColor}{RGB}{171,177,80}

\path[fill=fillColor] (114.48, 75.54) rectangle (121.23, 76.01);
\definecolor{fillColor}{RGB}{0,193,178}

\path[fill=fillColor] (122.52, 73.82) rectangle (129.27, 75.02);
\definecolor{fillColor}{RGB}{172,162,236}

\path[fill=fillColor] (122.52, 72.71) rectangle (129.27, 73.82);
\definecolor{fillColor}{RGB}{171,177,80}

\path[fill=fillColor] (122.52, 75.02) rectangle (129.27, 75.63);
\definecolor{fillColor}{RGB}{0,193,178}

\path[fill=fillColor] (130.82, 75.10) rectangle (137.57, 77.20);
\definecolor{fillColor}{RGB}{172,162,236}

\path[fill=fillColor] (130.82, 72.71) rectangle (137.57, 75.10);
\definecolor{fillColor}{RGB}{171,177,80}

\path[fill=fillColor] (130.82, 77.20) rectangle (137.57, 77.89);
\definecolor{fillColor}{RGB}{0,193,178}

\path[fill=fillColor] (139.12, 77.12) rectangle (145.87, 80.46);
\definecolor{fillColor}{RGB}{172,162,236}

\path[fill=fillColor] (139.12, 72.71) rectangle (145.87, 77.12);
\definecolor{fillColor}{RGB}{171,177,80}

\path[fill=fillColor] (139.12, 80.46) rectangle (145.87, 81.56);
\definecolor{fillColor}{RGB}{0,193,178}

\path[fill=fillColor] (146.89, 83.18) rectangle (153.64, 93.55);
\definecolor{fillColor}{RGB}{172,162,236}

\path[fill=fillColor] (146.89, 72.71) rectangle (153.64, 83.18);
\definecolor{fillColor}{RGB}{171,177,80}

\path[fill=fillColor] (146.89, 93.55) rectangle (153.64, 97.38);
\definecolor{fillColor}{RGB}{0,193,178}

\path[fill=fillColor] (155.19, 92.31) rectangle (161.94,100.95);
\definecolor{fillColor}{RGB}{172,162,236}

\path[fill=fillColor] (155.19, 72.71) rectangle (161.94, 92.31);
\definecolor{fillColor}{RGB}{171,177,80}

\path[fill=fillColor] (155.19,100.95) rectangle (161.94,105.23);
\definecolor{fillColor}{RGB}{0,193,178}

\path[fill=fillColor] (163.22, 96.55) rectangle (169.97,103.68);
\definecolor{fillColor}{RGB}{172,162,236}

\path[fill=fillColor] (163.22, 72.71) rectangle (169.97, 96.55);
\definecolor{fillColor}{RGB}{171,177,80}

\path[fill=fillColor] (163.22,103.68) rectangle (169.97,107.17);
\definecolor{fillColor}{RGB}{0,193,178}

\path[fill=fillColor] (171.52, 84.05) rectangle (178.27, 91.25);
\definecolor{fillColor}{RGB}{172,162,236}

\path[fill=fillColor] (171.52, 72.71) rectangle (178.27, 84.05);
\definecolor{fillColor}{RGB}{171,177,80}

\path[fill=fillColor] (171.52, 91.25) rectangle (178.27, 95.48);
\definecolor{fillColor}{RGB}{0,193,178}

\path[fill=fillColor] (179.56, 80.97) rectangle (186.31, 87.24);
\definecolor{fillColor}{RGB}{172,162,236}

\path[fill=fillColor] (179.56, 72.71) rectangle (186.31, 80.97);
\definecolor{fillColor}{RGB}{171,177,80}

\path[fill=fillColor] (179.56, 87.24) rectangle (186.31, 91.18);
\definecolor{fillColor}{RGB}{0,193,178}

\path[fill=fillColor] (187.86, 97.85) rectangle (194.61,109.09);
\definecolor{fillColor}{RGB}{172,162,236}

\path[fill=fillColor] (187.86, 72.71) rectangle (194.61, 97.85);
\definecolor{fillColor}{RGB}{171,177,80}

\path[fill=fillColor] (187.86,109.09) rectangle (194.61,112.87);
\definecolor{fillColor}{RGB}{0,193,178}

\path[fill=fillColor] (196.16, 94.67) rectangle (202.91,106.21);
\definecolor{fillColor}{RGB}{172,162,236}

\path[fill=fillColor] (196.16, 72.71) rectangle (202.91, 94.67);
\definecolor{fillColor}{RGB}{171,177,80}

\path[fill=fillColor] (196.16,106.21) rectangle (202.91,111.03);
\definecolor{fillColor}{RGB}{0,193,178}

\path[fill=fillColor] (204.20, 91.81) rectangle (210.94,103.40);
\definecolor{fillColor}{RGB}{172,162,236}

\path[fill=fillColor] (204.20, 72.71) rectangle (210.94, 91.81);
\definecolor{fillColor}{RGB}{171,177,80}

\path[fill=fillColor] (204.20,103.40) rectangle (210.94,107.44);
\definecolor{fillColor}{RGB}{0,193,178}

\path[fill=fillColor] (212.50, 91.35) rectangle (219.25,107.13);
\definecolor{fillColor}{RGB}{172,162,236}

\path[fill=fillColor] (212.50, 72.71) rectangle (219.25, 91.35);
\definecolor{fillColor}{RGB}{171,177,80}

\path[fill=fillColor] (212.50,107.13) rectangle (219.25,112.09);
\definecolor{fillColor}{RGB}{0,193,178}

\path[fill=fillColor] (220.53,101.46) rectangle (227.28,117.32);
\definecolor{fillColor}{RGB}{172,162,236}

\path[fill=fillColor] (220.53, 72.71) rectangle (227.28,101.46);
\definecolor{fillColor}{RGB}{171,177,80}

\path[fill=fillColor] (220.53,117.32) rectangle (227.28,121.67);
\definecolor{fillColor}{RGB}{0,193,178}

\path[fill=fillColor] (228.83,108.88) rectangle (235.58,118.33);
\definecolor{fillColor}{RGB}{172,162,236}

\path[fill=fillColor] (228.83, 72.71) rectangle (235.58,108.88);
\definecolor{fillColor}{RGB}{171,177,80}

\path[fill=fillColor] (228.83,118.33) rectangle (235.58,121.48);
\definecolor{fillColor}{RGB}{0,193,178}

\path[fill=fillColor] (237.13, 92.25) rectangle (243.88, 99.60);
\definecolor{fillColor}{RGB}{172,162,236}

\path[fill=fillColor] (237.13, 72.71) rectangle (243.88, 92.25);
\definecolor{fillColor}{RGB}{171,177,80}

\path[fill=fillColor] (237.13, 99.60) rectangle (243.88,102.11);
\definecolor{fillColor}{RGB}{0,193,178}

\path[fill=fillColor] (244.63, 93.12) rectangle (251.38, 99.80);
\definecolor{fillColor}{RGB}{172,162,236}

\path[fill=fillColor] (244.63, 72.71) rectangle (251.38, 93.12);
\definecolor{fillColor}{RGB}{171,177,80}

\path[fill=fillColor] (244.63, 99.80) rectangle (251.38,102.01);
\definecolor{fillColor}{RGB}{237,144,164}

\path[fill=fillColor] (244.63,102.01) rectangle (251.38,102.21);
\definecolor{fillColor}{RGB}{0,193,178}

\path[fill=fillColor] (252.93, 92.23) rectangle (259.68, 98.46);
\definecolor{fillColor}{RGB}{172,162,236}

\path[fill=fillColor] (252.93, 72.71) rectangle (259.68, 92.23);
\definecolor{fillColor}{RGB}{171,177,80}

\path[fill=fillColor] (252.93, 98.46) rectangle (259.68,100.06);
\definecolor{fillColor}{RGB}{237,144,164}

\path[fill=fillColor] (252.93,100.06) rectangle (259.68,100.24);
\definecolor{fillColor}{RGB}{0,193,178}

\path[fill=fillColor] (260.97, 96.90) rectangle (267.72,104.08);
\definecolor{fillColor}{RGB}{172,162,236}

\path[fill=fillColor] (260.97, 72.71) rectangle (267.72, 96.90);
\definecolor{fillColor}{RGB}{171,177,80}

\path[fill=fillColor] (260.97,104.08) rectangle (267.72,106.38);
\definecolor{fillColor}{RGB}{237,144,164}

\path[fill=fillColor] (260.97,106.38) rectangle (267.72,106.66);
\definecolor{fillColor}{RGB}{0,193,178}

\path[fill=fillColor] (269.27,116.49) rectangle (276.02,124.32);
\definecolor{fillColor}{RGB}{172,162,236}

\path[fill=fillColor] (269.27, 72.71) rectangle (276.02,116.49);
\definecolor{fillColor}{RGB}{171,177,80}

\path[fill=fillColor] (269.27,124.32) rectangle (276.02,126.32);
\definecolor{fillColor}{RGB}{237,144,164}

\path[fill=fillColor] (269.27,126.32) rectangle (276.02,126.60);
\definecolor{fillColor}{RGB}{0,193,178}

\path[fill=fillColor] (277.30,170.74) rectangle (284.05,180.13);
\definecolor{fillColor}{RGB}{172,162,236}

\path[fill=fillColor] (277.30, 72.71) rectangle (284.05,170.74);
\definecolor{fillColor}{RGB}{171,177,80}

\path[fill=fillColor] (277.30,180.13) rectangle (284.05,183.74);
\definecolor{fillColor}{RGB}{237,144,164}

\path[fill=fillColor] (277.30,183.74) rectangle (284.05,184.06);
\definecolor{fillColor}{RGB}{0,193,178}

\path[fill=fillColor] (285.60,133.79) rectangle (292.35,141.56);
\definecolor{fillColor}{RGB}{172,162,236}

\path[fill=fillColor] (285.60, 72.71) rectangle (292.35,133.79);
\definecolor{fillColor}{RGB}{171,177,80}

\path[fill=fillColor] (285.60,141.56) rectangle (292.35,144.61);
\definecolor{fillColor}{RGB}{237,144,164}

\path[fill=fillColor] (285.60,144.61) rectangle (292.35,144.95);
\definecolor{fillColor}{RGB}{0,193,178}

\path[fill=fillColor] (293.91,100.85) rectangle (300.65,107.89);
\definecolor{fillColor}{RGB}{172,162,236}

\path[fill=fillColor] (293.91, 72.71) rectangle (300.65,100.85);
\definecolor{fillColor}{RGB}{171,177,80}

\path[fill=fillColor] (293.91,107.89) rectangle (300.65,110.64);
\definecolor{fillColor}{RGB}{237,144,164}

\path[fill=fillColor] (293.91,110.64) rectangle (300.65,110.98);
\definecolor{fillColor}{RGB}{0,193,178}

\path[fill=fillColor] (301.94,108.19) rectangle (308.69,116.09);
\definecolor{fillColor}{RGB}{172,162,236}

\path[fill=fillColor] (301.94, 72.71) rectangle (308.69,108.19);
\definecolor{fillColor}{RGB}{171,177,80}

\path[fill=fillColor] (301.94,116.09) rectangle (308.69,119.17);
\definecolor{fillColor}{RGB}{237,144,164}

\path[fill=fillColor] (301.94,119.17) rectangle (308.69,119.59);
\definecolor{fillColor}{RGB}{0,193,178}

\path[fill=fillColor] (310.24, 99.28) rectangle (316.99,106.53);
\definecolor{fillColor}{RGB}{172,162,236}

\path[fill=fillColor] (310.24, 72.71) rectangle (316.99, 99.28);
\definecolor{fillColor}{RGB}{171,177,80}

\path[fill=fillColor] (310.24,106.53) rectangle (316.99,111.75);
\definecolor{fillColor}{RGB}{237,144,164}

\path[fill=fillColor] (310.24,111.75) rectangle (316.99,112.15);
\definecolor{fillColor}{RGB}{0,193,178}

\path[fill=fillColor] (318.28, 87.76) rectangle (325.02, 93.33);
\definecolor{fillColor}{RGB}{172,162,236}

\path[fill=fillColor] (318.28, 72.71) rectangle (325.02, 87.76);
\definecolor{fillColor}{RGB}{171,177,80}

\path[fill=fillColor] (318.28, 93.33) rectangle (325.02, 96.61);
\definecolor{fillColor}{RGB}{237,144,164}

\path[fill=fillColor] (318.28, 96.61) rectangle (325.02, 96.99);
\definecolor{fillColor}{RGB}{0,193,178}

\path[fill=fillColor] (326.58, 85.93) rectangle (333.33, 93.50);
\definecolor{fillColor}{RGB}{172,162,236}

\path[fill=fillColor] (326.58, 72.71) rectangle (333.33, 85.93);
\definecolor{fillColor}{RGB}{171,177,80}

\path[fill=fillColor] (326.58, 93.50) rectangle (333.33, 97.07);
\definecolor{fillColor}{RGB}{237,144,164}

\path[fill=fillColor] (326.58, 97.07) rectangle (333.33, 97.47);
\definecolor{fillColor}{RGB}{0,193,178}

\path[fill=fillColor] (334.88, 80.68) rectangle (341.63, 85.66);
\definecolor{fillColor}{RGB}{172,162,236}

\path[fill=fillColor] (334.88, 72.71) rectangle (341.63, 80.68);
\definecolor{fillColor}{RGB}{171,177,80}

\path[fill=fillColor] (334.88, 85.66) rectangle (341.63, 88.44);
\definecolor{fillColor}{RGB}{237,144,164}

\path[fill=fillColor] (334.88, 88.44) rectangle (341.63, 88.78);
\end{scope}
\begin{scope}
\path[clip] (  0.00,  0.00) rectangle (361.35,195.13);
\definecolor{drawColor}{gray}{0.30}

\node[text=drawColor,anchor=base east,inner sep=0pt, outer sep=0pt, scale=  0.88] at ( 38.00, 69.68) {0};

\node[text=drawColor,anchor=base east,inner sep=0pt, outer sep=0pt, scale=  0.88] at ( 38.00,111.23) {2,000};

\node[text=drawColor,anchor=base east,inner sep=0pt, outer sep=0pt, scale=  0.88] at ( 38.00,152.78) {4,000};
\end{scope}
\begin{scope}
\path[clip] (  0.00,  0.00) rectangle (361.35,195.13);
\definecolor{drawColor}{gray}{0.30}

\node[text=drawColor,anchor=base,inner sep=0pt, outer sep=0pt, scale=  0.88] at (134.19, 56.13) {2020};

\node[text=drawColor,anchor=base,inner sep=0pt, outer sep=0pt, scale=  0.88] at (232.21, 56.13) {2021};

\node[text=drawColor,anchor=base,inner sep=0pt, outer sep=0pt, scale=  0.88] at (329.95, 56.13) {2022};
\end{scope}
\begin{scope}
\path[clip] (  0.00,  0.00) rectangle (361.35,195.13);
\definecolor{drawColor}{RGB}{0,0,0}

\node[text=drawColor,anchor=base,inner sep=0pt, outer sep=0pt, scale=  1.10] at (199.40, 44.09) {Month};
\end{scope}
\begin{scope}
\path[clip] (  0.00,  0.00) rectangle (361.35,195.13);
\definecolor{drawColor}{RGB}{0,0,0}

\node[text=drawColor,rotate= 90.00,anchor=base,inner sep=0pt, outer sep=0pt, scale=  1.10] at ( 13.08,128.38) {Amount of BTC in pool};
\end{scope}
\begin{scope}
\path[clip] (  0.00,  0.00) rectangle (361.35,195.13);
\definecolor{drawColor}{RGB}{0,0,0}

\node[text=drawColor,anchor=base west,inner sep=0pt, outer sep=0pt, scale=  1.10] at ( 82.37, 14.44) {Samourai pool};
\end{scope}
\begin{scope}
\path[clip] (  0.00,  0.00) rectangle (361.35,195.13);
\definecolor{fillColor}{RGB}{237,144,164}

\path[fill=fillColor] (158.26, 11.71) rectangle (171.30, 24.74);
\end{scope}
\begin{scope}
\path[clip] (  0.00,  0.00) rectangle (361.35,195.13);
\definecolor{fillColor}{RGB}{171,177,80}

\path[fill=fillColor] (203.76, 11.71) rectangle (216.79, 24.74);
\end{scope}
\begin{scope}
\path[clip] (  0.00,  0.00) rectangle (361.35,195.13);
\definecolor{fillColor}{RGB}{0,193,178}

\path[fill=fillColor] (244.85, 11.71) rectangle (257.88, 24.74);
\end{scope}
\begin{scope}
\path[clip] (  0.00,  0.00) rectangle (361.35,195.13);
\definecolor{fillColor}{RGB}{172,162,236}

\path[fill=fillColor] (285.95, 11.71) rectangle (298.98, 24.74);
\end{scope}
\begin{scope}
\path[clip] (  0.00,  0.00) rectangle (361.35,195.13);
\definecolor{drawColor}{RGB}{0,0,0}

\node[text=drawColor,anchor=base west,inner sep=0pt, outer sep=0pt, scale=  0.88] at (177.51, 15.20) {0.001};
\end{scope}
\begin{scope}
\path[clip] (  0.00,  0.00) rectangle (361.35,195.13);
\definecolor{drawColor}{RGB}{0,0,0}

\node[text=drawColor,anchor=base west,inner sep=0pt, outer sep=0pt, scale=  0.88] at (223.00, 15.20) {0.01};
\end{scope}
\begin{scope}
\path[clip] (  0.00,  0.00) rectangle (361.35,195.13);
\definecolor{drawColor}{RGB}{0,0,0}

\node[text=drawColor,anchor=base west,inner sep=0pt, outer sep=0pt, scale=  0.88] at (264.10, 15.20) {0.05};
\end{scope}
\begin{scope}
\path[clip] (  0.00,  0.00) rectangle (361.35,195.13);
\definecolor{drawColor}{RGB}{0,0,0}

\node[text=drawColor,anchor=base west,inner sep=0pt, outer sep=0pt, scale=  0.88] at (305.19, 15.20) {0.5};
\end{scope}
\end{tikzpicture}

%% file: graphics/incoming_outgoing_entities.tex
% !TEX encoding = UTF-8 Unicode
\begin{tikzpicture}[x=1pt,y=1pt]
\definecolor{fillColor}{RGB}{255,255,255}
\path[use as bounding box,fill=fillColor,fill opacity=0.00] (0,0) rectangle (361.35,462.53);
\begin{scope}
\path[clip] ( 57.25,295.01) rectangle (350.35,434.37);
\definecolor{drawColor}{gray}{0.92}

\path[draw=drawColor,line width= 0.4pt,line join=round] ( 57.25,318.21) --
	(350.35,318.21);

\path[draw=drawColor,line width= 0.4pt,line join=round] ( 57.25,352.09) --
	(350.35,352.09);

\path[draw=drawColor,line width= 0.4pt,line join=round] ( 57.25,385.97) --
	(350.35,385.97);

\path[draw=drawColor,line width= 0.4pt,line join=round] ( 57.25,419.85) --
	(350.35,419.85);

\path[draw=drawColor,line width= 0.4pt,line join=round] (130.58,295.01) --
	(130.58,434.37);

\path[draw=drawColor,line width= 0.4pt,line join=round] (210.47,295.01) --
	(210.47,434.37);

\path[draw=drawColor,line width= 0.4pt,line join=round] (290.36,295.01) --
	(290.36,434.37);

\path[draw=drawColor,line width= 0.9pt,line join=round] ( 57.25,301.27) --
	(350.35,301.27);

\path[draw=drawColor,line width= 0.9pt,line join=round] ( 57.25,335.15) --
	(350.35,335.15);

\path[draw=drawColor,line width= 0.9pt,line join=round] ( 57.25,369.03) --
	(350.35,369.03);

\path[draw=drawColor,line width= 0.9pt,line join=round] ( 57.25,402.91) --
	(350.35,402.91);

\path[draw=drawColor,line width= 0.9pt,line join=round] ( 90.68,295.01) --
	( 90.68,434.37);

\path[draw=drawColor,line width= 0.9pt,line join=round] (170.47,295.01) --
	(170.47,434.37);

\path[draw=drawColor,line width= 0.9pt,line join=round] (250.47,295.01) --
	(250.47,434.37);

\path[draw=drawColor,line width= 0.9pt,line join=round] (330.25,295.01) --
	(330.25,434.37);
\definecolor{drawColor}{RGB}{237,144,164}

\path[draw=drawColor,line width= 0.9pt,line join=round] (110.36,301.35) --
	(116.91,301.35) --
	(116.91,304.23) --
	(123.69,304.23) --
	(123.69,304.95) --
	(130.25,304.95) --
	(130.25,310.28) --
	(137.02,310.28) --
	(137.02,306.43) --
	(143.80,306.43) --
	(143.80,307.36) --
	(150.36,307.36) --
	(150.36,311.72) --
	(157.13,311.72) --
	(157.13,308.04) --
	(163.69,308.04) --
	(163.69,308.53) --
	(170.47,308.53) --
	(170.47,311.52) --
	(177.24,311.52) --
	(177.24,316.04) --
	(183.58,316.04) --
	(183.58,357.97) --
	(190.36,357.97) --
	(190.36,358.77) --
	(196.92,358.77) --
	(196.92,346.62) --
	(203.69,346.62) --
	(203.69,351.33) --
	(210.25,351.33) --
	(210.25,346.02) --
	(217.03,346.02) --
	(217.03,356.62) --
	(223.80,356.62) --
	(223.80,367.50) --
	(230.36,367.50) --
	(230.36,360.07) --
	(237.13,360.07) --
	(237.13,375.82) --
	(243.69,375.82) --
	(243.69,371.23) --
	(250.47,371.23) --
	(250.47,345.80) --
	(257.24,345.80) --
	(257.24,330.83) --
	(263.36,330.83) --
	(263.36,346.09) --
	(270.14,346.09) --
	(270.14,335.10) --
	(276.70,335.10) --
	(276.70,353.89) --
	(283.47,353.89) --
	(283.47,352.13) --
	(290.03,352.13) --
	(290.03,387.01) --
	(296.81,387.01) --
	(296.81,370.58) --
	(303.58,370.58) --
	(303.58,359.42) --
	(310.14,359.42) --
	(310.14,372.59) --
	(316.92,372.59) --
	(316.92,394.30) --
	(323.48,394.30) --
	(323.48,365.72) --
	(330.25,365.72) --
	(330.25,380.75) --
	(337.03,380.75) --
	(337.03,382.99);

\path[draw=drawColor,line width= 0.9pt,dash pattern=on 2pt off 2pt ,line join=round] (110.36,301.35) --
	(116.91,301.35) --
	(116.91,304.23) --
	(123.69,304.23) --
	(123.69,304.94) --
	(130.25,304.94) --
	(130.25,310.28) --
	(137.02,310.28) --
	(137.02,306.43) --
	(143.80,306.43) --
	(143.80,307.36) --
	(150.36,307.36) --
	(150.36,311.72) --
	(157.13,311.72) --
	(157.13,308.04) --
	(163.69,308.04) --
	(163.69,308.53) --
	(170.47,308.53) --
	(170.47,311.52) --
	(177.24,311.52) --
	(177.24,316.03) --
	(183.58,316.03) --
	(183.58,357.97) --
	(190.36,357.97) --
	(190.36,358.77) --
	(196.92,358.77) --
	(196.92,346.62) --
	(203.69,346.62) --
	(203.69,351.33) --
	(210.25,351.33) --
	(210.25,346.02) --
	(217.03,346.02) --
	(217.03,356.59) --
	(223.80,356.59) --
	(223.80,367.47) --
	(230.36,367.47) --
	(230.36,360.06) --
	(237.13,360.06) --
	(237.13,375.81) --
	(243.69,375.81) --
	(243.69,371.22) --
	(250.47,371.22) --
	(250.47,345.80) --
	(257.24,345.80) --
	(257.24,330.83) --
	(263.36,330.83) --
	(263.36,346.09) --
	(270.14,346.09) --
	(270.14,335.09) --
	(276.70,335.09) --
	(276.70,353.88) --
	(283.47,353.88) --
	(283.47,352.13) --
	(290.03,352.13) --
	(290.03,387.01) --
	(296.81,387.01) --
	(296.81,370.58) --
	(303.58,370.58) --
	(303.58,359.42) --
	(310.14,359.42) --
	(310.14,372.59) --
	(316.92,372.59) --
	(316.92,394.30) --
	(323.48,394.30) --
	(323.48,365.72) --
	(330.25,365.72) --
	(330.25,380.60) --
	(337.03,380.60) --
	(337.03,382.60);
\definecolor{drawColor}{RGB}{0,193,178}

\path[draw=drawColor,line width= 0.9pt,line join=round] ( 70.57,301.37) --
	( 77.35,301.37) --
	( 77.35,313.84) --
	( 83.91,313.84) --
	( 83.91,320.52) --
	( 90.68,320.52) --
	( 90.68,348.26) --
	( 97.46,348.26) --
	( 97.46,338.85) --
	(103.58,338.85) --
	(103.58,368.90) --
	(110.36,368.90) --
	(110.36,352.53) --
	(116.91,352.53) --
	(116.91,346.29) --
	(123.69,346.29) --
	(123.69,361.85) --
	(130.25,361.85) --
	(130.25,366.66) --
	(137.02,366.66) --
	(137.02,375.83) --
	(143.80,375.83) --
	(143.80,368.64) --
	(150.36,368.64) --
	(150.36,382.38) --
	(157.13,382.38) --
	(157.13,373.55) --
	(163.69,373.55) --
	(163.69,379.38) --
	(170.47,379.38) --
	(170.47,406.69) --
	(177.24,406.69) --
	(177.24,405.28) --
	(183.58,405.28) --
	(183.58,409.88) --
	(190.36,409.88) --
	(190.36,417.35) --
	(196.92,417.35) --
	(196.92,428.04) --
	(203.69,428.04) --
	(203.69,409.48) --
	(210.25,409.48) --
	(210.25,404.66) --
	(217.03,404.66) --
	(217.03,412.07) --
	(223.80,412.07) --
	(223.80,412.07) --
	(230.36,412.07) --
	(230.36,413.39) --
	(237.13,413.39) --
	(237.13,418.86) --
	(243.69,418.86) --
	(243.69,420.93) --
	(250.47,420.93) --
	(250.47,401.87) --
	(257.24,401.87) --
	(257.24,394.50) --
	(263.36,394.50) --
	(263.36,395.62) --
	(270.14,395.62) --
	(270.14,369.10) --
	(276.70,369.10) --
	(276.70,372.11) --
	(283.47,372.11) --
	(283.47,361.19) --
	(290.03,361.19) --
	(290.03,379.26) --
	(296.81,379.26) --
	(296.81,373.55) --
	(303.58,373.55) --
	(303.58,367.66) --
	(310.14,367.66) --
	(310.14,358.40) --
	(316.92,358.40) --
	(316.92,355.32) --
	(323.48,355.32) --
	(323.48,365.52) --
	(330.25,365.52) --
	(330.25,360.06) --
	(337.03,360.06) --
	(337.03,354.90);

\path[draw=drawColor,line width= 0.9pt,dash pattern=on 2pt off 2pt ,line join=round] ( 70.57,301.35) --
	( 77.35,301.35) --
	( 77.35,307.57) --
	( 83.91,307.57) --
	( 83.91,311.42) --
	( 90.68,311.42) --
	( 90.68,317.07) --
	( 97.46,317.07) --
	( 97.46,308.01) --
	(103.58,308.01) --
	(103.58,311.62) --
	(110.36,311.62) --
	(110.36,313.84) --
	(116.91,313.84) --
	(116.91,317.39) --
	(123.69,317.39) --
	(123.69,314.23) --
	(130.25,314.23) --
	(130.25,313.22) --
	(137.02,313.22) --
	(137.02,315.47) --
	(143.80,315.47) --
	(143.80,315.44) --
	(150.36,315.44) --
	(150.36,314.80) --
	(157.13,314.80) --
	(157.13,313.58) --
	(163.69,313.58) --
	(163.69,314.02) --
	(170.47,314.02) --
	(170.47,318.62) --
	(177.24,318.62) --
	(177.24,317.04) --
	(183.58,317.04) --
	(183.58,319.54) --
	(190.36,319.54) --
	(190.36,317.72) --
	(196.92,317.72) --
	(196.92,320.47) --
	(203.69,320.47) --
	(203.69,318.00) --
	(210.25,318.00) --
	(210.25,320.54) --
	(217.03,320.54) --
	(217.03,319.35) --
	(223.80,319.35) --
	(223.80,318.51) --
	(230.36,318.51) --
	(230.36,320.65) --
	(237.13,320.65) --
	(237.13,322.10) --
	(243.69,322.10) --
	(243.69,323.34) --
	(250.47,323.34) --
	(250.47,321.40) --
	(257.24,321.40) --
	(257.24,321.98) --
	(263.36,321.98) --
	(263.36,328.79) --
	(270.14,328.79) --
	(270.14,320.15) --
	(276.70,320.15) --
	(276.70,318.91) --
	(283.47,318.91) --
	(283.47,317.26) --
	(290.03,317.26) --
	(290.03,321.02) --
	(296.81,321.02) --
	(296.81,319.88) --
	(303.58,319.88) --
	(303.58,319.23) --
	(310.14,319.23) --
	(310.14,317.79) --
	(316.92,317.79) --
	(316.92,317.12) --
	(323.48,317.12) --
	(323.48,320.67) --
	(330.25,320.67) --
	(330.25,320.33) --
	(337.03,320.33) --
	(337.03,320.27);
\end{scope}
\begin{scope}
\path[clip] (  0.00,  0.00) rectangle (361.35,462.53);
\definecolor{drawColor}{gray}{0.30}

\node[text=drawColor,anchor=base east,inner sep=0pt, outer sep=0pt, scale=  0.88] at ( 52.30,298.24) {0};

\node[text=drawColor,anchor=base east,inner sep=0pt, outer sep=0pt, scale=  0.88] at ( 52.30,332.12) {25,000};

\node[text=drawColor,anchor=base east,inner sep=0pt, outer sep=0pt, scale=  0.88] at ( 52.30,366.00) {50,000};

\node[text=drawColor,anchor=base east,inner sep=0pt, outer sep=0pt, scale=  0.88] at ( 52.30,399.88) {75,000};
\end{scope}
\begin{scope}
\path[clip] (  0.00,  0.00) rectangle (361.35,462.53);
\definecolor{drawColor}{gray}{0.30}

\node[text=drawColor,anchor=base,inner sep=0pt, outer sep=0pt, scale=  0.88] at ( 90.68,284.00) {2019};

\node[text=drawColor,anchor=base,inner sep=0pt, outer sep=0pt, scale=  0.88] at (170.47,284.00) {2020};

\node[text=drawColor,anchor=base,inner sep=0pt, outer sep=0pt, scale=  0.88] at (250.47,284.00) {2021};

\node[text=drawColor,anchor=base,inner sep=0pt, outer sep=0pt, scale=  0.88] at (330.25,284.00) {2022};
\end{scope}
\begin{scope}
\path[clip] (  0.00,  0.00) rectangle (361.35,462.53);
\definecolor{drawColor}{RGB}{0,0,0}

\node[text=drawColor,anchor=base,inner sep=0pt, outer sep=0pt, scale=  1.10] at (203.80,271.96) {Month};
\end{scope}
\begin{scope}
\path[clip] (  0.00,  0.00) rectangle (361.35,462.53);
\definecolor{drawColor}{RGB}{0,0,0}

\node[text=drawColor,rotate= 90.00,anchor=base,inner sep=0pt, outer sep=0pt, scale=  1.10] at ( 18.58,364.69) {Number of addresses/entities};
\end{scope}
\begin{scope}
\path[clip] (  0.00,  0.00) rectangle (361.35,462.53);
\definecolor{drawColor}{RGB}{0,0,0}

\node[text=drawColor,anchor=base west,inner sep=0pt, outer sep=0pt, scale=  1.32] at ( 57.25,442.44) {Pre-mixing};
\end{scope}
\begin{scope}
\path[clip] ( 57.25,102.31) rectangle (350.35,241.67);
\definecolor{drawColor}{gray}{0.92}

\path[draw=drawColor,line width= 0.4pt,line join=round] ( 57.25,128.87) --
	(350.35,128.87);

\path[draw=drawColor,line width= 0.4pt,line join=round] ( 57.25,169.40) --
	(350.35,169.40);

\path[draw=drawColor,line width= 0.4pt,line join=round] ( 57.25,209.94) --
	(350.35,209.94);

\path[draw=drawColor,line width= 0.4pt,line join=round] (130.58,102.31) --
	(130.58,241.67);

\path[draw=drawColor,line width= 0.4pt,line join=round] (210.47,102.31) --
	(210.47,241.67);

\path[draw=drawColor,line width= 0.4pt,line join=round] (290.36,102.31) --
	(290.36,241.67);

\path[draw=drawColor,line width= 0.9pt,line join=round] ( 57.25,108.60) --
	(350.35,108.60);

\path[draw=drawColor,line width= 0.9pt,line join=round] ( 57.25,149.14) --
	(350.35,149.14);

\path[draw=drawColor,line width= 0.9pt,line join=round] ( 57.25,189.67) --
	(350.35,189.67);

\path[draw=drawColor,line width= 0.9pt,line join=round] ( 57.25,230.21) --
	(350.35,230.21);

\path[draw=drawColor,line width= 0.9pt,line join=round] ( 90.68,102.31) --
	( 90.68,241.67);

\path[draw=drawColor,line width= 0.9pt,line join=round] (170.47,102.31) --
	(170.47,241.67);

\path[draw=drawColor,line width= 0.9pt,line join=round] (250.47,102.31) --
	(250.47,241.67);

\path[draw=drawColor,line width= 0.9pt,line join=round] (330.25,102.31) --
	(330.25,241.67);
\definecolor{drawColor}{RGB}{237,144,164}

\path[draw=drawColor,line width= 0.9pt,line join=round] (110.36,108.65) --
	(116.91,108.65) --
	(116.91,110.37) --
	(123.69,110.37) --
	(123.69,110.80) --
	(130.25,110.80) --
	(130.25,113.99) --
	(137.02,113.99) --
	(137.02,111.69) --
	(143.80,111.69) --
	(143.80,112.24) --
	(150.36,112.24) --
	(150.36,114.85) --
	(157.13,114.85) --
	(157.13,112.65) --
	(163.69,112.65) --
	(163.69,112.95) --
	(170.47,112.95) --
	(170.47,114.74) --
	(177.24,114.74) --
	(177.24,117.44) --
	(183.58,117.44) --
	(183.58,142.52) --
	(190.36,142.52) --
	(190.36,143.00) --
	(196.92,143.00) --
	(196.92,135.73) --
	(203.69,135.73) --
	(203.69,138.55) --
	(210.25,138.55) --
	(210.25,135.37) --
	(217.03,135.37) --
	(217.03,141.72) --
	(223.80,141.72) --
	(223.80,148.22) --
	(230.36,148.22) --
	(230.36,143.78) --
	(237.13,143.78) --
	(237.13,153.20) --
	(243.69,153.20) --
	(243.69,150.45) --
	(250.47,150.45) --
	(250.47,135.24) --
	(257.24,135.24) --
	(257.24,126.28) --
	(263.36,126.28) --
	(263.36,135.42) --
	(270.14,135.42) --
	(270.14,128.84) --
	(276.70,128.84) --
	(276.70,140.08) --
	(283.47,140.08) --
	(283.47,139.03) --
	(290.03,139.03) --
	(290.03,159.90) --
	(296.81,159.90) --
	(296.81,150.07) --
	(303.58,150.07) --
	(303.58,143.39) --
	(310.14,143.39) --
	(310.14,151.27) --
	(316.92,151.27) --
	(316.92,164.25) --
	(323.48,164.25) --
	(323.48,147.16) --
	(330.25,147.16) --
	(330.25,156.15) --
	(337.03,156.15) --
	(337.03,157.49);

\path[draw=drawColor,line width= 0.9pt,dash pattern=on 2pt off 2pt ,line join=round] (110.36,108.64) --
	(116.91,108.64) --
	(116.91,110.32) --
	(123.69,110.32) --
	(123.69,110.76) --
	(130.25,110.76) --
	(130.25,113.93) --
	(137.02,113.93) --
	(137.02,111.66) --
	(143.80,111.66) --
	(143.80,112.21) --
	(150.36,112.21) --
	(150.36,114.75) --
	(157.13,114.75) --
	(157.13,112.63) --
	(163.69,112.63) --
	(163.69,112.90) --
	(170.47,112.90) --
	(170.47,114.72) --
	(177.24,114.72) --
	(177.24,117.34) --
	(183.58,117.34) --
	(183.58,142.00) --
	(190.36,142.00) --
	(190.36,142.47) --
	(196.92,142.47) --
	(196.92,135.12) --
	(203.69,135.12) --
	(203.69,137.98) --
	(210.25,137.98) --
	(210.25,135.08) --
	(217.03,135.08) --
	(217.03,141.40) --
	(223.80,141.40) --
	(223.80,147.81) --
	(230.36,147.81) --
	(230.36,143.49) --
	(237.13,143.49) --
	(237.13,152.33) --
	(243.69,152.33) --
	(243.69,149.42) --
	(250.47,149.42) --
	(250.47,133.35) --
	(257.24,133.35) --
	(257.24,125.46) --
	(263.36,125.46) --
	(263.36,134.50) --
	(270.14,134.50) --
	(270.14,128.06) --
	(276.70,128.06) --
	(276.70,138.92) --
	(283.47,138.92) --
	(283.47,137.64) --
	(290.03,137.64) --
	(290.03,157.06) --
	(296.81,157.06) --
	(296.81,147.35) --
	(303.58,147.35) --
	(303.58,141.68) --
	(310.14,141.68) --
	(310.14,148.42) --
	(316.92,148.42) --
	(316.92,159.83) --
	(323.48,159.83) --
	(323.48,143.22) --
	(330.25,143.22) --
	(330.25,147.93) --
	(337.03,147.93) --
	(337.03,140.27);
\definecolor{drawColor}{RGB}{0,193,178}

\path[draw=drawColor,line width= 0.9pt,line join=round] ( 70.57,108.67) --
	( 77.35,108.67) --
	( 77.35,120.16) --
	( 83.91,120.16) --
	( 83.91,126.07) --
	( 90.68,126.07) --
	( 90.68,154.19) --
	( 97.46,154.19) --
	( 97.46,143.94) --
	(103.58,143.94) --
	(103.58,171.63) --
	(110.36,171.63) --
	(110.36,161.10) --
	(116.91,161.10) --
	(116.91,154.27) --
	(123.69,154.27) --
	(123.69,175.15) --
	(130.25,175.15) --
	(130.25,172.28) --
	(137.02,172.28) --
	(137.02,192.05) --
	(143.80,192.05) --
	(143.80,177.32) --
	(150.36,177.32) --
	(150.36,185.30) --
	(157.13,185.30) --
	(157.13,177.12) --
	(163.69,177.12) --
	(163.69,181.81) --
	(170.47,181.81) --
	(170.47,206.98) --
	(177.24,206.98) --
	(177.24,204.66) --
	(183.58,204.66) --
	(183.58,213.83) --
	(190.36,213.83) --
	(190.36,220.24) --
	(196.92,220.24) --
	(196.92,235.33) --
	(203.69,235.33) --
	(203.69,206.51) --
	(210.25,206.51) --
	(210.25,210.28) --
	(217.03,210.28) --
	(217.03,219.68) --
	(223.80,219.68) --
	(223.80,217.13) --
	(230.36,217.13) --
	(230.36,215.89) --
	(237.13,215.89) --
	(237.13,220.78) --
	(243.69,220.78) --
	(243.69,222.55) --
	(250.47,222.55) --
	(250.47,205.00) --
	(257.24,205.00) --
	(257.24,200.41) --
	(263.36,200.41) --
	(263.36,190.51) --
	(270.14,190.51) --
	(270.14,175.15) --
	(276.70,175.15) --
	(276.70,179.39) --
	(283.47,179.39) --
	(283.47,164.40) --
	(290.03,164.40) --
	(290.03,180.43) --
	(296.81,180.43) --
	(296.81,170.86) --
	(303.58,170.86) --
	(303.58,164.71) --
	(310.14,164.71) --
	(310.14,158.24) --
	(316.92,158.24) --
	(316.92,156.05) --
	(323.48,156.05) --
	(323.48,164.66) --
	(330.25,164.66) --
	(330.25,157.71) --
	(337.03,157.71) --
	(337.03,155.38);

\path[draw=drawColor,line width= 0.9pt,dash pattern=on 2pt off 2pt ,line join=round] ( 70.57,108.65) --
	( 77.35,108.65) --
	( 77.35,117.49) --
	( 83.91,117.49) --
	( 83.91,121.80) --
	( 90.68,121.80) --
	( 90.68,137.14) --
	( 97.46,137.14) --
	( 97.46,128.95) --
	(103.58,128.95) --
	(103.58,145.68) --
	(110.36,145.68) --
	(110.36,144.75) --
	(116.91,144.75) --
	(116.91,139.92) --
	(123.69,139.92) --
	(123.69,149.98) --
	(130.25,149.98) --
	(130.25,146.28) --
	(137.02,146.28) --
	(137.02,148.13) --
	(143.80,148.13) --
	(143.80,142.71) --
	(150.36,142.71) --
	(150.36,151.13) --
	(157.13,151.13) --
	(157.13,147.69) --
	(163.69,147.69) --
	(163.69,148.01) --
	(170.47,148.01) --
	(170.47,160.55) --
	(177.24,160.55) --
	(177.24,161.83) --
	(183.58,161.83) --
	(183.58,168.41) --
	(190.36,168.41) --
	(190.36,169.76) --
	(196.92,169.76) --
	(196.92,177.33) --
	(203.69,177.33) --
	(203.69,161.08) --
	(210.25,161.08) --
	(210.25,165.29) --
	(217.03,165.29) --
	(217.03,168.78) --
	(223.80,168.78) --
	(223.80,169.85) --
	(230.36,169.85) --
	(230.36,171.20) --
	(237.13,171.20) --
	(237.13,173.92) --
	(243.69,173.92) --
	(243.69,173.54) --
	(250.47,173.54) --
	(250.47,166.09) --
	(257.24,166.09) --
	(257.24,165.23) --
	(263.36,165.23) --
	(263.36,160.17) --
	(270.14,160.17) --
	(270.14,149.40) --
	(276.70,149.40) --
	(276.70,153.76) --
	(283.47,153.76) --
	(283.47,143.73) --
	(290.03,143.73) --
	(290.03,152.81) --
	(296.81,152.81) --
	(296.81,145.55) --
	(303.58,145.55) --
	(303.58,143.19) --
	(310.14,143.19) --
	(310.14,138.28) --
	(316.92,138.28) --
	(316.92,138.81) --
	(323.48,138.81) --
	(323.48,143.03) --
	(330.25,143.03) --
	(330.25,137.88) --
	(337.03,137.88) --
	(337.03,133.75);
\end{scope}
\begin{scope}
\path[clip] (  0.00,  0.00) rectangle (361.35,462.53);
\definecolor{drawColor}{gray}{0.30}

\node[text=drawColor,anchor=base east,inner sep=0pt, outer sep=0pt, scale=  0.88] at ( 52.30,105.57) {0};

\node[text=drawColor,anchor=base east,inner sep=0pt, outer sep=0pt, scale=  0.88] at ( 52.30,146.11) {50,000};

\node[text=drawColor,anchor=base east,inner sep=0pt, outer sep=0pt, scale=  0.88] at ( 52.30,186.64) {100,000};

\node[text=drawColor,anchor=base east,inner sep=0pt, outer sep=0pt, scale=  0.88] at ( 52.30,227.18) {150,000};
\end{scope}
\begin{scope}
\path[clip] (  0.00,  0.00) rectangle (361.35,462.53);
\definecolor{drawColor}{gray}{0.30}

\node[text=drawColor,anchor=base,inner sep=0pt, outer sep=0pt, scale=  0.88] at ( 90.68, 91.30) {2019};

\node[text=drawColor,anchor=base,inner sep=0pt, outer sep=0pt, scale=  0.88] at (170.47, 91.30) {2020};

\node[text=drawColor,anchor=base,inner sep=0pt, outer sep=0pt, scale=  0.88] at (250.47, 91.30) {2021};

\node[text=drawColor,anchor=base,inner sep=0pt, outer sep=0pt, scale=  0.88] at (330.25, 91.30) {2022};
\end{scope}
\begin{scope}
\path[clip] (  0.00,  0.00) rectangle (361.35,462.53);
\definecolor{drawColor}{RGB}{0,0,0}

\node[text=drawColor,anchor=base,inner sep=0pt, outer sep=0pt, scale=  1.10] at (203.80, 79.26) {Month};
\end{scope}
\begin{scope}
\path[clip] (  0.00,  0.00) rectangle (361.35,462.53);
\definecolor{drawColor}{RGB}{0,0,0}

\node[text=drawColor,rotate= 90.00,anchor=base,inner sep=0pt, outer sep=0pt, scale=  1.10] at ( 18.58,171.99) {Number of addresses/entities};
\end{scope}
\begin{scope}
\path[clip] (  0.00,  0.00) rectangle (361.35,462.53);
\definecolor{drawColor}{RGB}{0,0,0}

\node[text=drawColor,anchor=base west,inner sep=0pt, outer sep=0pt, scale=  1.32] at ( 57.25,249.73) {Post-mixing};
\end{scope}
\begin{scope}
\path[clip] (  0.00,  0.00) rectangle (361.35,462.53);
\definecolor{drawColor}{RGB}{0,0,0}

\node[text=drawColor,anchor=base west,inner sep=0pt, outer sep=0pt, scale=  1.10] at ( 96.67, 51.98) {Service};
\end{scope}
\begin{scope}
\path[clip] (  0.00,  0.00) rectangle (361.35,462.53);
\definecolor{drawColor}{RGB}{237,144,164}

\path[draw=drawColor,line width= 0.9pt,line join=round] ( 98.12, 38.18) -- (109.68, 38.18);
\end{scope}
\begin{scope}
\path[clip] (  0.00,  0.00) rectangle (361.35,462.53);
\definecolor{drawColor}{RGB}{0,193,178}

\path[draw=drawColor,line width= 0.9pt,line join=round] ( 98.12, 23.73) -- (109.68, 23.73);
\end{scope}
\begin{scope}
\path[clip] (  0.00,  0.00) rectangle (361.35,462.53);
\definecolor{drawColor}{RGB}{0,0,0}

\node[text=drawColor,anchor=base west,inner sep=0pt, outer sep=0pt, scale=  0.88] at (116.63, 35.15) {Samourai};
\end{scope}
\begin{scope}
\path[clip] (  0.00,  0.00) rectangle (361.35,462.53);
\definecolor{drawColor}{RGB}{0,0,0}

\node[text=drawColor,anchor=base west,inner sep=0pt, outer sep=0pt, scale=  0.88] at (116.63, 20.70) {Wasabi};
\end{scope}
\begin{scope}
\path[clip] (  0.00,  0.00) rectangle (361.35,462.53);
\definecolor{drawColor}{RGB}{0,0,0}

\node[text=drawColor,anchor=base west,inner sep=0pt, outer sep=0pt, scale=  1.10] at (174.82, 51.98) {Level};
\end{scope}
\begin{scope}
\path[clip] (  0.00,  0.00) rectangle (361.35,462.53);
\definecolor{drawColor}{RGB}{0,0,0}

\path[draw=drawColor,line width= 0.9pt,line join=round] (176.26, 38.18) -- (187.83, 38.18);
\end{scope}
\begin{scope}
\path[clip] (  0.00,  0.00) rectangle (361.35,462.53);
\definecolor{drawColor}{RGB}{0,0,0}

\path[draw=drawColor,line width= 0.9pt,dash pattern=on 2pt off 2pt ,line join=round] (176.26, 23.73) -- (187.83, 23.73);
\end{scope}
\begin{scope}
\path[clip] (  0.00,  0.00) rectangle (361.35,462.53);
\definecolor{drawColor}{RGB}{0,0,0}

\node[text=drawColor,anchor=base west,inner sep=0pt, outer sep=0pt, scale=  0.88] at (194.77, 35.15) {Number of addresses - Level 0};
\end{scope}
\begin{scope}
\path[clip] (  0.00,  0.00) rectangle (361.35,462.53);
\definecolor{drawColor}{RGB}{0,0,0}

\node[text=drawColor,anchor=base west,inner sep=0pt, outer sep=0pt, scale=  0.88] at (194.77, 20.70) {Number of entities - Level 1};
\end{scope}
\end{tikzpicture}

%% file: sections/6_discussion.tex
\section{Discussion}
\label{sec:discussion}

% Interpretation of results
Wasabi and Samourai offer users privacy behind the scenes via decentralized CoinJoin and have become, as our results show, an integral part of the Bitcoin ecosystem since Nov 2018. We have shown that it is possible to detect CoinJoin transactions with high accuracy using relatively simple algorithmic methods. We can also trace the flows of funds received by services directly or indirectly. Our analysis also reveals \nTotalMixedBTC mixed coins with a total value of \nTotalMixedUSD USD in the past and a mixing throughput of around \nMixedRecentBTC BTC or \nMixedRecentUSD USD within recent months. Furthermore, by attributing services on the input and output side of CoinJoin transactions, we found a lower bound of \nExchangesReceivingBTC entities controlled by cryptoasset exchanges and received coins from these wallets via one or two hops. Surprisingly, the amount of accepted CoinJoin transactions and mixed BTC are growing steadily despite increasingly tightening Anti-Money Laundering (AML) regulations. These regulatory efforts demand traceability of funds, which opposes the design goal of decentralized mixing services like Wasabi or Samourai.

% Limitations
The empirical results reported herein should be considered in light of some limitations. First, \Ficsor et al.~\cite{Ficsor:2021a} have recently published a generalization of the Chaumian CoinJoin named WabiSabi. WabiSabi serves as the basis for Wasabi Wallet 2.0, which has already seen TestNet releases and is expected to be fully released in 2022. It offers, among other features, support for arbitrarily variable CoinJoin amounts. Wasabi 2.0 is expected to limit the effectiveness of the discussed WCDH heuristic severely. Second, we point out that our attribution dataset, which identifies cryptoasset exchanges, is incomplete and that the numbers reported in our ecosystem analysis are, therefore, lower bounds. However, our analysis is easily reproducible with a more comprehensive attribution tag dataset, and it is even possible to name the involved exchanges, which we refrain from for ethical reasons. Therefore, another possible future direction is to run our analysis with a more comprehensive attribution tag dataset to obtain a complete picture of mixing activities in the Bitcoin ecosystem, make informed decisions, and assess compliance with AML regulation.

% Implications
CoinJoin transactions and software facilitating them are a double-edged sword, and the implications of our work very much depend on the perspective.

For privacy-seeking \textbf{end users}, wallets like Wasabi and Samourai are a practical, low-entry barrier solution to Bitcoin's anonymity problem. To the best of our knowledge, it is hardly possible to de-mix CoinJoins produced by these wallets. However, users should be aware that a priori perceived anonymity gains of such services is hindered because their transactions are visible on-chain, and cryptoasset tracing and tracking solutions can detect them. Also, pre-mixed and post-mixed addresses can be tracked, effectively reducing the anonymity guarantees provided by these mixing wallets. On the other hand, stricter regulations could require users to clearly explain the origin of their coins if they want to cash out coins at an AML-compliant exchange. Of course, this is more difficult to explain when CoinJoins are involved.

For \textbf{cryptoasset exchanges} our findings show that automatically detecting CoinJoin transactions created by two popular wallets is easily possible with relatively simple heuristics. Our results show that the number of transactions and mixed coins accepted is still growing, from which we can infer that acceptance of CoinJoins currently does not yet raise compliance issues, at least for the exchanges in our dataset, \nExchangesInTop of them ranking in the top \nExchangesTop by 24h trading volume\footnote{\url{https://coinmarketcap.com/rankings/exchanges/}}. However, this may change when the AML compliance rules, which apply to traditional financial service providers, are also enforced in the cryptocurrency area.

For \textbf{regulatory bodies} our results provide insight into the current use of two popular mixing services preventing traceability of funds. One could even imagine applying similar procedures to monitor the adoption of AML compliance in cryptoassets ecosystems without invading users' privacy. However, we also point out that imposing the traceability-of-funds requirement on cryptoasset exchanges essentially demands CoinJoin not to be part of a coin's lineage. In the long run, this might discourage users from using tools that protect their fundamental right to privacy in a completely transparent financial ecosystem.

%% file: sections/7_conclusions.tex
% !TeX root = ../main.tex

\section{Conclusions}
\label{sec:conclusions}

% Summary
We measured the adoption and privacy guarantees of two popular decentralized CoinJoin implementations: Wasabi and Samourai wallet. We found algorithmic methods that can yield transactions generated by these wallets with high accuracy and constructed a transaction dataset spanning the entire history of these wallets up to the end of our observation period on 2022-02-28. We then quantified the lower-bound volumes of coins mixed by these wallets and showed they have been adopted over time and are not yet affected by current regulatory efforts. 
However, we also show that the anonymity guarantees are much lower than assumed when considering the flow of funds before the pre-mix and after the post-mix wallets. Overall, our work contributes empirical evidence to a highly controversial discussion in the field of tension between the legitimate right to privacy and the need for transparency and traceability to mitigate the abuse of cryptoassets for illegitimate purposes.

% Take away and future work
Our work currently focuses on Bitcoin only. Future work could expand our methodology to account-model ledgers like Ethereum and more closely investigate emerging mixing services such as Tornado-Cash. Also, the entire Decentralized Finance (DeFi) ecosystem offers tremendous opportunities for increasing anonymity by tunneling cryptoassets through DeFi protocol compositions, which are currently hardly understood.

%% file: sections/8_appendix.tex
% !TeX root = ../main.tex

\section{Appendix}
\label{sec:appendix}

\subsection{Fresh inputs}\label{sec:appendix_fresh_inputs}

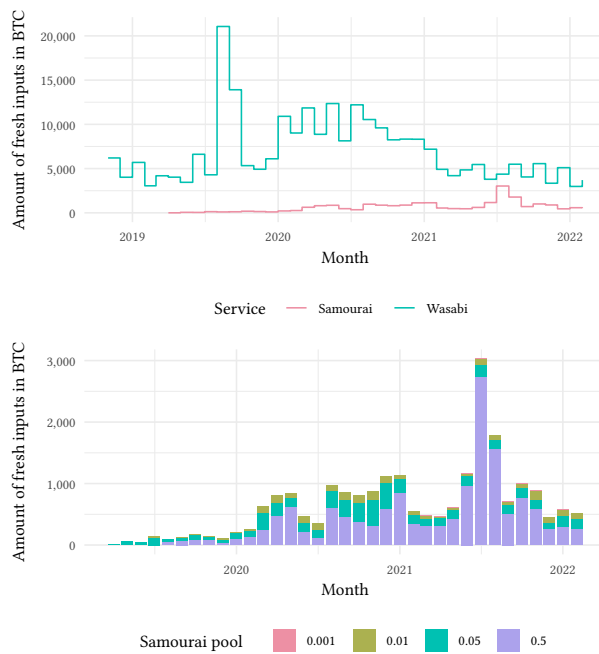
\begin{figure}
    \centering
    \resizebox{\columnwidth}{!}{\input{graphics/fresh_inputs.tex}}
    \caption{Monthly amount of fresh BTC entering Wasabi and Samourai (top),
             as well as by Samourai pools (bottom).}
    \label{fig:freshbtc}
\end{figure}

As with mix outputs, we can also compare the fresh inputs for both services by
analyzing CoinJoin inputs that are not the outputs of previous CoinJoins in
order to find fresh inputs entering either Wasabi or Samourai.
Figure~\ref{fig:freshbtc} shows the development of fresh inputs for Wasabi and
all Samourai pools. For Wasabi, there are is huge spike in fresh inputs in
August \& September 2019. This is in line with the spike in mixed outputs
leaving Wasabi (cf.\ Figure~\ref{fig:cj-mixed-amount}). For Samourai, the
amount of fresh BTC also co-evolves very closely to the amount of mixed outputs.

\subsection{Manual Wasabi CoinJoin Validation}\label{sec:appendix_wasabi_manual}

We conducted the following Wasabi CoinJoin transactions:
\begin{itemize}
    \item \txhash{508e59614f442d35a1d901972ab4b4baa5e223251d90c0183a20e23dcf5343ab}
    \item \txhash{fd2b5239c022d60d20da88f3c9288c518713646125f4628cdbc234146e839d17}
    \item \txhash{6da17d7f3519e3218ed898ec9e6d4abcaec7226f3ec98513fea8a54c63d471c5}
    \item \txhash{c6ef8546c4d4b404b40988478a2fd16a81556ac030ca2e956db3ce41883089c8}
    \item \txhash{cb64224eb96ca22118d1af44d3f13a95b844ea93c13e099f78c7307097a9a37d}
    \item \txhash{9a5e5bf667f38affe21203cf51b4ddc6baee0b9e30542c1e9d3152e5f4989aa0}
    \item \txhash{130764d7a801d6f73ba3fec03ab969fbabe3f585b4636ff9ff68b8719e19ab09}
    \item \txhash{dcd5d22043c48f989984827560f4803bb37c75b6afb5adf35caac5edb43612e5}
    \item \txhash{1f6a3288f60c1e23690c8a836d1743f807c53fcb8ea24803f9166b02c1b43720}
    \item \txhash{3a03cfd29bb02a228b1af287963e643925aca8c33b0ada323fda8ad94be567a4}
    \item \txhash{eb58313d1dbca66973ff7490a7d69d924e6fc964ef404f42cb924fc9dac6813f}
\end{itemize}

\subsection{Manual Samourai CoinJoin Validation}\label{sec:appendix_samourai_manual}

We issued the following Tx0 transactions:
\begin{itemize}
    \item \txhash{c61527d6174858699690ebf1db60912a08edff175b5de77c77035ad740e2ec02} % normal 0.001
    \item \txhash{a3f2cc5e220dde7dadf50f6b3c854e9f8d0e064331df86d21bfcb91a01cdc559} % normal 0.05
    \item \txhash{79ecae4659bfb940b369057bfe4109ba30b6d8d15b42c0028dde59fd155a6c99} % normal 0.01
    \item \txhash{2893cf9ec37ab64809e9dfbb665b04e5d7d857ddbb50ea2b7d6fc1708132bb07} % high 0.001
    \item \txhash{abb222724d61a34775609b6d1fd124310f85a43d9cac7f837c8ec052df8f983b} % low 0.001
\end{itemize}

The mix outputs of these Tx0 transactions were used in the following Whirlpool
CoinJoin transactions:
\begin{itemize}
    \item \txhash{970bca671619b3c0572c9adaf34f7075e5558d004b66baade10130d4f525de7c}
    \item \txhash{e55b71aa734227dbff42869ae70b1bec76d87151de5a7380e0b73d85383f29f7}
    \item \txhash{933cd23bad3c014571ec2790c49f95b03a288da11546487eeac3d8910e860b54}
    \item \txhash{c6c1f655dd54969ed8664d7cd6136ffb70214385e1b5dd86567ece8ae75106aa}
    \item \txhash{f0cc7e1f8c68db6c60471f4331223b0499037e278b0c94c7a7a8637f1df7cc59}
    \item \txhash{d12d1475eef88e5ee63b2cb8d54244aa883c93dc3b96120d0a37d353182d4e2a}
    \item \txhash{264cd4f585dc03c6d37e1d809d5a7082ef630e7a4c50e0e90bb0f417822b4f35}
    \item \txhash{f64ca6a9728152bf1da35d8b6b3bffa999e7c4ffcf7d9903c4de148cf94e0912}
    \item \txhash{5334b15344417b6d33375f6266053918160dcd00aa95c0ef0e1b15ca2bd241f2}
    \item \txhash{7531eee3a77ac460471faf22b27729f303358fbed6f3d527118ab94202453010}
    \item \txhash{130a1fc346ba72e68d1574546a8f4cca1a9f6deb09f4bb8a05abb97b91e955e9}
    \item \txhash{20e8630e029583ccdb1e1478c9c0238d8743442476832c4071871f20fd9af8cc}
    \item \txhash{9bacbc7275e38be43d23def78c8c5406f06024d5e1314f7950ee581643b75f55}
    \item \txhash{d82685cd6f60b85c4b8c5cc39e9fd8b9bef44c52c43dcfa1970dc95080ae9ddf}
    \item \txhash{66f3164335185ced1bcab582116434e271a8f8837b79d89ca799256bd15b3105}
    \item \txhash{5f8fc7232ab8d66820f6043010f36d5663cf452f7efae9acbf3708f8dea8af86}
    \item \txhash{76b7c6f5ebbf5e35edb89a73557238e034d3161f2a34c3c9df21c1fb1b9904c8}
    \item \txhash{e9b39d693015aeffef35a1cb75b1e4a972dc860a6c93bcdd718b83354dbb5b8a}
    \item \txhash{5f348a29a5e38e97753b8f1e8959d575b4e1c85f174f67c9b86aaefa594620a7}
    \item \txhash{8f9f4b3550783eb0a5672e793e082dd48ec89338a5845e0ba7c74eb52ff988d8}
    \item \txhash{4897e73f2d7d90aadbf42888a7b054d5c8d7f24a804681d1d98ae227edf34e8d}
    \item \txhash{c76365c9e9174b351449d1739d158cb1fc06b60bf3e713280889587a1b1b6479}
    \item \txhash{535d4dca58e3d4e2d9b66128e0a585be31a7f69d44f5888cb97da4eada4735ff}
\end{itemize}

%% file: graphics/fresh_inputs.tex
% !TEX encoding = UTF-8 Unicode
\begin{tikzpicture}[x=1pt,y=1pt]
\definecolor{fillColor}{RGB}{255,255,255}
\path[use as bounding box,fill=fillColor,fill opacity=0.00] (0,0) rectangle (361.35,390.26);
\begin{scope}
\path[clip] ( 52.85,262.27) rectangle (350.35,379.26);
\definecolor{drawColor}{gray}{0.92}

\path[draw=drawColor,line width= 0.4pt,line join=round] ( 52.85,280.20) --
	(350.35,280.20);

\path[draw=drawColor,line width= 0.4pt,line join=round] ( 52.85,305.45) --
	(350.35,305.45);

\path[draw=drawColor,line width= 0.4pt,line join=round] ( 52.85,330.70) --
	(350.35,330.70);

\path[draw=drawColor,line width= 0.4pt,line join=round] ( 52.85,355.95) --
	(350.35,355.95);

\path[draw=drawColor,line width= 0.4pt,line join=round] (121.81,262.27) --
	(121.81,379.26);

\path[draw=drawColor,line width= 0.4pt,line join=round] (205.02,262.27) --
	(205.02,379.26);

\path[draw=drawColor,line width= 0.4pt,line join=round] (288.22,262.27) --
	(288.22,379.26);

\path[draw=drawColor,line width= 0.9pt,line join=round] ( 52.85,267.58) --
	(350.35,267.58);

\path[draw=drawColor,line width= 0.9pt,line join=round] ( 52.85,292.83) --
	(350.35,292.83);

\path[draw=drawColor,line width= 0.9pt,line join=round] ( 52.85,318.08) --
	(350.35,318.08);

\path[draw=drawColor,line width= 0.9pt,line join=round] ( 52.85,343.33) --
	(350.35,343.33);

\path[draw=drawColor,line width= 0.9pt,line join=round] ( 52.85,368.58) --
	(350.35,368.58);

\path[draw=drawColor,line width= 0.9pt,line join=round] ( 80.26,262.27) --
	( 80.26,379.26);

\path[draw=drawColor,line width= 0.9pt,line join=round] (163.36,262.27) --
	(163.36,379.26);

\path[draw=drawColor,line width= 0.9pt,line join=round] (246.68,262.27) --
	(246.68,379.26);

\path[draw=drawColor,line width= 0.9pt,line join=round] (329.77,262.27) --
	(329.77,379.26);
\definecolor{drawColor}{RGB}{237,144,164}

\path[draw=drawColor,line width= 0.9pt,line join=round] (100.75,267.59) --
	(107.58,267.59) --
	(107.58,267.86) --
	(114.64,267.86) --
	(114.64,267.81) --
	(121.47,267.81) --
	(121.47,268.24) --
	(128.52,268.24) --
	(128.52,268.09) --
	(135.58,268.09) --
	(135.58,268.19) --
	(142.41,268.19) --
	(142.41,268.50) --
	(149.47,268.50) --
	(149.47,268.31) --
	(156.30,268.31) --
	(156.30,268.10) --
	(163.36,268.10) --
	(163.36,268.66) --
	(170.41,268.66) --
	(170.41,268.88) --
	(177.01,268.88) --
	(177.01,270.75) --
	(184.07,270.75) --
	(184.07,271.64) --
	(190.90,271.64) --
	(190.90,271.88) --
	(197.96,271.88) --
	(197.96,269.95) --
	(204.79,269.95) --
	(204.79,269.34) --
	(211.85,269.34) --
	(211.85,272.50) --
	(218.90,272.50) --
	(218.90,271.95) --
	(225.73,271.95) --
	(225.73,271.64) --
	(232.79,271.64) --
	(232.79,271.99) --
	(239.62,271.99) --
	(239.62,273.27) --
	(246.68,273.27) --
	(246.68,273.32) --
	(253.73,273.32) --
	(253.73,270.32) --
	(260.11,270.32) --
	(260.11,269.99) --
	(267.17,269.99) --
	(267.17,269.90) --
	(274.00,269.90) --
	(274.00,270.67) --
	(281.05,270.67) --
	(281.05,273.45) --
	(287.88,273.45) --
	(287.88,282.88) --
	(294.94,282.88) --
	(294.94,276.58) --
	(302.00,276.58) --
	(302.00,271.14) --
	(308.83,271.14) --
	(308.83,272.64) --
	(315.88,272.64) --
	(315.88,272.07) --
	(322.71,272.07) --
	(322.71,269.86) --
	(329.77,269.86) --
	(329.77,270.50) --
	(336.83,270.50) --
	(336.83,270.22);
\definecolor{drawColor}{RGB}{0,193,178}

\path[draw=drawColor,line width= 0.9pt,line join=round] ( 66.38,298.90) --
	( 73.21,298.90) --
	( 73.21,287.89) --
	( 80.26,287.89) --
	( 80.26,296.35) --
	( 87.32,296.35) --
	( 87.32,283.00) --
	( 93.69,283.00) --
	( 93.69,288.73) --
	(100.75,288.73) --
	(100.75,287.93) --
	(107.58,287.93) --
	(107.58,284.99) --
	(114.64,284.99) --
	(114.64,300.99) --
	(121.47,300.99) --
	(121.47,289.28) --
	(128.52,289.28) --
	(128.52,373.94) --
	(135.58,373.94) --
	(135.58,337.83) --
	(142.41,337.83) --
	(142.41,294.51) --
	(149.47,294.51) --
	(149.47,292.45) --
	(156.30,292.45) --
	(156.30,298.42) --
	(163.36,298.42) --
	(163.36,322.66) --
	(170.41,322.66) --
	(170.41,313.14) --
	(177.01,313.14) --
	(177.01,327.45) --
	(184.07,327.45) --
	(184.07,312.39) --
	(190.90,312.39) --
	(190.90,329.95) --
	(197.96,329.95) --
	(197.96,308.67) --
	(204.79,308.67) --
	(204.79,329.22) --
	(211.85,329.22) --
	(211.85,320.83) --
	(218.90,320.83) --
	(218.90,316.07) --
	(225.73,316.07) --
	(225.73,309.25) --
	(232.79,309.25) --
	(232.79,309.65) --
	(239.62,309.65) --
	(239.62,309.58) --
	(246.68,309.58) --
	(246.68,303.89) --
	(253.73,303.89) --
	(253.73,292.40) --
	(260.11,292.40) --
	(260.11,288.79) --
	(267.17,288.79) --
	(267.17,292.08) --
	(274.00,292.08) --
	(274.00,295.16) --
	(281.05,295.16) --
	(281.05,286.73) --
	(287.88,286.73) --
	(287.88,289.65) --
	(294.94,289.65) --
	(294.94,295.32) --
	(302.00,295.32) --
	(302.00,288.05) --
	(308.83,288.05) --
	(308.83,295.66) --
	(315.88,295.66) --
	(315.88,284.47) --
	(322.71,284.47) --
	(322.71,293.35) --
	(329.77,293.35) --
	(329.77,282.64) --
	(336.83,282.64) --
	(336.83,286.27);
\end{scope}
\begin{scope}
\path[clip] (  0.00,  0.00) rectangle (361.35,390.26);
\definecolor{drawColor}{gray}{0.30}

\node[text=drawColor,anchor=base east,inner sep=0pt, outer sep=0pt, scale=  0.88] at ( 47.90,264.55) {0};

\node[text=drawColor,anchor=base east,inner sep=0pt, outer sep=0pt, scale=  0.88] at ( 47.90,289.80) {5,000};

\node[text=drawColor,anchor=base east,inner sep=0pt, outer sep=0pt, scale=  0.88] at ( 47.90,315.05) {10,000};

\node[text=drawColor,anchor=base east,inner sep=0pt, outer sep=0pt, scale=  0.88] at ( 47.90,340.30) {15,000};

\node[text=drawColor,anchor=base east,inner sep=0pt, outer sep=0pt, scale=  0.88] at ( 47.90,365.55) {20,000};
\end{scope}
\begin{scope}
\path[clip] (  0.00,  0.00) rectangle (361.35,390.26);
\definecolor{drawColor}{gray}{0.30}

\node[text=drawColor,anchor=base,inner sep=0pt, outer sep=0pt, scale=  0.88] at ( 80.26,251.26) {2019};

\node[text=drawColor,anchor=base,inner sep=0pt, outer sep=0pt, scale=  0.88] at (163.36,251.26) {2020};

\node[text=drawColor,anchor=base,inner sep=0pt, outer sep=0pt, scale=  0.88] at (246.68,251.26) {2021};

\node[text=drawColor,anchor=base,inner sep=0pt, outer sep=0pt, scale=  0.88] at (329.77,251.26) {2022};
\end{scope}
\begin{scope}
\path[clip] (  0.00,  0.00) rectangle (361.35,390.26);
\definecolor{drawColor}{RGB}{0,0,0}

\node[text=drawColor,anchor=base,inner sep=0pt, outer sep=0pt, scale=  1.10] at (201.60,239.22) {Month};
\end{scope}
\begin{scope}
\path[clip] (  0.00,  0.00) rectangle (361.35,390.26);
\definecolor{drawColor}{RGB}{0,0,0}

\node[text=drawColor,rotate= 90.00,anchor=base,inner sep=0pt, outer sep=0pt, scale=  1.10] at ( 18.58,320.76) {Amount of fresh inputs in BTC};
\end{scope}
\begin{scope}
\path[clip] (  0.00,  0.00) rectangle (361.35,390.26);
\definecolor{drawColor}{RGB}{0,0,0}

\node[text=drawColor,anchor=base west,inner sep=0pt, outer sep=0pt, scale=  1.10] at (127.13,209.57) {Service};
\end{scope}
\begin{scope}
\path[clip] (  0.00,  0.00) rectangle (361.35,390.26);
\definecolor{drawColor}{RGB}{237,144,164}

\path[draw=drawColor,line width= 0.9pt,line join=round] (168.01,213.36) -- (179.57,213.36);
\end{scope}
\begin{scope}
\path[clip] (  0.00,  0.00) rectangle (361.35,390.26);
\definecolor{drawColor}{RGB}{0,193,178}

\path[draw=drawColor,line width= 0.9pt,line join=round] (229.66,213.36) -- (241.22,213.36);
\end{scope}
\begin{scope}
\path[clip] (  0.00,  0.00) rectangle (361.35,390.26);
\definecolor{drawColor}{RGB}{0,0,0}

\node[text=drawColor,anchor=base west,inner sep=0pt, outer sep=0pt, scale=  0.88] at (186.52,210.33) {Samourai};
\end{scope}
\begin{scope}
\path[clip] (  0.00,  0.00) rectangle (361.35,390.26);
\definecolor{drawColor}{RGB}{0,0,0}

\node[text=drawColor,anchor=base west,inner sep=0pt, outer sep=0pt, scale=  0.88] at (248.17,210.33) {Wasabi};
\end{scope}
\begin{scope}
\path[clip] ( 52.85, 72.64) rectangle (350.35,189.63);
\definecolor{drawColor}{gray}{0.92}

\path[draw=drawColor,line width= 0.4pt,line join=round] ( 52.85, 95.50) --
	(350.35, 95.50);

\path[draw=drawColor,line width= 0.4pt,line join=round] ( 52.85,130.60) --
	(350.35,130.60);

\path[draw=drawColor,line width= 0.4pt,line join=round] ( 52.85,165.69) --
	(350.35,165.69);

\path[draw=drawColor,line width= 0.4pt,line join=round] ( 93.01, 72.64) --
	( 93.01,189.63);

\path[draw=drawColor,line width= 0.4pt,line join=round] (186.20, 72.64) --
	(186.20,189.63);

\path[draw=drawColor,line width= 0.4pt,line join=round] (279.26, 72.64) --
	(279.26,189.63);

\path[draw=drawColor,line width= 0.9pt,line join=round] ( 52.85, 77.96) --
	(350.35, 77.96);

\path[draw=drawColor,line width= 0.9pt,line join=round] ( 52.85,113.05) --
	(350.35,113.05);

\path[draw=drawColor,line width= 0.9pt,line join=round] ( 52.85,148.14) --
	(350.35,148.14);

\path[draw=drawColor,line width= 0.9pt,line join=round] ( 52.85,183.24) --
	(350.35,183.24);

\path[draw=drawColor,line width= 0.9pt,line join=round] (139.60, 72.64) --
	(139.60,189.63);

\path[draw=drawColor,line width= 0.9pt,line join=round] (232.79, 72.64) --
	(232.79,189.63);

\path[draw=drawColor,line width= 0.9pt,line join=round] (325.73, 72.64) --
	(325.73,189.63);
\definecolor{fillColor}{RGB}{237,144,164}

\path[fill=fillColor] (244.61, 94.47) rectangle (251.02, 94.70);

\path[fill=fillColor] (252.50, 93.93) rectangle (258.92, 94.12);

\path[fill=fillColor] (260.14, 99.14) rectangle (266.55, 99.44);

\path[fill=fillColor] (268.03,118.53) rectangle (274.45,118.78);

\path[fill=fillColor] (275.67,184.04) rectangle (282.09,184.31);

\path[fill=fillColor] (283.56,140.26) rectangle (289.98,140.54);

\path[fill=fillColor] (291.46,102.43) rectangle (297.87,102.71);

\path[fill=fillColor] (299.09,112.74) rectangle (305.51,113.11);

\path[fill=fillColor] (306.99,108.86) rectangle (313.40,109.20);

\path[fill=fillColor] (314.62, 93.50) rectangle (321.04, 93.82);

\path[fill=fillColor] (322.52, 97.89) rectangle (328.93, 98.26);

\path[fill=fillColor] (330.41, 95.93) rectangle (336.83, 96.34);
\definecolor{fillColor}{RGB}{171,177,80}

\path[fill=fillColor] ( 74.01, 79.79) rectangle ( 80.43, 79.90);

\path[fill=fillColor] ( 81.91, 79.24) rectangle ( 88.32, 79.57);

\path[fill=fillColor] ( 89.55, 81.98) rectangle ( 95.96, 82.58);

\path[fill=fillColor] ( 97.44, 81.01) rectangle (103.85, 81.49);

\path[fill=fillColor] (105.33, 81.53) rectangle (111.75, 82.18);

\path[fill=fillColor] (112.97, 83.25) rectangle (119.39, 84.33);

\path[fill=fillColor] (120.86, 82.40) rectangle (127.28, 83.01);

\path[fill=fillColor] (128.50, 80.90) rectangle (134.92, 81.61);

\path[fill=fillColor] (136.39, 84.62) rectangle (142.81, 85.50);

\path[fill=fillColor] (144.29, 86.01) rectangle (150.70, 87.00);

\path[fill=fillColor] (151.67, 96.12) rectangle (158.09,100.02);

\path[fill=fillColor] (159.56,101.85) rectangle (165.98,106.15);

\path[fill=fillColor] (167.20,104.55) rectangle (173.62,107.84);

\path[fill=fillColor] (175.10, 90.38) rectangle (181.51, 94.44);

\path[fill=fillColor] (182.73, 86.38) rectangle (189.15, 90.18);

\path[fill=fillColor] (190.63,108.55) rectangle (197.04,112.17);

\path[fill=fillColor] (198.52,103.53) rectangle (204.94,108.36);

\path[fill=fillColor] (206.16,102.11) rectangle (212.58,106.21);

\path[fill=fillColor] (214.05,103.71) rectangle (220.47,108.58);

\path[fill=fillColor] (221.69,113.17) rectangle (228.11,117.52);

\path[fill=fillColor] (229.58,115.17) rectangle (236.00,117.85);

\path[fill=fillColor] (237.48, 95.09) rectangle (243.89, 97.02);

\path[fill=fillColor] (244.61, 92.77) rectangle (251.02, 94.47);

\path[fill=fillColor] (252.50, 92.93) rectangle (258.92, 93.93);

\path[fill=fillColor] (260.14, 97.46) rectangle (266.55, 99.14);

\path[fill=fillColor] (268.03,117.26) rectangle (274.45,118.53);

\path[fill=fillColor] (275.67,180.93) rectangle (282.09,184.04);

\path[fill=fillColor] (283.56,137.88) rectangle (289.98,140.26);

\path[fill=fillColor] (291.46,100.39) rectangle (297.87,102.43);

\path[fill=fillColor] (299.09,110.47) rectangle (305.51,112.74);

\path[fill=fillColor] (306.99,103.27) rectangle (313.40,108.86);

\path[fill=fillColor] (314.62, 90.51) rectangle (321.04, 93.50);

\path[fill=fillColor] (322.52, 94.55) rectangle (328.93, 97.89);

\path[fill=fillColor] (330.41, 92.65) rectangle (336.83, 95.93);
\definecolor{fillColor}{RGB}{0,193,178}

\path[fill=fillColor] ( 66.38, 77.96) rectangle ( 72.79, 78.01);

\path[fill=fillColor] ( 74.01, 77.96) rectangle ( 80.43, 79.79);

\path[fill=fillColor] ( 81.91, 77.96) rectangle ( 88.32, 79.24);

\path[fill=fillColor] ( 89.55, 77.96) rectangle ( 95.96, 81.98);

\path[fill=fillColor] ( 97.44, 79.55) rectangle (103.85, 81.01);

\path[fill=fillColor] (105.33, 80.27) rectangle (111.75, 81.53);

\path[fill=fillColor] (112.97, 80.75) rectangle (119.39, 83.25);

\path[fill=fillColor] (120.86, 80.33) rectangle (127.28, 82.40);

\path[fill=fillColor] (128.50, 78.91) rectangle (134.92, 80.90);

\path[fill=fillColor] (136.39, 81.33) rectangle (142.81, 84.62);

\path[fill=fillColor] (144.29, 82.20) rectangle (150.70, 86.01);

\path[fill=fillColor] (151.67, 86.12) rectangle (158.09, 96.12);

\path[fill=fillColor] (159.56, 94.50) rectangle (165.98,101.85);

\path[fill=fillColor] (167.20, 99.17) rectangle (173.62,104.55);

\path[fill=fillColor] (175.10, 84.91) rectangle (181.51, 90.38);

\path[fill=fillColor] (182.73, 81.94) rectangle (189.15, 86.38);

\path[fill=fillColor] (190.63, 98.77) rectangle (197.04,108.55);

\path[fill=fillColor] (198.52, 93.73) rectangle (204.94,103.53);

\path[fill=fillColor] (206.16, 90.86) rectangle (212.58,102.11);

\path[fill=fillColor] (214.05, 88.78) rectangle (220.47,103.71);

\path[fill=fillColor] (221.69, 98.14) rectangle (228.11,113.17);

\path[fill=fillColor] (229.58,107.16) rectangle (236.00,115.17);

\path[fill=fillColor] (237.48, 89.63) rectangle (243.89, 95.09);

\path[fill=fillColor] (244.61, 88.29) rectangle (251.02, 92.77);

\path[fill=fillColor] (252.50, 88.35) rectangle (258.92, 92.93);

\path[fill=fillColor] (260.14, 92.30) rectangle (266.55, 97.46);

\path[fill=fillColor] (268.03,111.67) rectangle (274.45,117.26);

\path[fill=fillColor] (275.67,173.64) rectangle (282.09,180.93);

\path[fill=fillColor] (283.56,132.72) rectangle (289.98,137.88);

\path[fill=fillColor] (291.46, 95.68) rectangle (297.87,100.39);

\path[fill=fillColor] (299.09,104.70) rectangle (305.51,110.47);

\path[fill=fillColor] (306.99, 98.07) rectangle (313.40,103.27);

\path[fill=fillColor] (314.62, 87.05) rectangle (321.04, 90.51);

\path[fill=fillColor] (322.52, 88.06) rectangle (328.93, 94.55);

\path[fill=fillColor] (330.41, 86.96) rectangle (336.83, 92.65);
\definecolor{fillColor}{RGB}{172,162,236}

\path[fill=fillColor] ( 97.44, 77.96) rectangle (103.85, 79.55);

\path[fill=fillColor] (105.33, 77.96) rectangle (111.75, 80.27);

\path[fill=fillColor] (112.97, 77.96) rectangle (119.39, 80.75);

\path[fill=fillColor] (120.86, 77.96) rectangle (127.28, 80.33);

\path[fill=fillColor] (128.50, 77.96) rectangle (134.92, 78.91);

\path[fill=fillColor] (136.39, 77.96) rectangle (142.81, 81.33);

\path[fill=fillColor] (144.29, 77.96) rectangle (150.70, 82.20);

\path[fill=fillColor] (151.67, 77.96) rectangle (158.09, 86.12);

\path[fill=fillColor] (159.56, 77.96) rectangle (165.98, 94.50);

\path[fill=fillColor] (167.20, 77.96) rectangle (173.62, 99.17);

\path[fill=fillColor] (175.10, 77.96) rectangle (181.51, 84.91);

\path[fill=fillColor] (182.73, 77.96) rectangle (189.15, 81.94);

\path[fill=fillColor] (190.63, 77.96) rectangle (197.04, 98.77);

\path[fill=fillColor] (198.52, 77.96) rectangle (204.94, 93.73);

\path[fill=fillColor] (206.16, 77.96) rectangle (212.58, 90.86);

\path[fill=fillColor] (214.05, 77.96) rectangle (220.47, 88.78);

\path[fill=fillColor] (221.69, 77.96) rectangle (228.11, 98.14);

\path[fill=fillColor] (229.58, 77.96) rectangle (236.00,107.16);

\path[fill=fillColor] (237.48, 77.96) rectangle (243.89, 89.63);

\path[fill=fillColor] (244.61, 77.96) rectangle (251.02, 88.29);

\path[fill=fillColor] (252.50, 77.96) rectangle (258.92, 88.35);

\path[fill=fillColor] (260.14, 77.96) rectangle (266.55, 92.30);

\path[fill=fillColor] (268.03, 77.96) rectangle (274.45,111.67);

\path[fill=fillColor] (275.67, 77.96) rectangle (282.09,173.64);

\path[fill=fillColor] (283.56, 77.96) rectangle (289.98,132.72);

\path[fill=fillColor] (291.46, 77.96) rectangle (297.87, 95.68);

\path[fill=fillColor] (299.09, 77.96) rectangle (305.51,104.70);

\path[fill=fillColor] (306.99, 77.96) rectangle (313.40, 98.07);

\path[fill=fillColor] (314.62, 77.96) rectangle (321.04, 87.05);

\path[fill=fillColor] (322.52, 77.96) rectangle (328.93, 88.06);

\path[fill=fillColor] (330.41, 77.96) rectangle (336.83, 86.96);
\end{scope}
\begin{scope}
\path[clip] (  0.00,  0.00) rectangle (361.35,390.26);
\definecolor{drawColor}{gray}{0.30}

\node[text=drawColor,anchor=base east,inner sep=0pt, outer sep=0pt, scale=  0.88] at ( 47.90, 74.93) {0};

\node[text=drawColor,anchor=base east,inner sep=0pt, outer sep=0pt, scale=  0.88] at ( 47.90,110.02) {1,000};

\node[text=drawColor,anchor=base east,inner sep=0pt, outer sep=0pt, scale=  0.88] at ( 47.90,145.11) {2,000};

\node[text=drawColor,anchor=base east,inner sep=0pt, outer sep=0pt, scale=  0.88] at ( 47.90,180.20) {3,000};
\end{scope}
\begin{scope}
\path[clip] (  0.00,  0.00) rectangle (361.35,390.26);
\definecolor{drawColor}{gray}{0.30}

\node[text=drawColor,anchor=base,inner sep=0pt, outer sep=0pt, scale=  0.88] at (139.60, 61.63) {2020};

\node[text=drawColor,anchor=base,inner sep=0pt, outer sep=0pt, scale=  0.88] at (232.79, 61.63) {2021};

\node[text=drawColor,anchor=base,inner sep=0pt, outer sep=0pt, scale=  0.88] at (325.73, 61.63) {2022};
\end{scope}
\begin{scope}
\path[clip] (  0.00,  0.00) rectangle (361.35,390.26);
\definecolor{drawColor}{RGB}{0,0,0}

\node[text=drawColor,anchor=base,inner sep=0pt, outer sep=0pt, scale=  1.10] at (201.60, 49.59) {Month};
\end{scope}
\begin{scope}
\path[clip] (  0.00,  0.00) rectangle (361.35,390.26);
\definecolor{drawColor}{RGB}{0,0,0}

\node[text=drawColor,rotate= 90.00,anchor=base,inner sep=0pt, outer sep=0pt, scale=  1.10] at ( 18.58,131.13) {Amount of fresh inputs in BTC};
\end{scope}
\begin{scope}
\path[clip] (  0.00,  0.00) rectangle (361.35,390.26);
\definecolor{drawColor}{RGB}{0,0,0}

\node[text=drawColor,anchor=base west,inner sep=0pt, outer sep=0pt, scale=  1.10] at ( 84.57, 19.94) {Samourai pool};
\end{scope}
\begin{scope}
\path[clip] (  0.00,  0.00) rectangle (361.35,390.26);
\definecolor{fillColor}{RGB}{237,144,164}

\path[fill=fillColor] (160.46, 17.21) rectangle (173.50, 30.24);
\end{scope}
\begin{scope}
\path[clip] (  0.00,  0.00) rectangle (361.35,390.26);
\definecolor{fillColor}{RGB}{171,177,80}

\path[fill=fillColor] (205.96, 17.21) rectangle (218.99, 30.24);
\end{scope}
\begin{scope}
\path[clip] (  0.00,  0.00) rectangle (361.35,390.26);
\definecolor{fillColor}{RGB}{0,193,178}

\path[fill=fillColor] (247.05, 17.21) rectangle (260.08, 30.24);
\end{scope}
\begin{scope}
\path[clip] (  0.00,  0.00) rectangle (361.35,390.26);
\definecolor{fillColor}{RGB}{172,162,236}

\path[fill=fillColor] (288.15, 17.21) rectangle (301.18, 30.24);
\end{scope}
\begin{scope}
\path[clip] (  0.00,  0.00) rectangle (361.35,390.26);
\definecolor{drawColor}{RGB}{0,0,0}

\node[text=drawColor,anchor=base west,inner sep=0pt, outer sep=0pt, scale=  0.88] at (179.71, 20.70) {0.001};
\end{scope}
\begin{scope}
\path[clip] (  0.00,  0.00) rectangle (361.35,390.26);
\definecolor{drawColor}{RGB}{0,0,0}

\node[text=drawColor,anchor=base west,inner sep=0pt, outer sep=0pt, scale=  0.88] at (225.20, 20.70) {0.01};
\end{scope}
\begin{scope}
\path[clip] (  0.00,  0.00) rectangle (361.35,390.26);
\definecolor{drawColor}{RGB}{0,0,0}

\node[text=drawColor,anchor=base west,inner sep=0pt, outer sep=0pt, scale=  0.88] at (266.29, 20.70) {0.05};
\end{scope}
\begin{scope}
\path[clip] (  0.00,  0.00) rectangle (361.35,390.26);
\definecolor{drawColor}{RGB}{0,0,0}

\node[text=drawColor,anchor=base west,inner sep=0pt, outer sep=0pt, scale=  0.88] at (307.39, 20.70) {0.5};
\end{scope}
\end{tikzpicture}